\documentclass[useAMS,usenatbib]{mn2e}

\DeclareSymbolFont{cmletters}{OML}{cmm}{m}{it}
\DeclareMathSymbol{v}{\mathalpha}{cmletters}{"76}

\usepackage{amsmath}
\usepackage{amssymb}
\usepackage{epsfig}
\usepackage{graphicx}
\usepackage{ifthen}
\usepackage{latexsym}
\usepackage{rotating}
\usepackage{subfigure}
\usepackage{times,epsf}
\usepackage{txfonts}
\usepackage{varioref}
\usepackage{verbatim}
\usepackage{url}
\usepackage{color}
\usepackage{array}
\usepackage[dvipsnames]{xcolor}
\usepackage[T1]{fontenc}

\newcolumntype{M}{>{$\vcenter\bgroup\hbox\bgroup}c<{\egroup\egroup$}}

\newcommand{\be}{\begin{equation}}
\newcommand{\ee}{\end{equation}}
\newcommand{\bea}{\begin{eqnarray}}
\newcommand{\eea}{\end{eqnarray}}

\title[Outflows from accreting BHs]{Energy, momentum and mass outflows
  and feedback from thick accretion discs around rotating black holes}
\author[A. S\k{a}dowski, R. Narayan, R. Penna, Y. Zhu] {Aleksander
  S\k{a}dowski$^1$\footnotemark[1], Ramesh
  Narayan$^{1}$\footnotemark[1], Robert Penna$^{1}$\footnotemark[1],
  Yucong Zhu$^{1}$\thanks{E-mail: asadowski@cfa.harvard.edu (AS);
    rnarayan@cfa.harvard.edu (RN); rpenna@cfa.harvard.edu (RP);
    yzhu@cfa.harvard.edu (YZ)} \\ $^1$ Harvard-Smithsonian Center for
  Astrophysics, 60 Garden St., Cambridge, MA 02134, USA }

\begin{document}

\maketitle

\label{firstpage}

\begin{abstract}

Using long-duration general relativistic magnetohydrodynamic
simulations of radiatively inefficient accretion discs, the energy,
momentum and mass outflow rates from such systems are estimated.
Outflows occur via two fairly distinct modes: a relativistic jet and a
sub-relativistic wind. The jet power depends strongly on the black
hole spin and on the magnetic flux at the horizon.  Unless these are
very small, the energy output in the jet dominates over that in the
wind.  For a rapidly spinning black hole accreting in the magnetically
arrested limit, it is confirmed that jet power exceeds the total rate
of accretion of rest mass energy. However, because of strong
collimation, the jet probably does not have a significant feedback
effect on its immediate surroundings.  The power in the wind is more
modest and shows a weaker dependence on black hole spin and magnetic
flux.  Nevertheless, because the wind subtends a large solid angle, it
is expected to provide efficient feedback on a wide range of scales
inside the host galaxy.  Empirical formulae are obtained for the
energy and momentum outflow rates in the jet and the wind.

\end{abstract}


\section{Introduction}
\label{introduction}

\subsection{Feedback}

Black hole (BH) accretion discs are some of the most energetic objects
in the Universe \citep{frank-book, kato-book}.  Geometrically thin
discs \citep{ss73,nov73} are radiatively efficient and convert about
10\% of the rest mass energy of the accreting gas into
radiation. Geometrically thick discs, on the other hand, are advection
dominated accretion flows \citep[ADAFs, ][]{nar94,nar95a,abramowicz-adafs,nm08} and
produce little radiation relative to their mass accretion
rates. Instead, they produce outflows in the form of jets and winds
which carry huge amounts of energy, mass and momentum. This is the
topic of the present paper.

Energy and momentum outflow, and to a lesser extent mass outflow, can
affect the BH's surroundings. This effect is most profound in the case
of supermassive BHs (SMBHs) in the centers of galaxies, where a number
of observations suggest the existence of strong``feedback'' effects
from the SMBH accretion disc on the evolution of the entire host
galaxy. The best-known evidence for such coupling is the celebrated
$M_{\rm BH}$-$\sigma_{\rm bulge}$ relation between the mass $M_{\rm
  BH}$ of the SMBH and the velocity dispersion $\sigma_{\rm bulge}$ of
its host galaxy bulge \citep{FerrareseMerritt00, GebhardtEtAl00,
  Gultekin09,kormendyho-13}, and the earlier \citet{MagorrianEtAl98} relation
between $M_{\rm BH}$ and the bulge mass $M_{\rm bulge}$.  Given the
very large size ratio $\sim10^8$ and mass ratio $\sim10^3$ between the
galaxy bulge and the SMBH, it is reasonable to assume that the
coupling occurs via feedback of energy or momentum
\citep[eg.,][]{SilkRees98, King03, king-05, King10, HopkinsEtAl09}.

A second line of evidence is the observed exponential cutoff in the
number density of galaxies at the high mass/luminosity end
\citep{Schechter76}, even though there is no cutoff at the equivalent mass
scale in the distribution of dark matter halos.  One of the mechanisms
invoked to explain the galaxy cutoff is a reduction in the star
formation rate in massive systems due to inefficient cooling
\citep{WhiteRees78, WhiteFrenk91}. However, cooling effects alone are
insufficient \citep{ThoulWeinberg95}, and attempts to fit observed
luminosity functions need to include additional feedback
processes \citep[e.g.,][]{BensonEtAl03, CrotonEtAl06}.  The most
viable process is expulsion of gas from young galaxies in superwinds
as a result of feedback from supernovae and/or agtive galactic nuclei (AGN).  Observations of
starburst galaxies \citep[e.g.,][]{McNamaraEtAl06}, and midly-relativistic 
winds in AGN \citep{tombesi+10a,tombesi+10b} confirm this
picture.

Yet another puzzling aspect of the galaxy population is the fact that
the most massive galaxies, typically ellipticals in clusters, are made
of the oldest stars \citep{BenderSaglia99}.  This ``downsizing'' is
counterintuitive, since it seems to conflict with hierarchical growth
of structure in a CDM cosmogony, where massive dark halos assemble at
lower redshift than lower mass halos
\citep[e.g.,][]{LaceyCole93}. Again, feedback provides a plausible
mechanism; it prevents significant accretion in massive galaxies, thus
suppressing star formation at late times \citep{BensonEtAl03,
  Springel+05}.

Finally, the observed X-ray emission in the centers of galaxy clusters
implies a cooling time much shorter than the age of the system,
suggesting that gas at the centers of these clusters must condense and
turn into stars; however, there is no observational evidence for star
formation at the required level \citep{FabianEtAl01, PetersonEtAl03,
  KaastraEtAl04, CrotonEtAl06}.  While thermal conduction may be part
of the explanation \citep[e.g.,][]{narayanmedvedev-01}, an important clue
comes from the observation \citep{BurnsGregoryHolman81} that every
cluster with a strong cooling flow also contains an active SMBH in a
central radio galaxy.  This suggests that energy feedback from the
SMBH keeps the cluster gas hot \citep{CiottiOstriker01,
  BruggenKaiser02, RuszkowskiBegelman02, ChurazovEtAl04}.

The evidence summarized above indicates that feedback from accreting
SMBHs plays a crucial role not only on galaxy scales but even on the
scale of galaxy clusters.  Two kinds of SMBH feedback are discussed in
the literature (see \citealt{Fabian12}, \citealt{kormendyho-13} for reviews).  One occurs in
the ``quasar mode'' when the SMBH accretes at a good fraction
($\sim0.1$ to $>1$) of the Eddington rate and deposits energy in its
surroundings either directly through radiation or via radiatively
driven winds. The second kind of feedback takes place in the ``radio mode''
\citep{CrotonEtAl06} or ``maintenance mode'' \citep{Hopkins10}. Here,
accretion occurs via an ADAF and the radiative luminosity is
low. Hence, feedback is almost entirely in the form of mechanical
energy and momentum. This ADAF-specific form of feedback is the topic
we wish to investigate.

Maintenance mode feedback has been (i) invoked for the ``cooling flow
problem'' in galaxy clusters \citep{CiottiOstriker01, BruggenKaiser02,
  RuszkowskiBegelman02, ChurazovEtAl04, gaspari+11}, (ii) included in
semi-empirical models of galaxy formation \citep{CrotonEtAl06,
  BestEtAl05, HopkinsEtAl06, SomervilleEtAl08}, and (iii) modeled via
simple prescriptions in gas dynamical computer simulations of
galaxy/cluster formation in the universe \citep{DiMatteo+05, Springel+05,
  CoxEtAl06, ciotti2010, novak+11, scannapieco2012}. However, all these efforts
are highly empirical since nobody knows exactly how much mechanical
energy or momentum flows out from an accreting SMBH. The general
practice is to employ the Bondi model \citep{Springel+05}, or some
variant of it \citep{DeBuhr2010}, to relate the energy or momentum
output from a SMBH to boundary conditions in the surrounding ISM.
However, whether or not the Bondi model is a reasonable description of
accretion from an external medium is still very much in debate (see,
e.g., \citealt{igu02,NF11}).
Recently, \cite{gaspari+13} carried out an in-depth investigation 
and showed that
cold gas condenses out of the hot phase via
  nonlinear thermal instabilities. As a result the cold filaments/clouds
  collide inelastically and boost the accretion rate, making the
  Bondi model very unrealistic.

A final important question is the following: Which is more important,
energy feedback or momentum feedback?  The former is traditionally
considered in cosmological simulations
\citep[e.g.,][]{Springel+05,DiMatteo+05}, but the latter may also be
important \citep{King03, King10, DeBuhr2010, ostriker+10}. For a given energy flux,
the momentum flux is smallest in the case of a relativistic jet and
largest for a non-relativistic wind. Thus the dependence of momentum
feedback efficiency on parameters such as the mass accretion rate or the BH spin could be quite different compared to
energy feedback efficiency. In this work we estimate both efficiency
factors for ADAFs using numerical simulations.

\subsection{Blandford-Znajek}
\label{s.bz}

It is a remarkable prediction of general relativity that magnetic field
lines threading a black hole can extract the hole's rotational energy
\citep{1975PhRvD..12.2959R}.
Rapidly rotating black holes can drive powerful jets.  In the standard
black hole jet model, the jet power scales as 
\begin{equation}
\label{eq.omhor}
P_{\rm jet}\sim \Omega_{\rm H}^2
\Phi_{\rm BH}^2\,, \qquad
\Omega_{\rm H}=a_*/2r_{\rm H}\,, \qquad
r_{\rm
  H}=1+\sqrt{1-a_*^2}\,,
\end{equation}
where $\Omega_{\rm H}$ is the angular velocity of the outer BH horizon
with radius $r_{\rm H}$, and $\Phi_{\rm BH}$ is the magnetic flux
threading the horizon \citep{blandfordznajek,
  macdonald_thorne_bh_forcefree_1982, Phinney82, tpm86}.  The main predictions of
the Blandford-Znajek jet model are supported by GRMHD simulations
\citep{kom01, hirose04, dhkh05, 2005ApJ...630L...5M, hk06,
  2008ApJ...678.1180B, 2010ApJ...711...50T, tchekh+11,
  Tchekhovskoy+12b}.  The Blandford-Znajek (BZ) model is a close
cousin of the \citet{gol69} model for pulsar magnetospheres, a
relationship which becomes particularly clear in the membrane
formulation of the BZ model
\citep{macdonald_thorne_bh_forcefree_1982,tpm86}.  Recently it has
been observed that the scaling of jet power with black hole spin in
galactic X-ray binaries is consistent with the BZ model
\citep{2012MNRAS.419L..69N,2013ApJ...762..104S}.

One of the aims of the present paper is to check how well jets and
winds in simulated ADAFs agree with the scaling shown in equation (\ref{eq.omhor}).

\subsection{Previous work}

Outflows of mass and energy are multi-dimensional and are best studied
with numerical simulations. Multi-dimensional numerical hydro- and
magnetohydro-dynamical simulations of hot accretion discs have been
performed for more than a decade.  Already early works based on
pseudo-Newtonian codes with purely hydrodynamic visosity showed that a
significant fraction of the inflowing mass near the equatorial plane
can flow out along the poles \citep{SPB99,IA99,IA00}. 
\citet*[][see also earlier work by \citealt{progabegelman-03}, and \citealt{janiuk+09}]{LOS12} 
ran a set of
hydrodynamical axisymmetric simulations of low-angular momentum gas.
For their viscous models they observed conical
outflows almost balancing inflow.

Pseudo-Newtonian
MHD simulations have been then performed by a number of authors
\citep{Machida+00,Machida+01,sto01,haw02,igu03}. Outflows were observed
and it was claimed that the initial configuration of the magnetic
field may play an important role in determining the mass outflow rate.
On the contrary, in a series of numerical MHD simulations,
\citet{Pen+03} and \citet{Pang+11} found little evidence for either
outflows or convection.  Even though the entropy gradient was unstable
the gas was apparently prevented from becoming convective by the
magnetic field. Recently, \cite{YWB12a} and \cite{YWB12b} carried out 2D
hydrodynamical and MHD simulations of ADAFs which cover a very large range of
radius and show fairly strong outflows.

Beginning with the work of \cite{dev03}, accretion flows
have been studied using general relativistic magneto-hydrodynamic
(GRMHD) codes. The authors observed two kinds of outflows:
bipolar unbound jets and bound coronal flow.  The coronal flow
supplied gas and magnetic field to the coronal envelope, but
apparently did not have sufficient energy to escape to infinity.  The
jets on the other hand were relativistic and escaped easily, though
carrying very little mass. Jets have been studied in detail by a
number of authors \citep{MG04,dhkh05,McKinney06}.
\cite{tchekh+11} simulated a strongly magnetized disc around a rapidly
spinning BH, and obtained very powerful jets with energy efficiency
$\eta>100\%$, i.e., jet power greater than 100\% of $\dot{M}_{\rm
  BH}c^2$, where $\dot{M}_{\rm BH}$ is the mass accretion rate on to
the BH. Their work showed beyond doubt that at least some part of the
jet power had to be extracted from the spin energy of the BH.  

Recently, \cite{MTB+13} have studied a large set of thick accretion disc
simulations both for rotating and non-rotating BHs. They found that 
models initiated with poloidal magnetic field
showed mass loss both in the jet and a magnetized wind.
The energy reaching large radii was dominated by the power produced 
via the magnetic flux penetrating the horizon as in the BZ mechanism.

\subsection{This work}

The present paper attempts to obtain via global general relativistic
magnetohydrodynamic (GRMHD) simulations quantitative estimates of the
amount of energy, mass and momentum that flow out from an ADAF around
a BH.  In a previous paper \citep{narayan+12a}, we studied mass
outflow from an ADAF around a non-rotating BH, and showed that a wind
flows out at relatively large radii, $r\gtrsim 40$\footnote{All
  distances are given in units of $R_G=GM_{\rm BH}/c^2$, where
  $M_{\rm BH}$ is the BH mass. Thus, the dimensionless radius $r$ is
  related to the dimensional radius $R$ by $r\equiv R/R_G$. Time is
  similarly given in units of $GM_{\rm BH}/c^3$. We adopt $G=c=1$ and choose the
  $(-+++)$ signature for the metric.}. Here we extend our analysis to
accretion flows around {\it rotating} BHs. In the process we show that
there are two kinds of outflows in such systems: (i) relatively slow
winds at larger radii, similar to the winds studied in
\citet{narayan+12a}, and (ii) relativistic jets which flow out from
close to the BH and are uniquely associated with spinning holes. We
study in detail the energy, mass and momentum in these two kinds of
outflow.  In an accompanying paper \citep{penna+13}, we discuss the
physics of relativistic jets and demonstrate that our numerical
simulations strongly validate the \citet{blandfordznajek} mechanism.

The paper is organized as follows. In Section~\ref{s.method} we
introduce the numerical scheme we used to simulate the discs. In
Section~\ref{s.results} we discuss outflows emerging from them.  In
particular, in Sections \ref{s.energy}, \ref{s.massloss} and
\ref{s.momentum} we present radial profiles of outflowing energy, mass
and momentum, respectively, and in Section~\ref{s.fits} we give approximate formulae
for the corresponding fluxes.  In Section~\ref{s.thin} we compare the
efficiencies of generating outflows by thick and thin discs.  We
conclude with Section~\ref{s.discussion} discussing implications of
our results.

\section{Numerical setup}
\label{s.method}

We performed seven simulations of radiatively inefficient accretion
flows around spinning and non-spinning BHs, as listed in
Table~\ref{t.models}.  Following the methods described in
\citet{narayan+12a}, two distinct initial conditions were used for the
seed magnetic field threading the initial gas torus: (i) In one set of
simulations the seed field consisted of multiple poloidal loops of
magnetic field with changing polarity (multi-loop or SANE, which stands for ``Standard
And Normal Evolution''). (ii) In the other set of simulations the
initial field was a single poloidal loop threading the entire torus
(single-loop or MAD, ``Magnetically Arrested Disc'').  SANE runs are
designed such that relatively little magnetic flux accumulates around
the BH. In contrast, MAD runs quickly saturate at the maximum allowed
magnetic flux on the BH for the given mass accretion rate; the
back-reaction of the saturated field causes the accretion flow to be
magnetically arrested and to settle down to the MAD state
\citep{narayan+mad}.

All the simulations were performed using the GRMHD code \texttt{HARM}
\citep{gammie+harm}. The coordinates, initial torus, and magnetic
field were set up following \cite{narayan+12a}. We used modified
Kerr-Shield horizon penetrating coordinates which covered uniformly the
full range of azimuthal and polar angles.  The radial size of cells increased
exponentially with radius, with the innermost radius chosen
to fit six cells inside the BH horizon. The outermost radius was set
to $r=10000$. The adopted resolutions are given in
Table~\ref{t.models}. The initial torus of gas \citep[set up
  following][]{pennatorus+13} was threaded either by multiple and 
counter-orientated, or single poloidal magnetic field loops for SANE
and MAD models, respectively.

Two of the simulations discussed
here have BH spin $a/M\equiv a_*=0$, and are the same as the ones
described in \cite{narayan+12a}, except that the $a_*=0$ MAD model has
been run up to a time $t_{\rm final} = 200000$ (to match the
corresponding SANE run), which is twice as long as in the previous
work. The runs with spinning BHs ($a_* = 0.7$, 0.9, 0.98) are all new.

To validate if the magneto-rotational instability (MRI) is resolved
we calculate the parameters,
\begin{equation}
Q_{\hat{\theta}} = \frac{2 \pi}{\Omega dx^{\hat{\theta}}}
          \frac{|b^{\hat{\theta}}|}{\sqrt{4\pi \rho}}, \quad 
Q_{\hat{\phi}} = \frac{2 \pi}{\Omega dx^{\hat{\phi}}}
          \frac{|b^{\hat{\phi}}|}{\sqrt{4\pi \rho}},
\end{equation}
where $dx^{\hat{i}}$ (grid cell size) and $b^{\hat{i}}$ (the magnetic field) 
are evaluated in the orthonormal fluid frame, $\Omega$ is the angular velocity,
and $\rho$ is the gas density. For SANE runs, 
the gas inside
$r=100$ and within one density scale height of the midplane has
$Q_{\hat{\theta}}$ and $Q_{\hat{\phi}}$ between $10-20$, for all of
the time chunks except the first one (see below for the definition of time chunks).  The MAD simulations have $Q_{\hat{\theta}}>100$ and
$Q_{\hat{\phi}}\sim 50$ in the same regions. Therefore, 
MRI is properly resolved in both cases \citep{HGK11}.

\begin{figure*}
  \centering
\subfigure{\includegraphics[width=.45\textwidth]{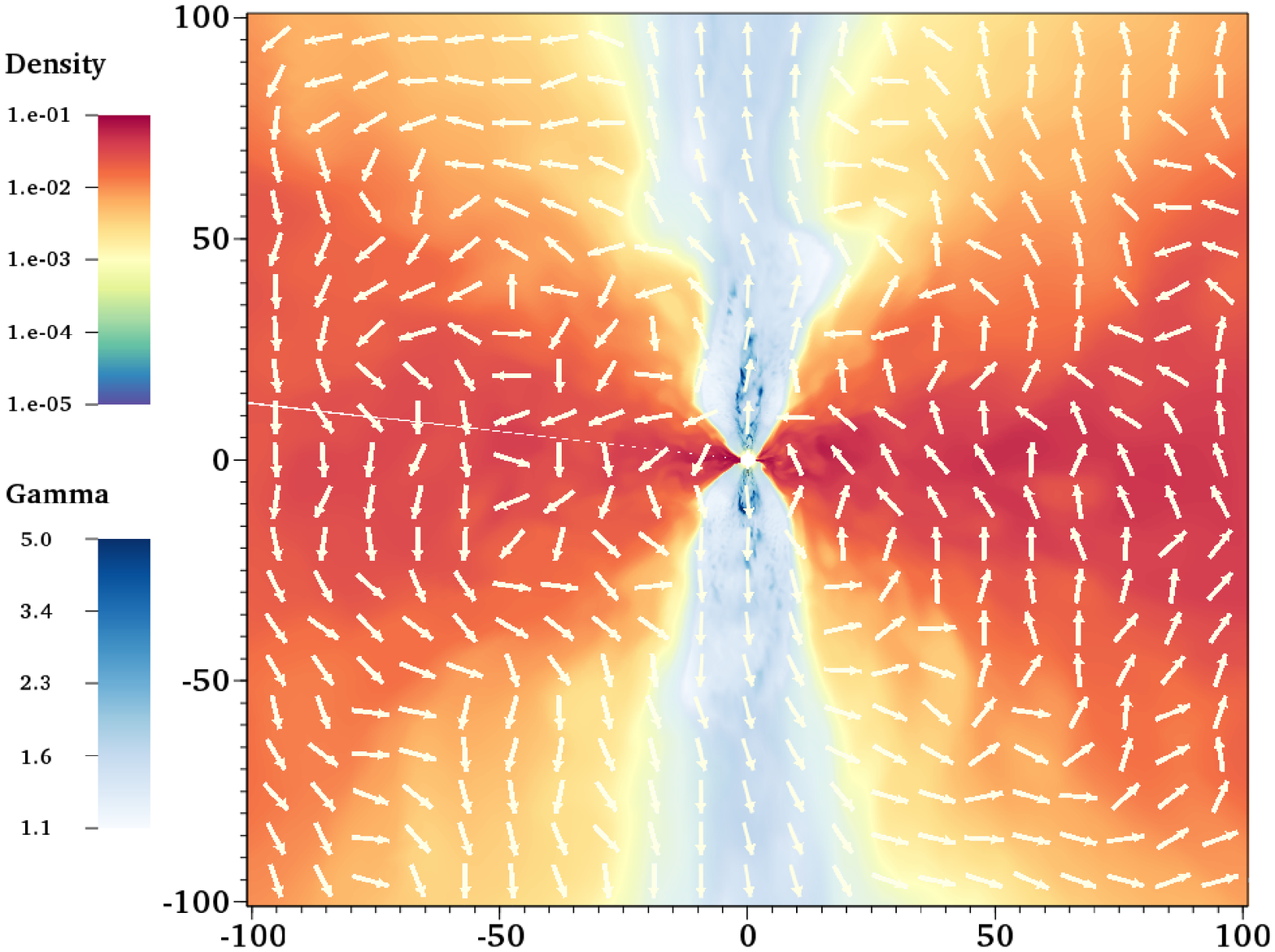}}
\subfigure{\includegraphics[width=.45\textwidth]{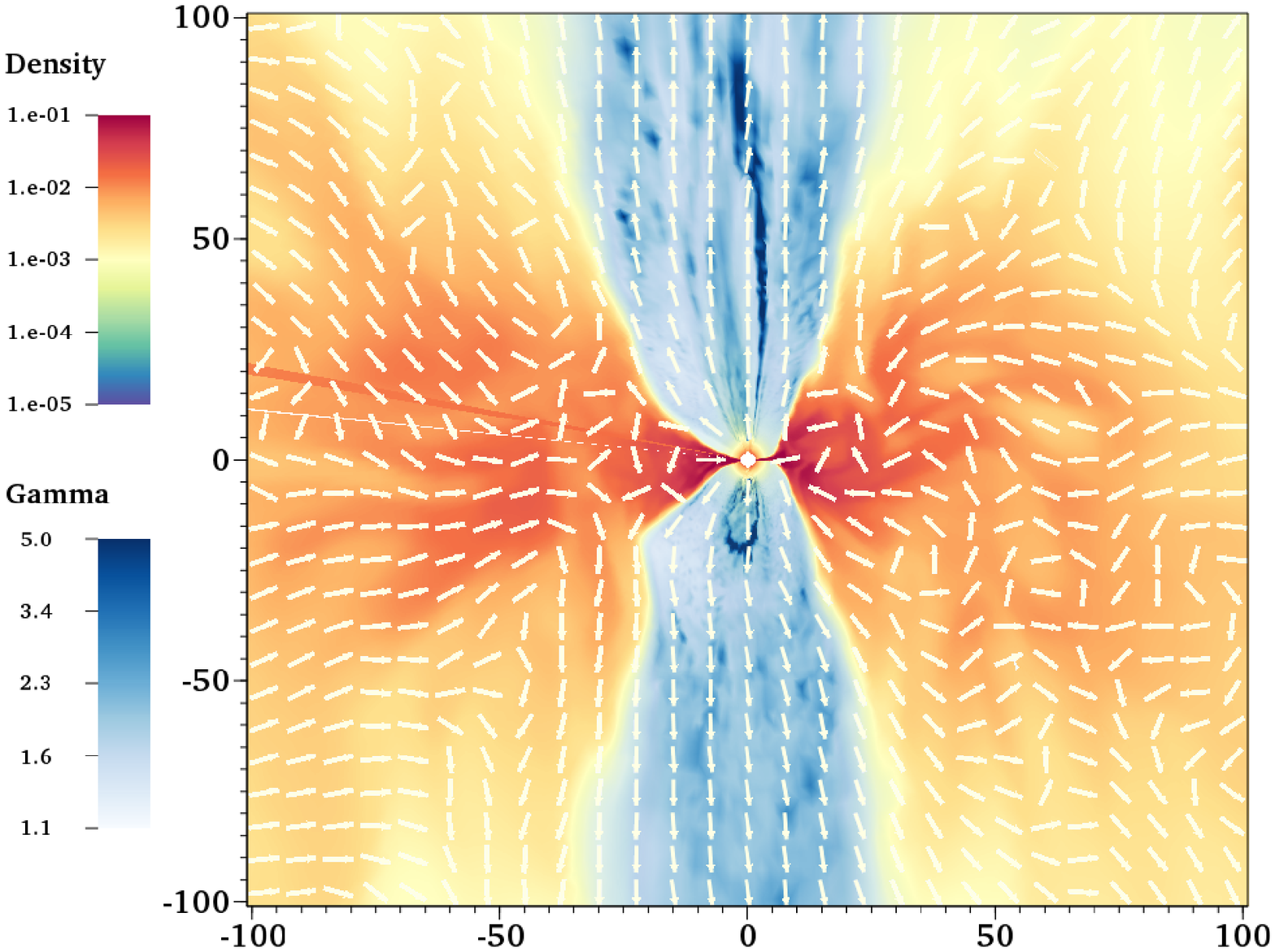}}
\caption{Snapshots of $a_*=0.7$ SANE (left) and MAD (right panel)
  models at $t=90000$. Reddish and blueish colors show the density and
  the Lorentz gamma factor $u^t$, respectively, in the poloidal
  ($r,\theta$) plane for an arbitrary value of the azimuthal angle
  $\phi$. Arrows show the direction of gas velocity in that plane.} 
  \label{f.dengam}
\end{figure*}

Figure~\ref{f.dengam} shows snapshots of density and Lorentz factor
$u^t$ for the $a_*=0.7$ SANE and MAD runs. In both models, the disc is
geometrically thick (root mean square $h/r\approx0.4$) and
turbulent. However, gas escapes in two fairly distinct structures: a
fast collimated laminar flow along the axis, which we refer to as the
``jet'', and a slower outflow covering a wider range of angles, which
we call the ``wind''. We focus on these two components in the rest of
the paper. Compared to the SANE run, the outflowing jet velocity is
much higher in the MAD simulation, sometimes exceeding $u^t=5$
($v_{\rm jet}>0.98c$).

\begin{table}
\caption{Disc models }
\label{t.models}
\centering\begin{tabular}{@{}ccccc}
\hline
 Model & BH spin & Initial & Resolution & $t_{\rm final}$ \\
& ($a_*$) & magnetic field & ($r$,$\theta$,$\phi$) &\\
\hline
 $a_*=0.0$ SANE   & $0.0$    & multi-loop     & 256x128x64 & 200000\\
 $a_*=0.7$ SANE   & $0.7$    & multi-loop     & 256x128x64 & 100000\\ 
 $a_*=0.9$ SANE   & $0.9$    & multi-loop     & 256x128x64 & 50000\\ 
 $a_*=0.98$ SANE  & $0.98$   & multi-loop     & 256x128x64 & 25000\\ 
 $a_*=0.0$ MAD    & $0.0$    & single-loop     & 264x126x60 & 200000\\
 $a_*=0.7$ MAD    & $0.7$    & single-loop     & 264x126x60 & 100000\\ 
 $a_*=0.9$ MAD    & $0.9$    & single-loop     & 264x126x60 & 50000\\
\hline
\end{tabular}
 \end{table}

\section{Analysis of Simulation Output}
\label{s.results}

\subsection{Averaging}

As in our previous work \citep{narayan+12a}, we used time-averaged
disc properties to extract radial profiles of quantities of
interest. The time-averaging was done over logarithmically increasing
chunks of time as listed in Table~\ref{t.chunks}, with each successive
time chunk being twice as long as the previous one. The two $a_*=0$
simulations were run up to $t_{\rm final}=200000$, allowing us to
compute time averages for chunks T1--T6.  The shorter $a_*=0.7$ runs
only went up to chunk T5, the $a_*=0.9$ runs reached chunk T4, while
the $a_*=0.98$ run stopped at chunk T3.\footnote{The physical wall
  time of all the simulations was comparable because of the smaller
  horizon radius of spinning BHs, which required a correspondingly
  smaller time step.}  Apart from averaging over time, we also
averaged the data over azimuth, and symmetrized it with respect to the
equatorial plane.  

\begin{table}
\begin{minipage}{\columnwidth}
\caption{Time chunks}
\centering\begin{tabular}{ccc}

\hline
Chunk & Time Range  & $t_{\rm chunk}$ \\ 
\hline
T1 & 3000--6000 & 3000  \\
T2 & 6000--12000 & 6000  \\
T3 & 12000--25000 & 13000  \\
T4 & 25000--50000 & 25000  \\
T5 & 50000--100000 & 50000 \\
T6 & 100000--200000 & 100000  \\
\hline

\end{tabular}
\label{t.chunks}
\end{minipage}
\end{table}

\subsection{Quantites of interest}
\label{s.quantities}

We are interested in estimating the amount of mass, energy and
momentum flowing out of the accretion disc. The radial flux of rest
mass is given by, 
\be
\dot m=\rho u^r,
\label{e.mdot}
\ee where a positive (negative)
sign indicates that matter flows away from (towards) the BH.  The
total energy flux is 
\be
\label{e.Trt}
\dot e_{\rm tot} = -T^r_t,
\ee 
where $T^r_t$ is
the ($r,t$) component of the magnetohydrodynamical stress energy
tensor \citep[e.g.,][]{mtw}, \be T_t^r = (\rho+\Gamma u +b^2)u^ru_t -
b^r b_t.
\ee and $b^\mu$ is the magnetic field four-vector
\citep[e.g.,][]{gammie+harm}. The negative sign in
equation~(\ref{e.Trt}) is because $u_t$ is negative; thus, a positive
value of $\dot{e}_{\rm tot}$ means that total energy flows outward,
and vice versa.  Note that $T^r_t$ represents the total energy
transported by the fluid and the magnetic field, including the rest
mass energy of the gas.  However, this is not very convenient when
considering energy at ``infinity'' since the rest mass energy plays no
role in feedback. Therefore, we consider a different measure of energy
flux in which we eliminate the rest mass energy, 
\be
\dot e = \dot e_{\rm tot}-\dot m=-T_t^r
- \rho u^r,
\label{e.edot}
\ee 
which we hereafter refer to as ``the energy flux''.
Positive values of $\dot e$ correspond to energy lost from the system,
i.e., energy flows out into the surrounding medium.


 Integrating over $\theta$ and $\phi$ and normalizing by
the net mass flow rate at $r=10$ (to avoid numerical issues near the
horizon) we obtain the normalized radial profiles of mass and energy
flow. The mass accretion rate is thus \be \dot{M}(r) = \frac{1}{|\dot
  M_{\rm net}|}\int_\theta\int_\phi \dot m \,dA_{\theta\phi},
\label{eq:Mdot}
\ee and the two energy loss rates are \be
\dot{E}_{\rm tot}(r) = \frac{1}{|\dot M_{\rm
    net}|}\int_\theta\int_\phi \dot e_{\rm tot} \,dA_{\theta\phi},
\label{eq:Etotdot}
\ee
\be
\dot{E}(r) = \frac{1}{|\dot M_{\rm net}|}\int_\theta\int_\phi \dot e \,dA_{\theta\phi},
\label{eq:Edot}
\ee where $dA_{\theta\phi}=\sqrt{-g}d\theta d\phi$ is an area element
in the $\theta$-$\phi$ plane, \be\dot M_{\rm net}=\int_\theta\int_\phi
\dot m(r=10) \,dA_{\theta\phi},\ee and signs have been chosen so as to
make the integrated fluxes positive for outflow. The integrands in the
above integrals correspond to time-averages over the duration of the
time chunk of interest. The $\phi$ integral is over the range 0 to
$2\pi$, while the range of the $\theta$ integral depends on the
quantity of interest. When we wish to calculate the net outflow of
mass or energy, we integrate over the full range $\theta =
0-\pi$. When we are interested in outflow in the jet or the wind, we
limit the $\theta$ range accordingly, as described in the next two
subsections.  From the integrated rate of outflow of mass and energy,
we calculate the integrated momentum in the outflow using the
relativistic formula, \be
\dot{P}(r) = \sqrt{(\dot{E}_{\rm tot}(r))^2 -  (\dot{M}(r))^2}.
\label{eq:Pdot}
\ee

Finally, we quantify the magnetic field strength at the BH horizon by
means of the magnetic flux parameter \citep{tchekh+11},
\begin{equation}
\Phi_{\rm BH}(t) = \frac{1}{2\sqrt{\dot{M}}} \int_\theta\int_\phi
|B^r(r_{\rm H},t)|\,dA_{\theta\phi},
\label{e.phibh}
\end{equation}
where $B^r$ is the radial component of the magnetic field and $r_{\rm
  H}$ is the radius of the horizon. The integral is over the whole
sphere, and the factor of $1/2$ converts the result to one hemisphere.
An accretion flow with geometrical thickness $h/r\approx 0.4$
transitions to the MAD state once $\Phi_{\rm BH}$ reaches $\sim50$
\citep{tchekh+11,Tchekhovskoy+12b}. As we show later, the three MAD runs
reach this limit quickly and remain there, whereas the four SANE runs
are for the most part well below this limit.

\subsection{The outflow criterion}
\label{s.mu}

Since the simulations extend over only a finite range of radius, we
need a criterion to decide whether a particular parcel of fluid
can escape to infinity. The quantity we use to determine this is the
Bernoulli parameter $\mu$ \citep{narayan+12a},
\begin{equation}
\mu = -\frac{T_t^r\rho u^r\sqrt{g_{rr}}+T_t^\theta\rho u^\theta\sqrt{g_{\theta\theta}}}
{(\rho u^r)^2g_{rr} + (\rho u^\theta)^2g_{\theta\theta}}-1. \label{eq:mu}
\end{equation}
To understand this expression, note that the quantities $\dot{e}$ and
$\dot{m}$ introduced earlier correspond to the radial components of
the respective fluxes. Correspondingly, there are $\theta$ components
of these fluxes, and the two together may be viewed as two-vectors
$\vec{\dot{e}}_p$ and $\vec{\dot{m}}_p$ in the poloidal $r\theta$
plane. We see then that $\mu$ is equal to $\vec{\dot{e}}_p \cdot
\vec{\dot{m}}_p/\vec{\dot{m}}_p \cdot \vec{\dot{m}}_p$, i.e., it is
the flux of energy parallel to the flow streamline. This quantity has
to be positive for gas to be able to escape to infinity.

\begin{figure}
  \centering
\includegraphics[height=.95\columnwidth]{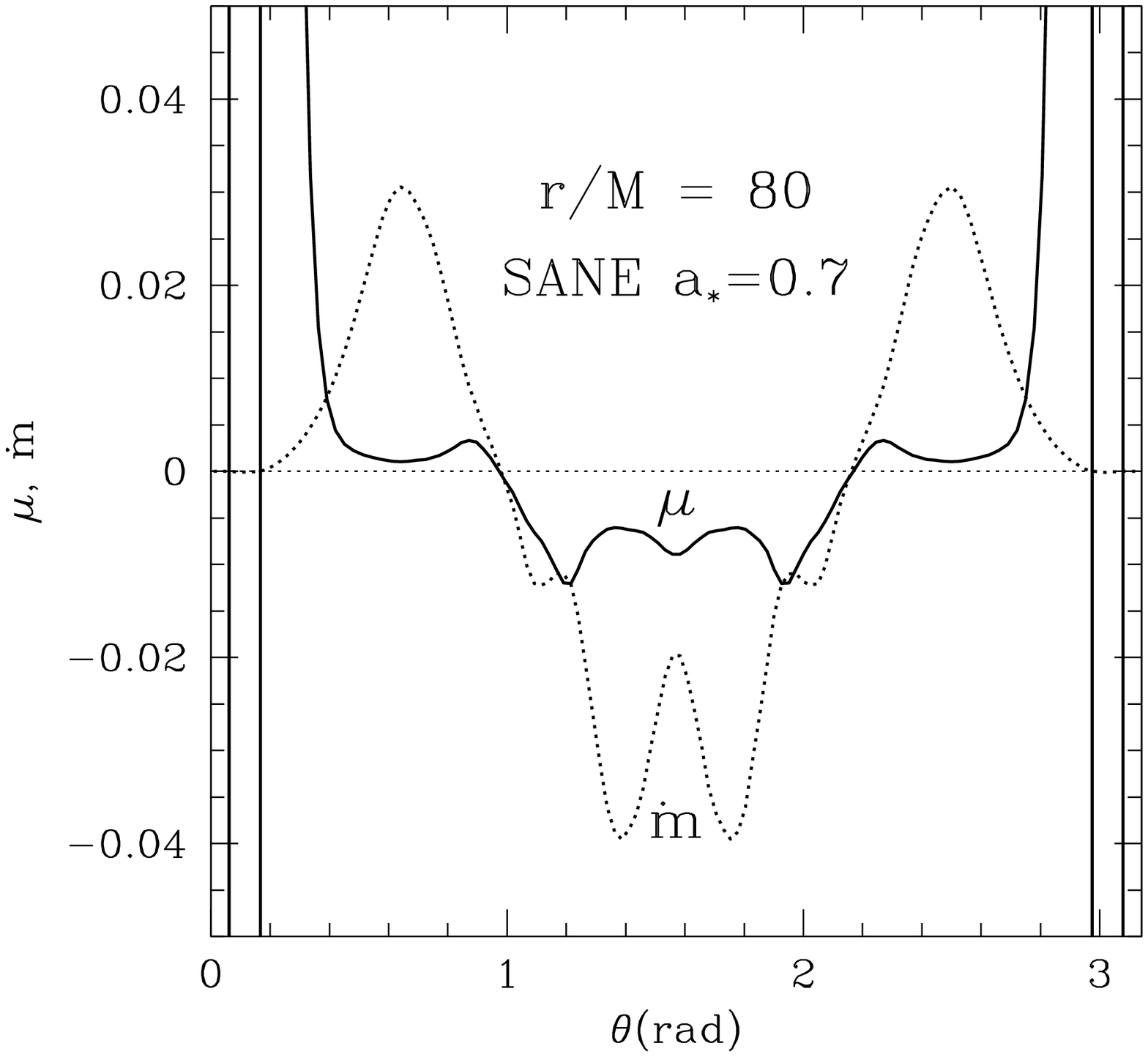}\vspace{-1.25cm}
\includegraphics[height=.95\columnwidth]{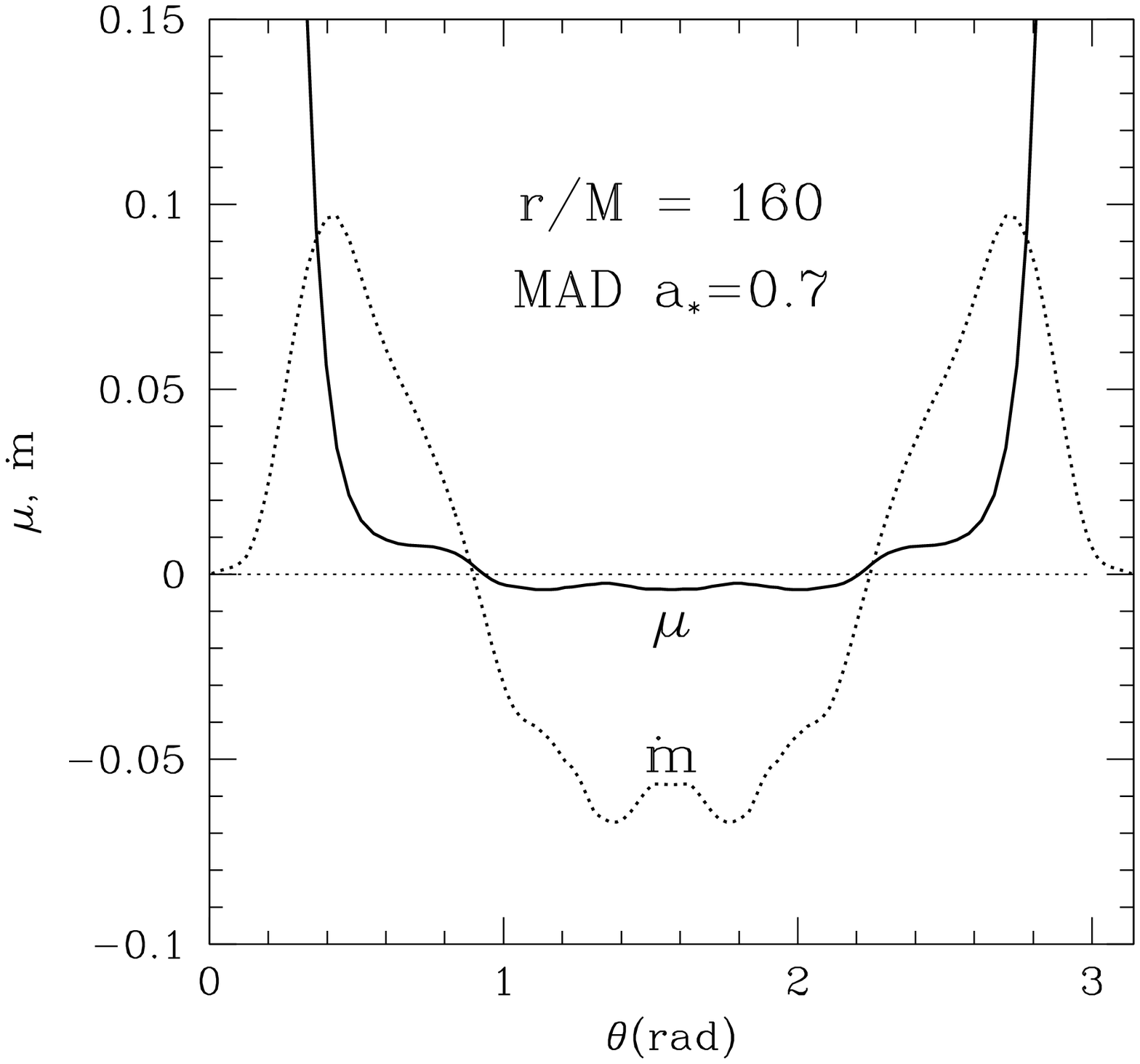}\vspace{-.5cm}
\caption{Vertical profiles of $\mu$ and $\dot m$ for $a_*=0.7$ SANE
  (top panel) and MAD (bottom) simulations. The profiles were
  calculated at $r=80$ and $r=160$, respectively.}
  \label{f.bemutheta}
\end{figure}

Figure~\ref{f.bemutheta} shows profiles of $\mu$ versus $\theta$ for
SANE and MAD runs with BH spin $a_*=0.7$. (This is just an example;
other spins give similar results.) We see that gas with $\theta$
values within $\sim1$\,rad of the two poles has a positive value of
$\mu$ and can escape to infinity, while gas closer to the equatorial
plane has negative $\mu$ and cannot escape. The figure also shows
profiles of the radial mass flux $\dot{m}$, which changes sign at
practically the same values of $\theta$ as $\mu$. In other words, gas
that has a positive value of $\mu$ has an outward-pointed velocity (it
is escaping from the BH), while gas with negative $\mu$ has an
inward-pointed velocity (it is accreting on to the BH).  This confirms
that the Bernoulli parameter $\mu$ is an excellent diagnostic of
outflows in our simulations.

To be safe, in this paper we consider gas to be outflowing to infinity
only if the conditions $\mu>0$ and $\dot{m}>0$ are both
satisfied. However, as Fig.~\ref{f.bemutheta} shows, the two
conditions are practically degenerate, so we could equally well have
used just one of them.

\subsection{Three regions: Disc, wind, jet}
\label{s.jet}

We identify the region of the solution where the gas flows in
($u^r<0$) as the ``disc''. In our radiatively inefficient ADAF-like
simulations, the disc region is geometrically thick and extends over a
range of $\theta \sim \pm0.6$\,rad around the mid-plane (see
Fig.~\ref{f.bemutheta}). Outside the disc zone we have outflowing gas,
which we further subdivide into two components, a slowly-moving
``wind'' and a rapidly-moving ``jet''. The distinction between wind
and jet is motivated by the shapes of the $\mu$ and $\dot{m}$ profiles
in Fig.~\ref{f.bemutheta}. In both the SANE and MAD simulations we see
that for $\theta$ values within about 0.4\,rad of the poles, $\mu$ is
large and $\dot{m}$ is small. The outflowing material here is clearly
relativistic and has a large energy per unit mass; we call it a
``jet''. The rest of the outflowing region, which lies in between the
jet and the disc, has a low value of $\mu$ and larger $\dot{m}$.  This
is a slowly-moving outflow, which we call a ``wind''.

How do we determine the boundary between the jet and the wind?
Unfortunately, there is no unique way of separating these regions
since the transition is smooth and the regions of high mass flux
(wind) and high energy flux (jet) overlap each other. One possibility
is to follow streamlines of the poloidal energy flux $T^p_t$, and to
identify all outflowing streamlines anchored on the horizon as the jet
and the remaining outflowing streamlines that originate in the disc as
the wind \cite[compare the bottom-most right panel of
  Fig.~\ref{f.streams} and see][]{penna+13}. This is similar to the
approach used by \cite{tchekh+11} and is a convenient way of
distinguishing the part of the outflow that is powered by the BH from
that which is powered by the disc. However, magnetic fields extract
rotational energy from the BH at all latitudes \citep{penna+13}, even
in the disc region where the inflowing rest mass flux overwhelms this
effect and causes the flux of total energy to be negative.

\begin{figure}
  \centering
\renewcommand{\arraystretch}{0.5}
\begin{tabular}{M}
 $\dot e = |T^r_t+\rho u^r|$  \\
\includegraphics[height=.295\textwidth]{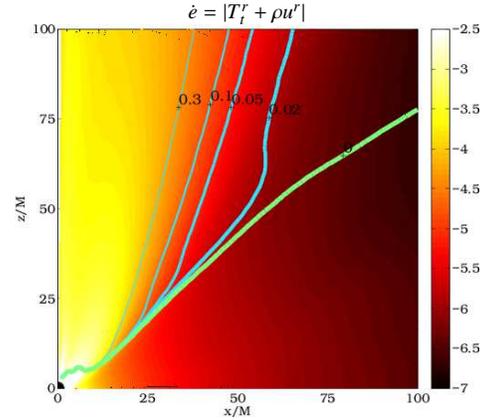}
\end{tabular}
\caption{Contours of $\mu$ in the poloidal plane for the $a_*=0.7$ MAD
  model plotted on top of the magnitude of the radial
  flux of energy $\dot e$. Five contours are plotted corresponding to $\mu=0.3$
  (thinnest blue line), $0.1$, $0.05$, $0.02$ (thickest blue line),
  and $0.0$ (green line).}
  \label{f.a7MAD_mu}
\end{figure}

Another approach, which we follow in this paper, is to choose a
critical value of $\mu$, or equivalently a critical value of the
``velocity at infinity'', and to identify all outflowing gas with
$\mu$ larger than this critical value as the jet and the rest as the
wind.  We choose \be \mu_{\rm crit}=0.05,
\label{mujet}
\ee which at infinity corresponds to $\beta_{\rm crit}=v_{\rm crit}/c
\approx 0.3$, a reasonable demarcation point between jet and wind. In
Fig.~\ref{f.a7MAD_mu} we plot five contours of $\mu$ on top of the
outflowing energy fluxes for the $a_*=0.7$
MAD simulation. Our particular choice of $\mu_{\rm crit}=0.05$
reasonably separates the region of high energy flux,
which corresponds to the jet, and the region of low energy flux, 
which corresponds to the wind.

\subsection{Radius of inflow equilibrium}
\label{s.radius}

Since each of our simulations has been run for only a finite
duration, the solutions reach inflow equlibrium only within a limited
volume. Gas outside this volume has not had enough time to be
influenced by the accretion flow structure near the horizon.  To
quantify the range of radii over which the disc solution is reliable,
we adopt the ``loose'' equilibrium criterion from \cite{narayan+12a},
i.e, we search for an inflow equilibrium radius $r_{\rm eq}$ which satisfies 
\be 
\label{eq.rconv}
r_{\rm  eq}=|v^r(r_{\rm eq})| t_{\rm chunk}, 
\ee 
where $v^r(r_{\rm
  eq})$ is the density-weighted average velocity at a given radius, and
$t_{\rm chunk}$ is the duration of the last time chunk for the
particular simulation. We carry out this calculation separately for
the disc, the wind, and the jet. In the case of the wind and jet $v^r(r_{\rm eq})$ is
outward, so technically $r_{\rm eq}$ is the outflow equilibrium radius, but the
principle remains the same.

Table~\ref{t.convradii} gives values of the limiting inflow/outflow
equilibrium radii $r_{\rm eq}$ for each of the runs for each of the
three regions.  These radii decrease with increasing BH spin, since
the durations of the simulations become shorter. In all cases, the
wind region reaches equilibrium to larger radii than the disc
since the radial velocity of the gas is larger.  Similarly, since the
jet has a relativistic velocity, this region reaches equlibrium
to very large radii (essentially the entire domain of the simulation).
Another systematic effect is that the MAD simulations, because of
their larger radial velocities \citep[see][]{narayan+12a}, are in
  inflow equlibrium over a significantly larger volume compared to
the SANE simulations.

\begin{table}
\caption{Inflow/outflow equlibrium radii}
\label{t.convradii}
\centering\begin{tabular}{@{}cccc}
\hline
 Model & Disc & Wind & Jet \\
\hline
 $a_*=0.0$ SANE & 110 & 210 & --- \\
 $a_*=0.7$ SANE & 70 & 130 & 21000 \\
  $a_*=0.9$ SANE & 50 & 110 & 16000 \\
  $a_*=0.98$ SANE & 50 & 80 & 8000 \\
 $a_*=0.0$ MAD & 340 & 720 & 65000 \\
 $a_*=0.7$ MAD & 160 & 320 & 29000 \\
$a_*=0.9$ MAD & 140 & 260 & 13000\\
\hline
\end{tabular}
 \end{table}

\section{Results}

\subsection{Accretion rate and magnetic flux}
\label{s.flux}

\begin{figure}
  \centering
\includegraphics[height=.95\columnwidth,angle=270]{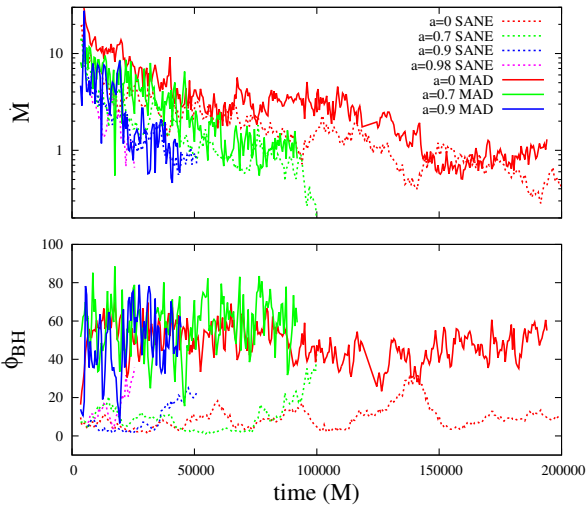}
\caption{Accretion rate $\dot{M}$ into the BH (top) and magnetic flux
  $\Phi_{\rm BH}$ threading the BH horizon (bottom) versus time for
  SANE (dotted lines) and MAD (solid lines) models with different BH spins.}
  \label{f.fluxes_aN}
\end{figure}

All the simulations were initialized with an equilibrium gas torus
threaded by a weak poloidal magnetic field.  Once the
magneto-rotational instability (MRI) develops, gas accretes towards
the BH and the inner regions of the torus are depleted of matter. The
accretion rate on the BH thus decreases with time, the variation being
determined by the density profile assumed in the initial torus
\citep{narayan+12a}. The upper panel of Fig.~\ref{f.fluxes_aN} shows
the mass accretion rate $\dot{M}$ on the BH versus time for all the
radiatively inefficient models studied here. Solid and dotted lines
correspond to MAD and SANE models, respectively. For a given BH spin,
the accretion rate evolution is roughly the same for SANE and MAD
models. However, at any given time, the higher the spin, the lower is
the accretion rate. This appears to be because of the increasing mass
loss rate (discussed in Section.~\ref{s.massloss}).

By monitoring the magnetic flux $\Phi_{\rm BH}$ threading the BH
horizon (eq.~\ref{e.phibh}) we can evaluate whether a particular
simulation is in the SANE or MAD state. The bottom panel of
Fig.~\ref{f.fluxes_aN} shows the evolution of $\Phi_{\rm BH}$ for each
of the models.  The magnetic flux at the horizon for the MAD runs
(solid lines) remains always near $\Phi_{\rm BH}\sim50$, showing that
the flux has saturated at the maximum allowed value, as appropriate
for the MAD state \citep{tchekh+11}. In contrast, the SANE models are
characterized by lower values of $\Phi_{\rm BH}$. However, once in a
while even these simulations show $\Phi_{\rm BH}$ approaching the MAD
limit, though the flux subsequently falls as an oppositely polarized
magnetic loop reaches the BH. For all the three SANE models with
non-zero BH spin, $\Phi_{\rm BH}$ approaches the saturation value
appropriate to the MAD state near the end of the simulation. Thus,
despite our efforts to avoid the MAD state in our SANE simulations,
the accretion flow apparently has a tendency to be pushed towards the
MAD limit.

\begin{figure*}
  \centering
\renewcommand{\arraystretch}{0.5}
\begin{tabular}{MMMM}
& $|\dot m|$ &$|\dot e|$ & $|\dot e_{\rm tot}|$  \\
\begin{sideways}$a_*=0.7$ SANE\end{sideways}&
\subfigure{\includegraphics[width=.3\textwidth]{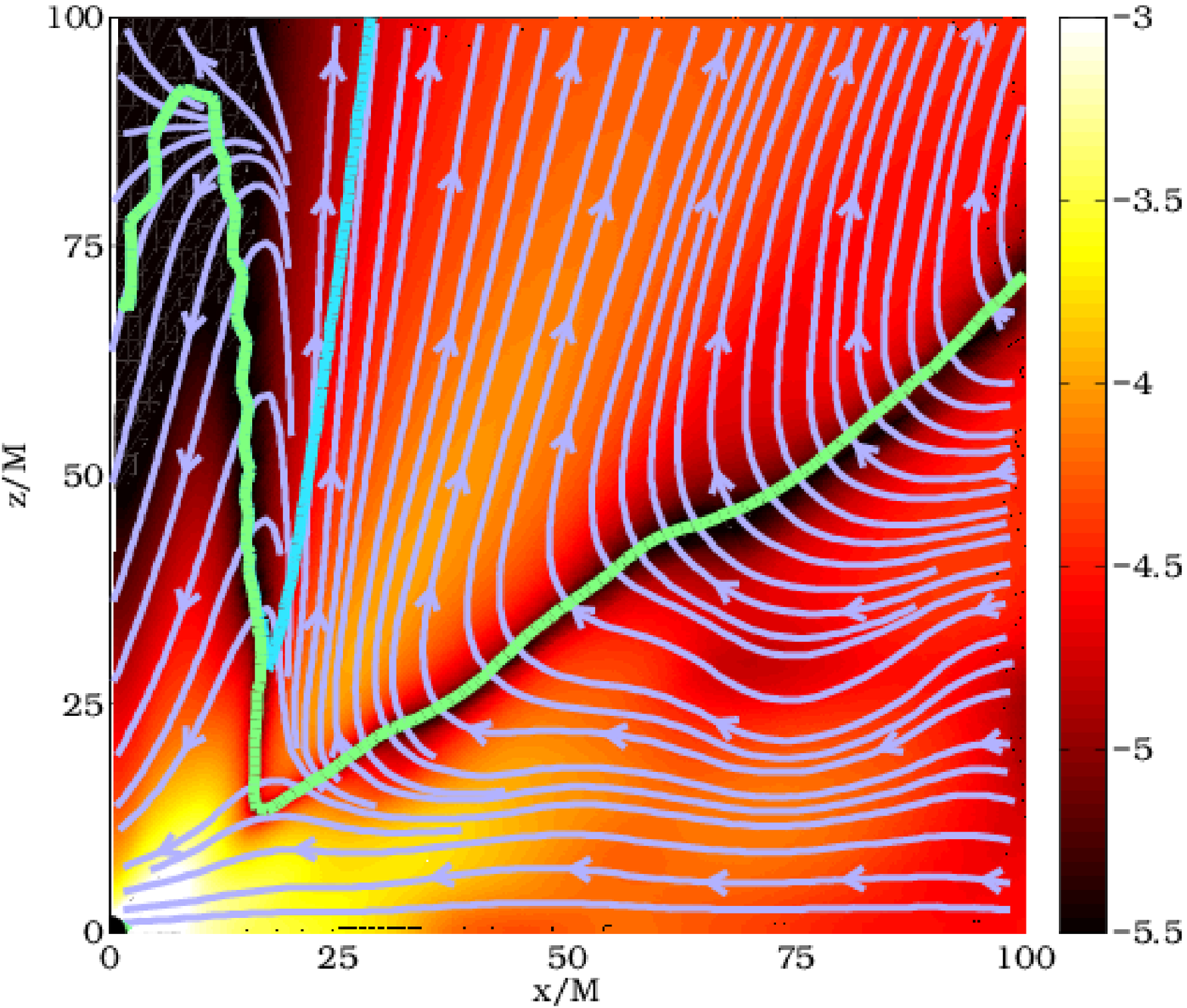}}&
\subfigure{\includegraphics[width=.3\textwidth]{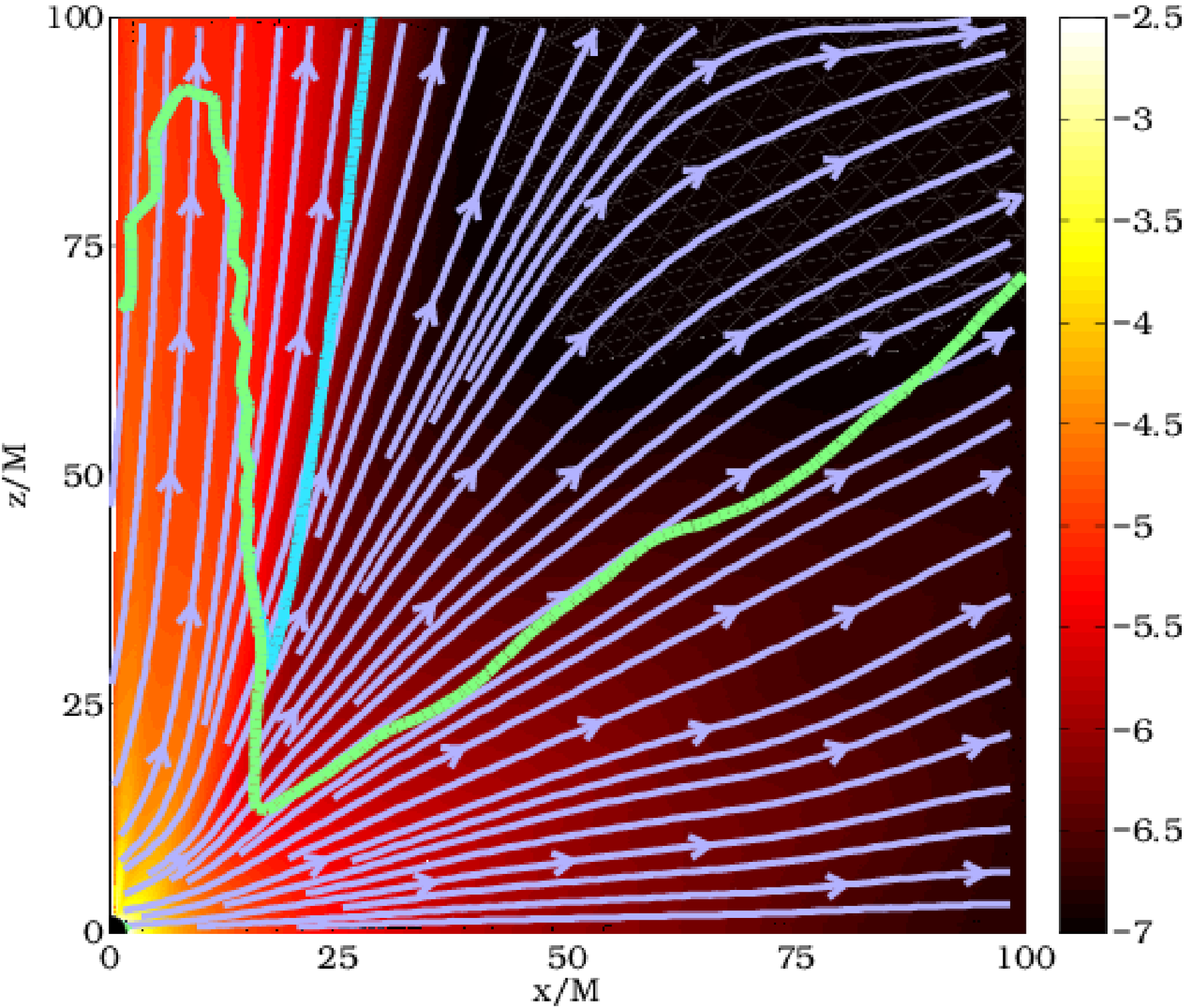}}&
\subfigure{\includegraphics[width=.3\textwidth]{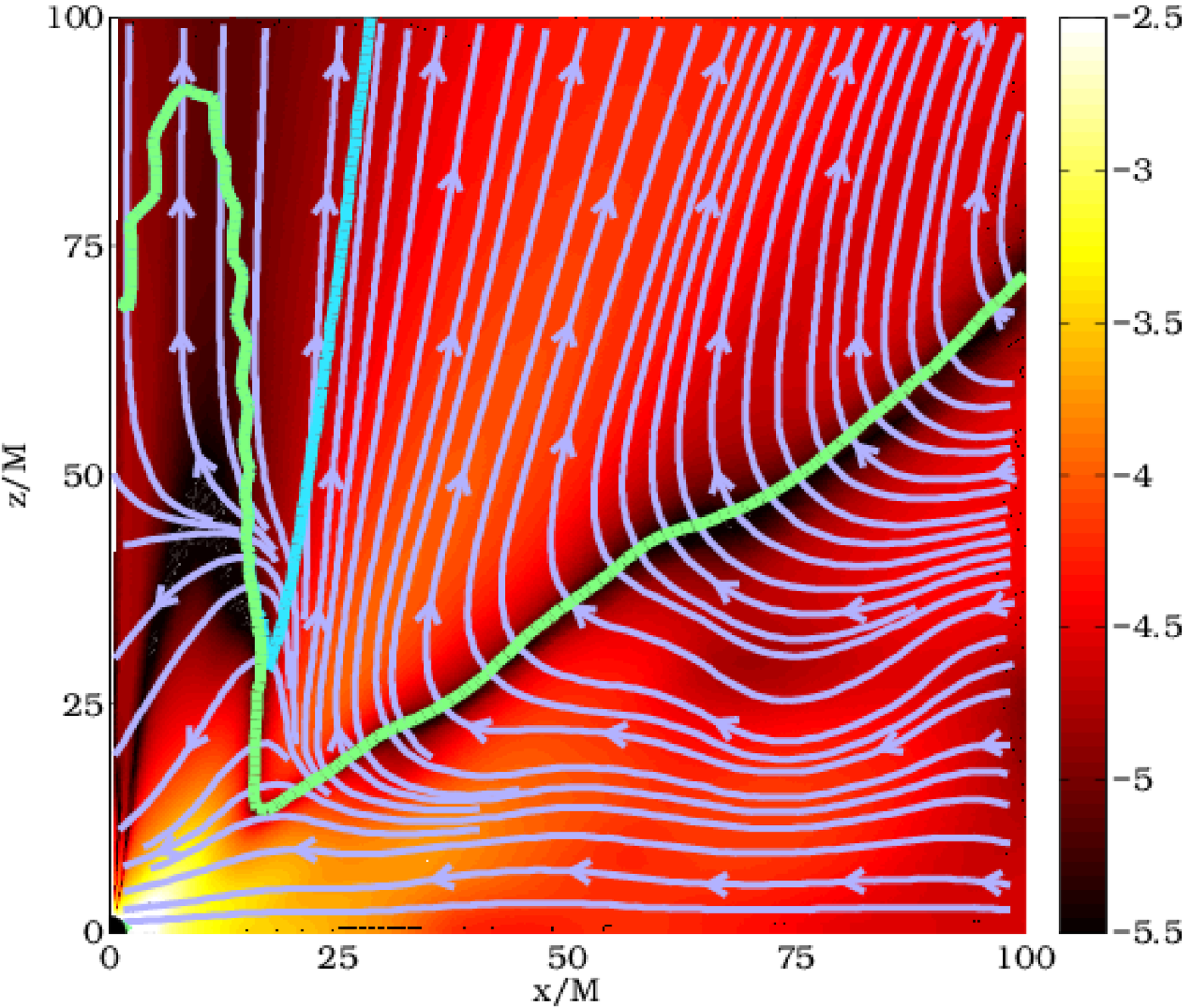}}\\
\begin{sideways}$a_*=0.7$ MAD\end{sideways}&
\subfigure{\includegraphics[width=.3\textwidth]{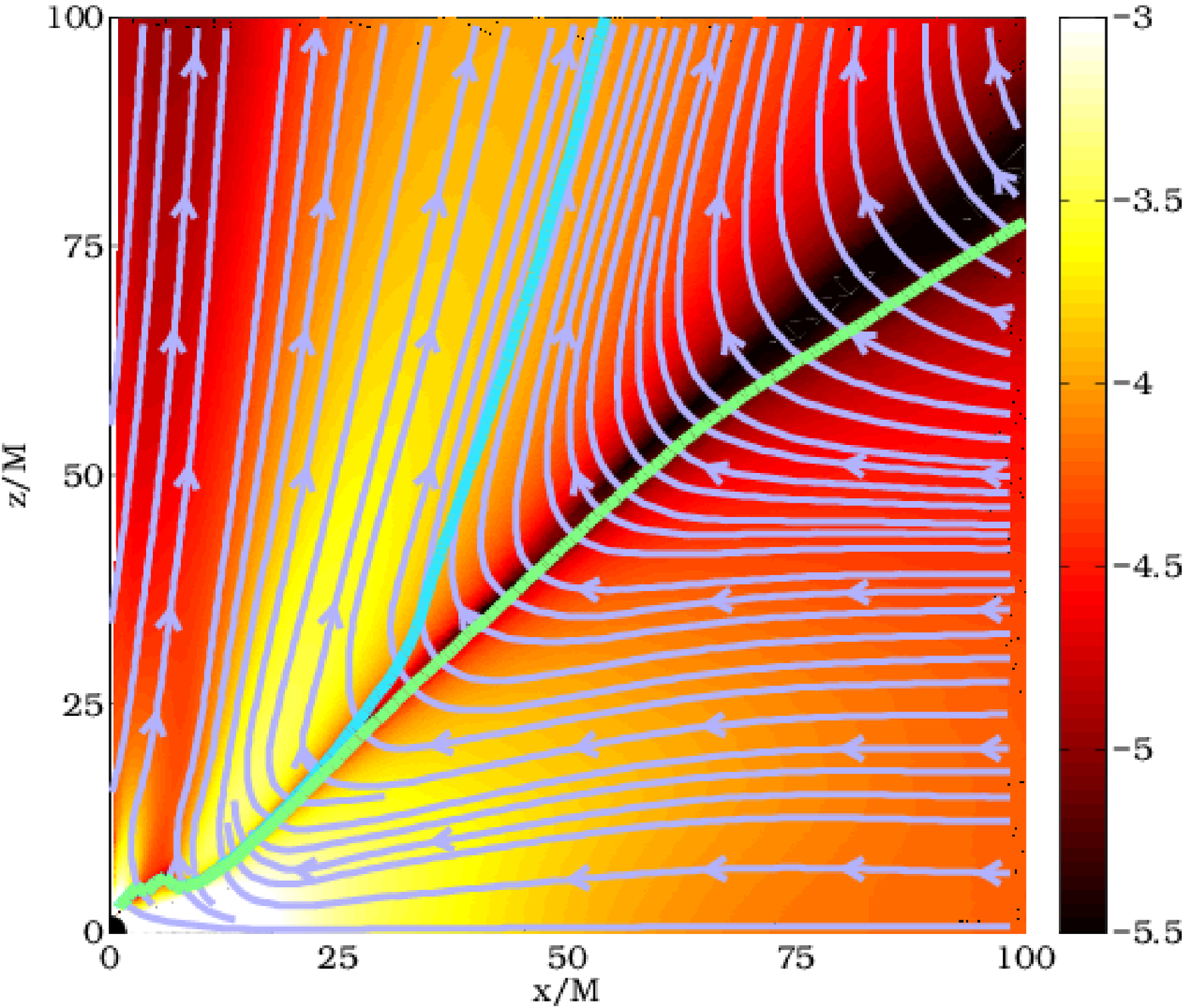}}&
\subfigure{\includegraphics[width=.3\textwidth]{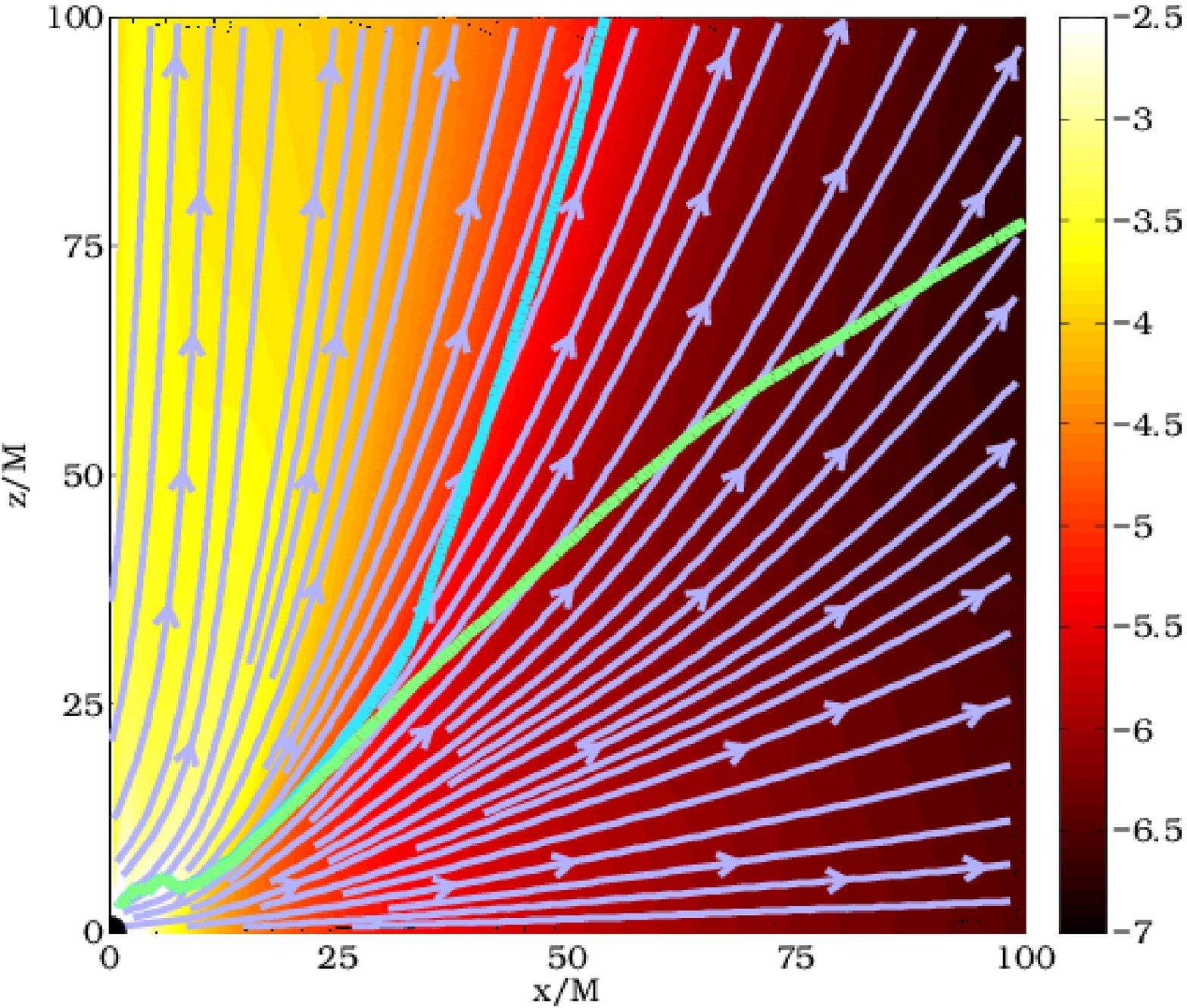}}&
\subfigure{\includegraphics[width=.3\textwidth]{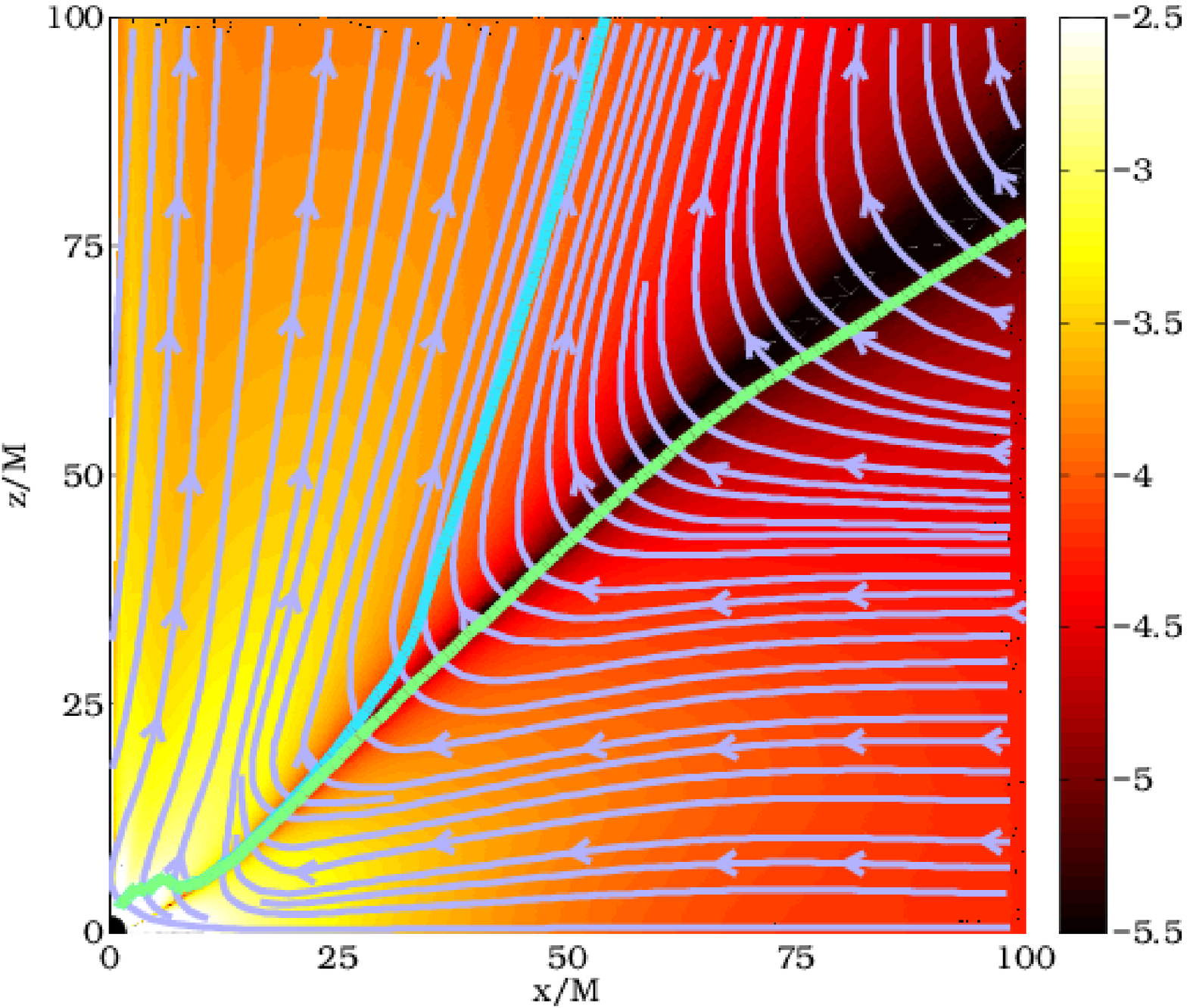}}\\
\begin{sideways}$a_*=0.7$ SANE (zoom in)\end{sideways}&
\subfigure{\includegraphics[width=.3\textwidth]{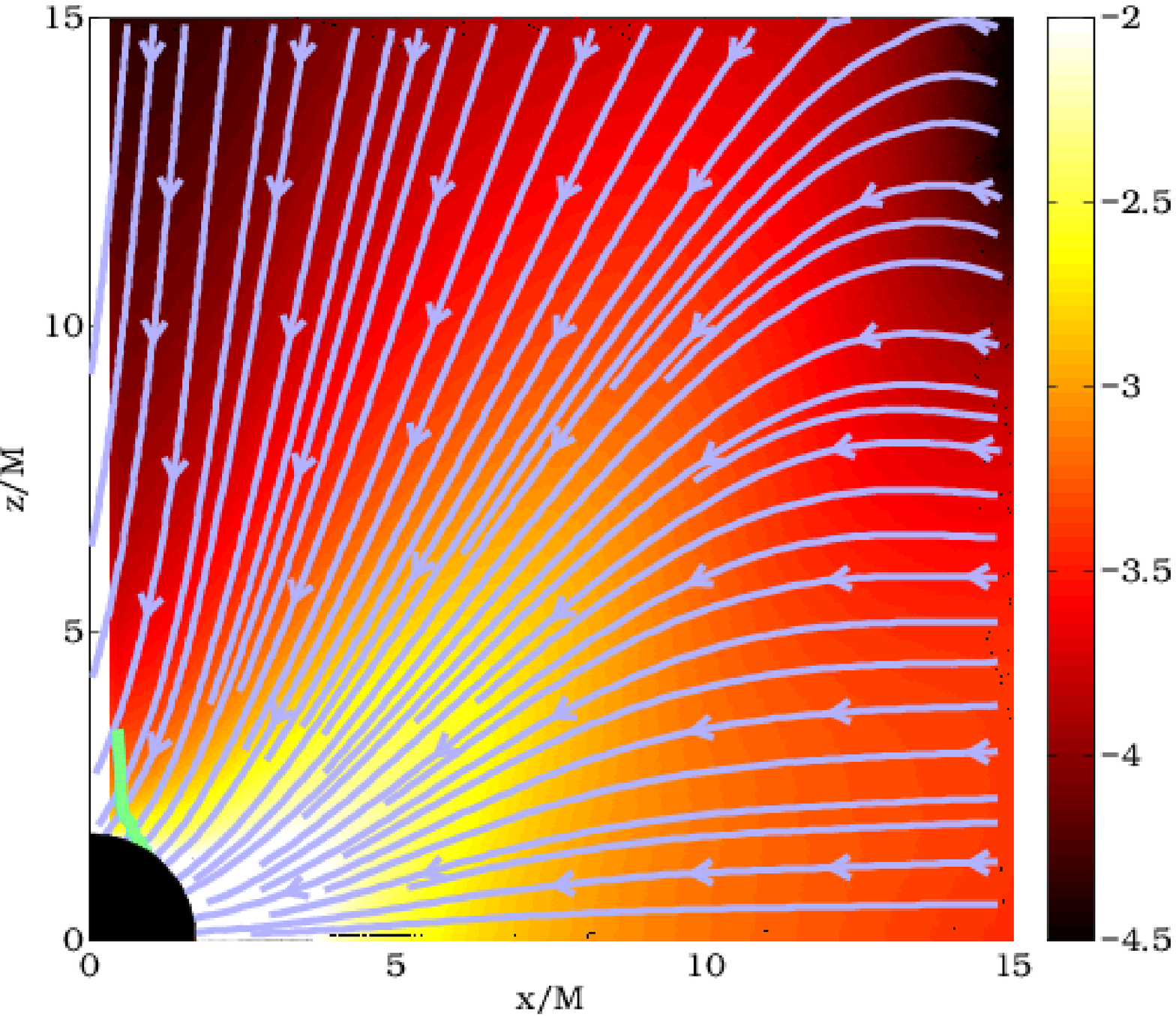}}&
\subfigure{\includegraphics[width=.3\textwidth]{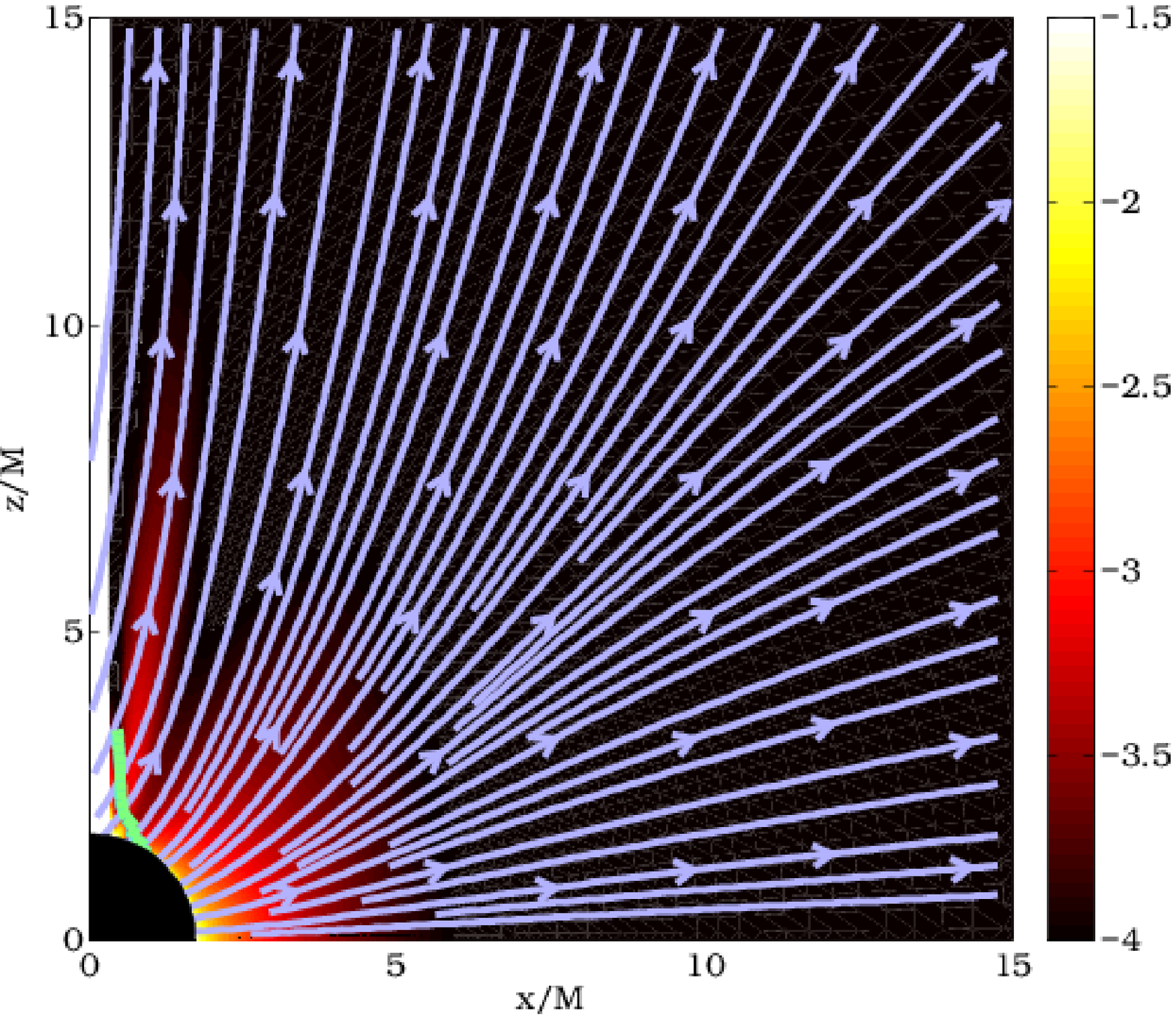}}&\
\subfigure{\includegraphics[width=.3\textwidth]{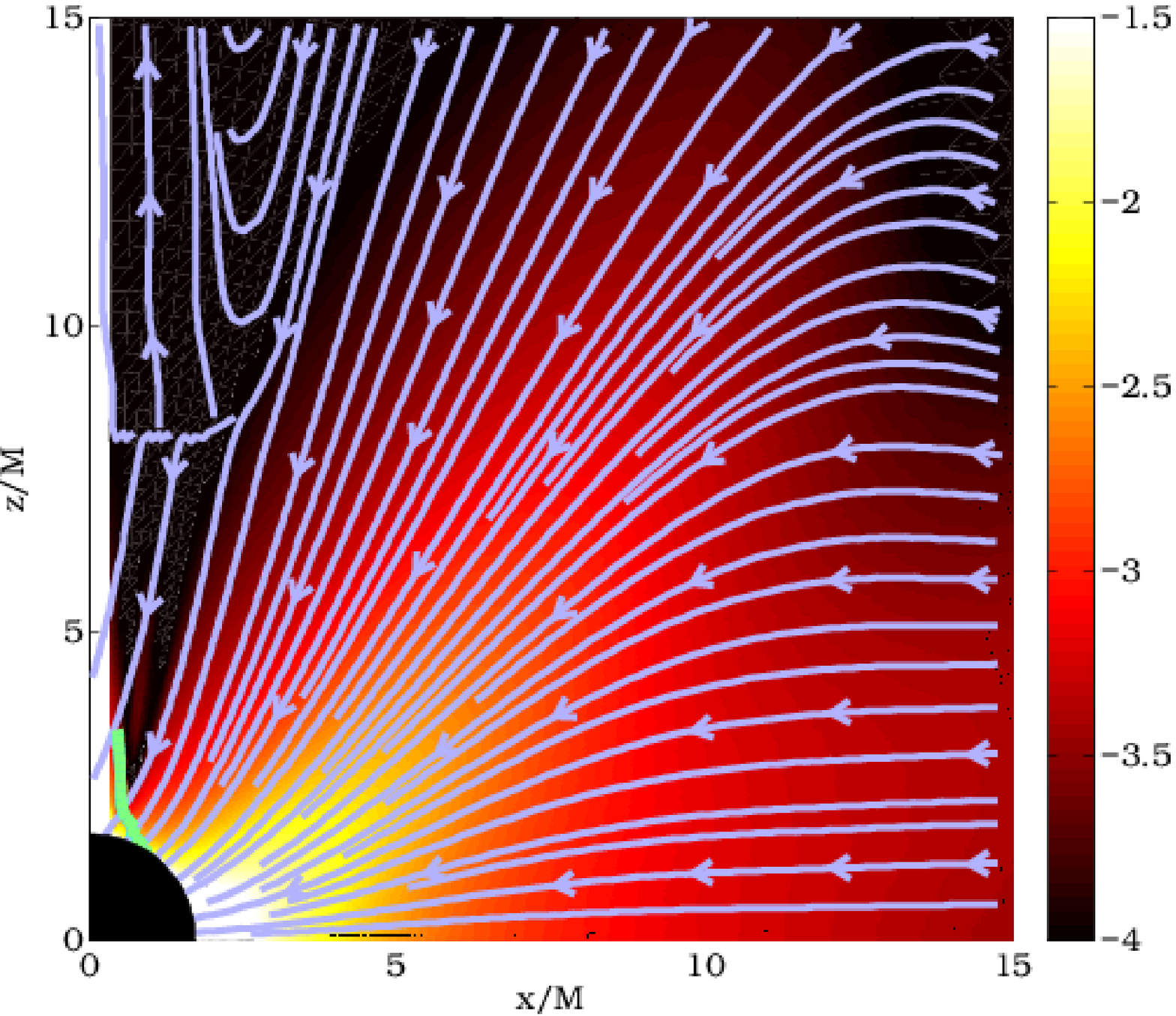}}\\
\begin{sideways}$a_*=0.7$ MAD (zoom in)\end{sideways}&
\subfigure{\includegraphics[width=.3\textwidth]{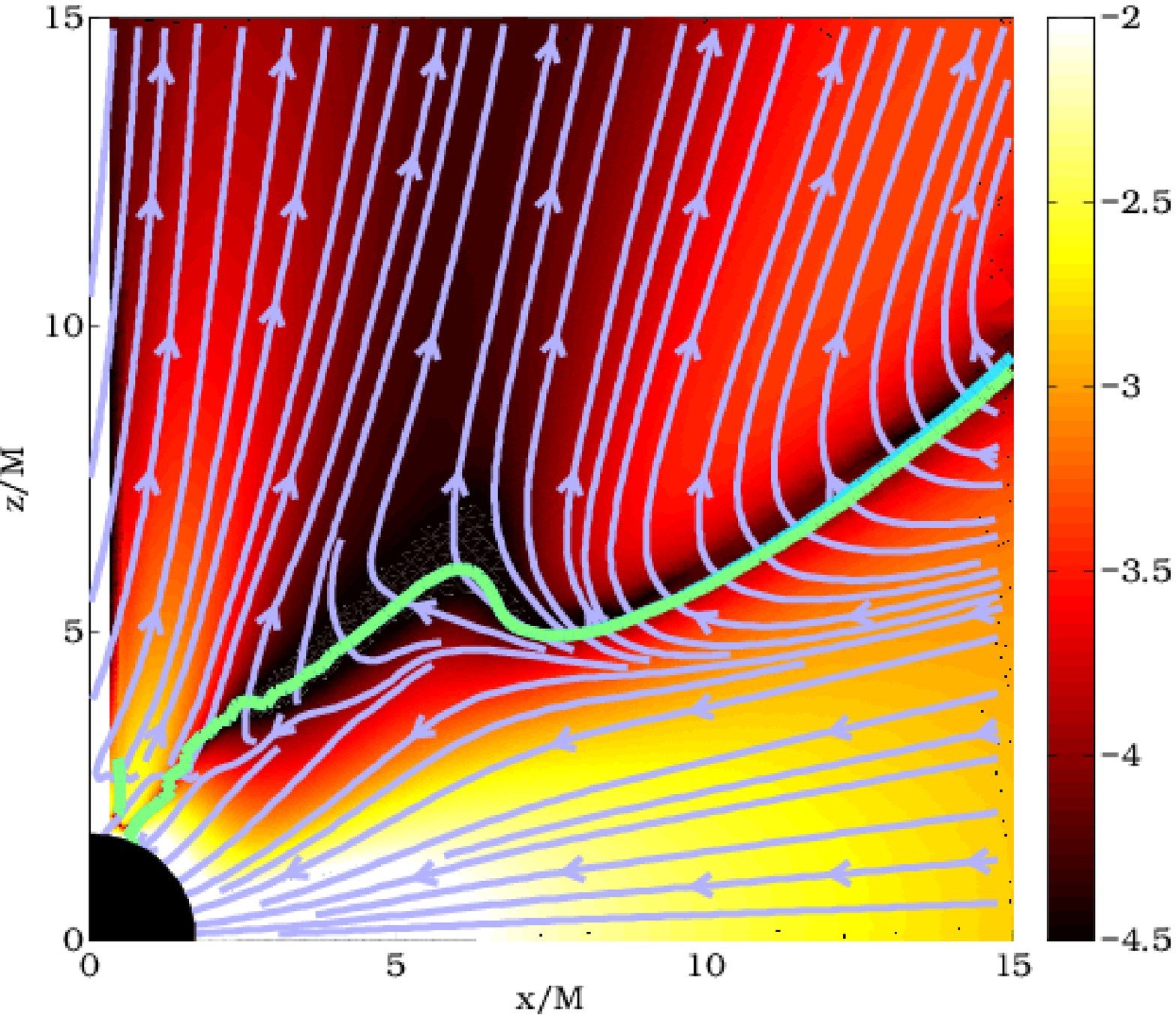}}&
\subfigure{\includegraphics[width=.3\textwidth]{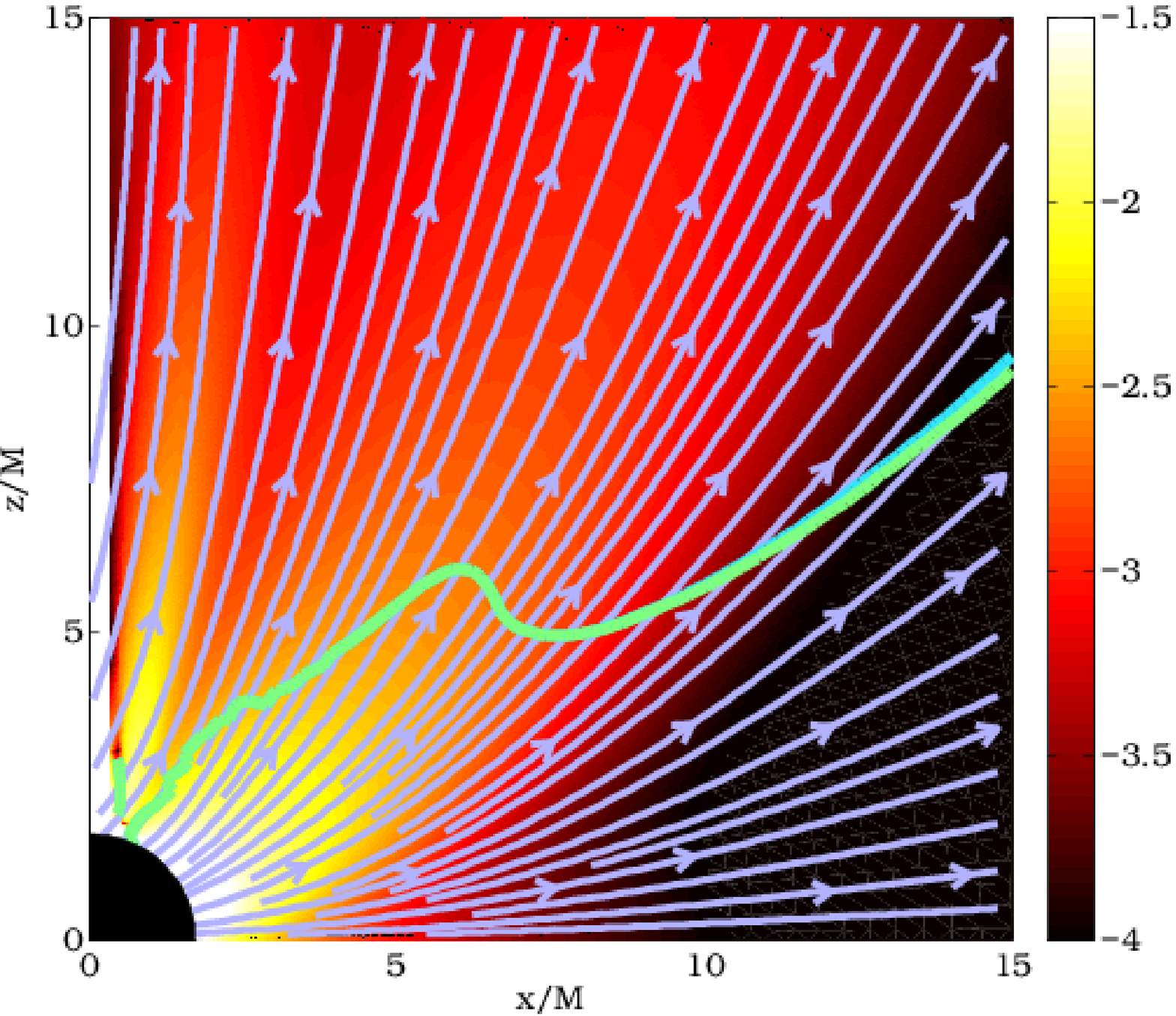}}&\
\subfigure{\includegraphics[width=.3\textwidth]{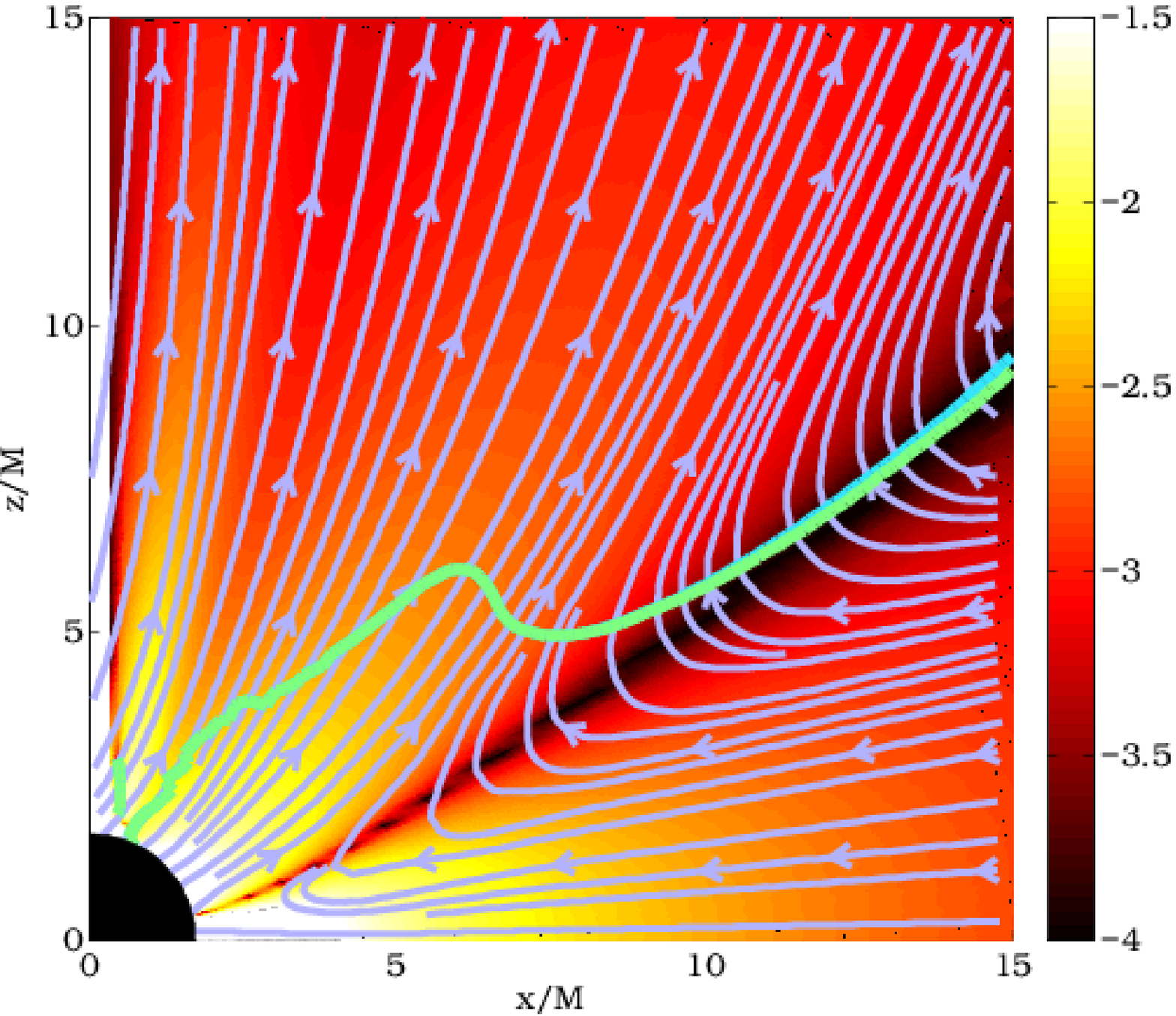}}
\end{tabular}
\caption{ Rest mass flux ($\rho u^r$, left), energy flux ($-T^r_t-\rho
  u^r$, middle), and total energy flux ($-T^r_t$, right column) in the poloidal plane for models $a_*=0.7$ SANE and
  $a_*=0.7$ MAD. Colors show the logarithm of magnitude of the radial
  flux while streamlines show the flux direction on the poloidal
  plane. Green contour shows $\mu=0$ and separates the outflow
  region from the inflowing disc region. Blue contour stands for $\mu=0.05$ and separates the jet and
  wind regions. The third and fourth rows zoom in on the innermost region.}
  \label{f.streams}
\end{figure*}

\subsection{Structure of the outflow regions}
\label{s.structure}

\begin{figure*}
  \centering
\includegraphics[width=.74\textwidth]{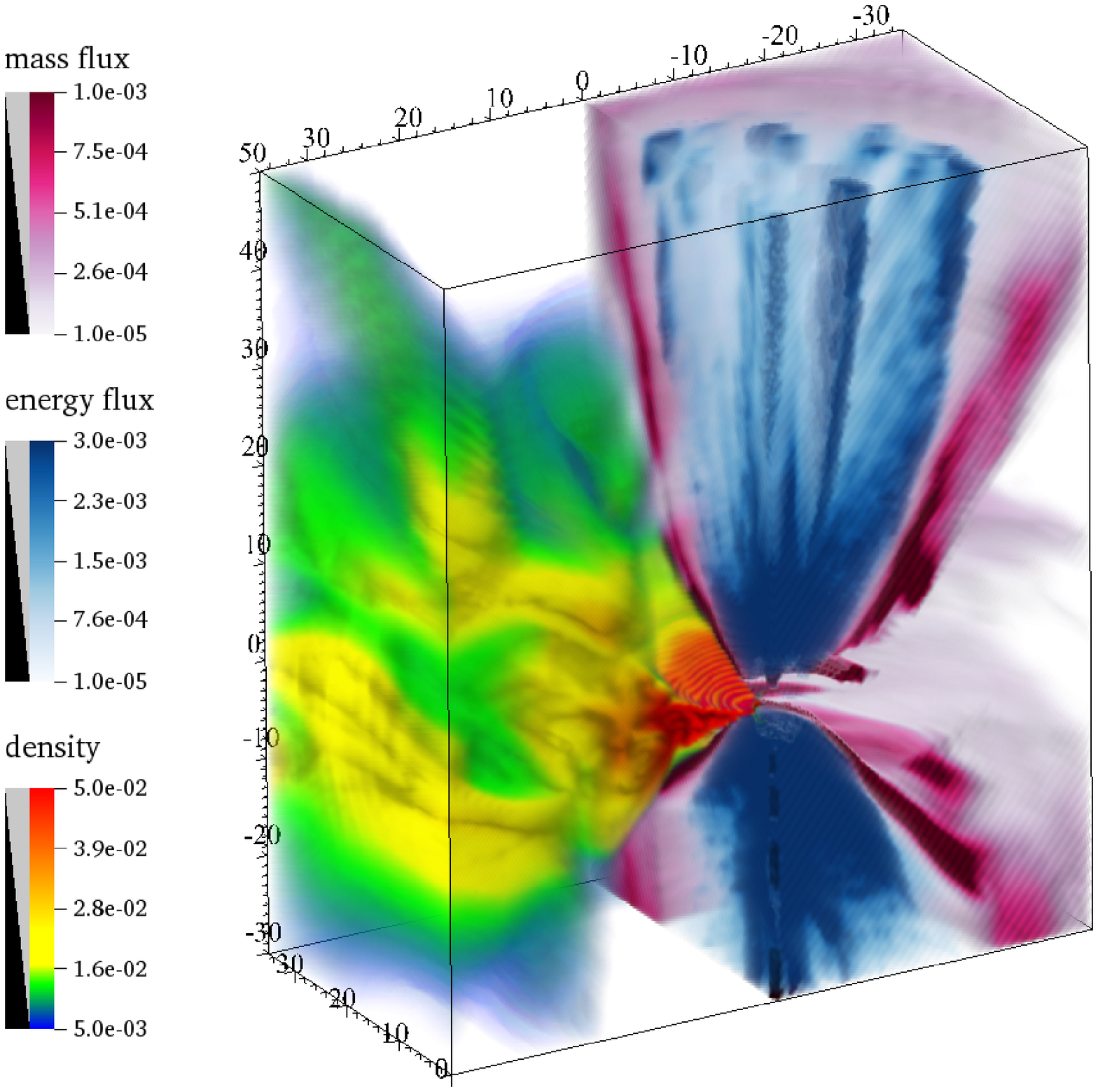}
\caption{3D visualisation of a snapshot of the $a_*=0.7$ MAD model. The yellow-green
colors in the left part of the plot show the spatial distribution of mass density.
The magenta colors in the right part correspond to the magnitude of the rest
mass flux $\rho u^r$ while the blue colors show the magnitude of the 
energy flux $T_t^r+\rho u^r$.}
  \label{f.fluxes3D}
\end{figure*}

In this section we discuss the general properties of the outflow
regions.  Fig.~\ref{f.streams} shows the magnitude of mass and energy
fluxes ($\dot m$, $\dot e$ and $\dot e_{\rm tot}$) for $a_*=0.7$ SANE and $a_*=0.7$ MAD models
together with corresponding streamlines in the poloidal plane. The
blue contours denote the border between wind and jet regions
(eq.~\ref{mujet}) while the green contours separate the outflow and
inflow regions.

The left column shows the rest mass flux. At large radii mass flows
mostly inside the bulk of the disc.  However, the streamlines clearly
show that the inflowing accretion rate is not constant in all models
--- some rest mass is lost from the inflow region and forms the
wind. \cite{narayan+12a} have shown that for $a_*=0$ models such a magnetically-driven wind from the
accretion disc itself does not extend all the
way towards the BH but stops around $r=40$.

The mass outflows are enhanced for rotating BHs. For the $a_*=0.7$
SANE model the extra energy flux along the polar axis
(middle panel) supresses accretion in the polar region. However, the
accretion flow wants to deposit the same amount of gas as in the
non-rotating case (locally the disc does not feel the impact of spin
at large radii). The surplus of gas has to find its way out of the
system and flows out in the wind.  For the $a_*=0.7$ MAD model the
energy flux along the axis is so strong that it not only reduces
the solid angle available for the inflow but also directly drives the
outflow, converting its magnetic energy to kinetic energy of gas. It
results in a strong, ``cocoon''-shaped, mass outflow originating in
the inner region ($10<r<20$) which significantly decreases the
inflowing accretion rate and contributes to the jet region. The amount
of mass lost in this region may be comparable to the net mass
accretion rate $\dot{M}_{\rm BH}$ on the BH.  Note that the region of
increased mass outflow close to the polar axis and near the BH is a
result of numerical corrections (floors) imposed to retain reasonable
ratios of energy densities and make the code stable, and therefore
should be regarded as unphysical. Fortunately, the amplitudes of all
the fluxes emerging from that region are negligible when compared to
the outflows at $r>10$.

The middle column of Fig.~\ref{f.streams} shows the magnitude of the
energy flux ($\dot e$). The outflowing energy is highly concentrated
in the jet region and is much higher for the MAD model. The
  magnetic jet is more collimated than the mass loaded jet --- most of
  the rest mass in this simulation flows out along a conical surface
  with opening angle $\sim 25$ deg.  The locations of both jets are
  visualised in Fig.~\ref{f.fluxes3D} showing the density
distribution in the left part of the plot and outflows of rest mass
and energy in the right part of the plot with magenta and blue,
respectively.  It is clear that the strong energy flux region is
surrounded by the region where the mass loss is most efficient.

The right column in Fig.~\ref{f.streams} shows the magnitude and
streamlines of the total energy flux $\dot{e}_{\rm tot}$, i.e., the
sum of the first two columns. Inside the disc the inflowing stream
of rest mass dominates and the net energy flux points inward. In
contrast, in the wind and jet the total energy flux is positive.

The solid angles of the jet regions in the $a_*=0.7$ simulations are
$\Omega_{\rm jet}\lesssim 1.0$\,sr for the SANE and MAD models,
respectively. The corresponding solid angles covered by the wind are
$\Omega_{\rm wind} \gtrsim 5.0$\,sr. The
wind thus subtends a significantly larger solid angle.

\begin{figure}
  \centering
\subfigure{\includegraphics[height=.45\textwidth,angle=270]{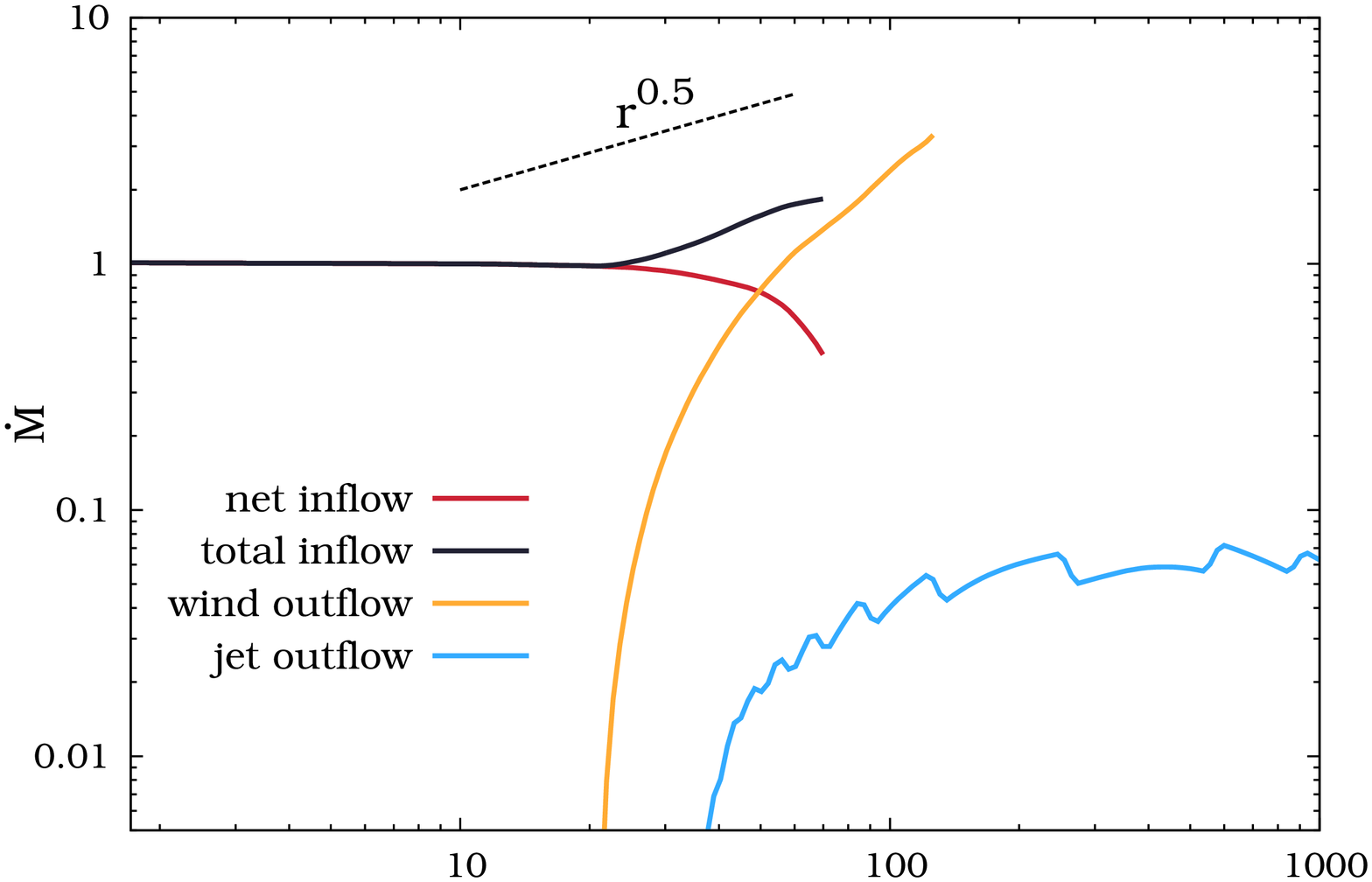}}
\subfigure{\includegraphics[height=.45\textwidth,angle=270]{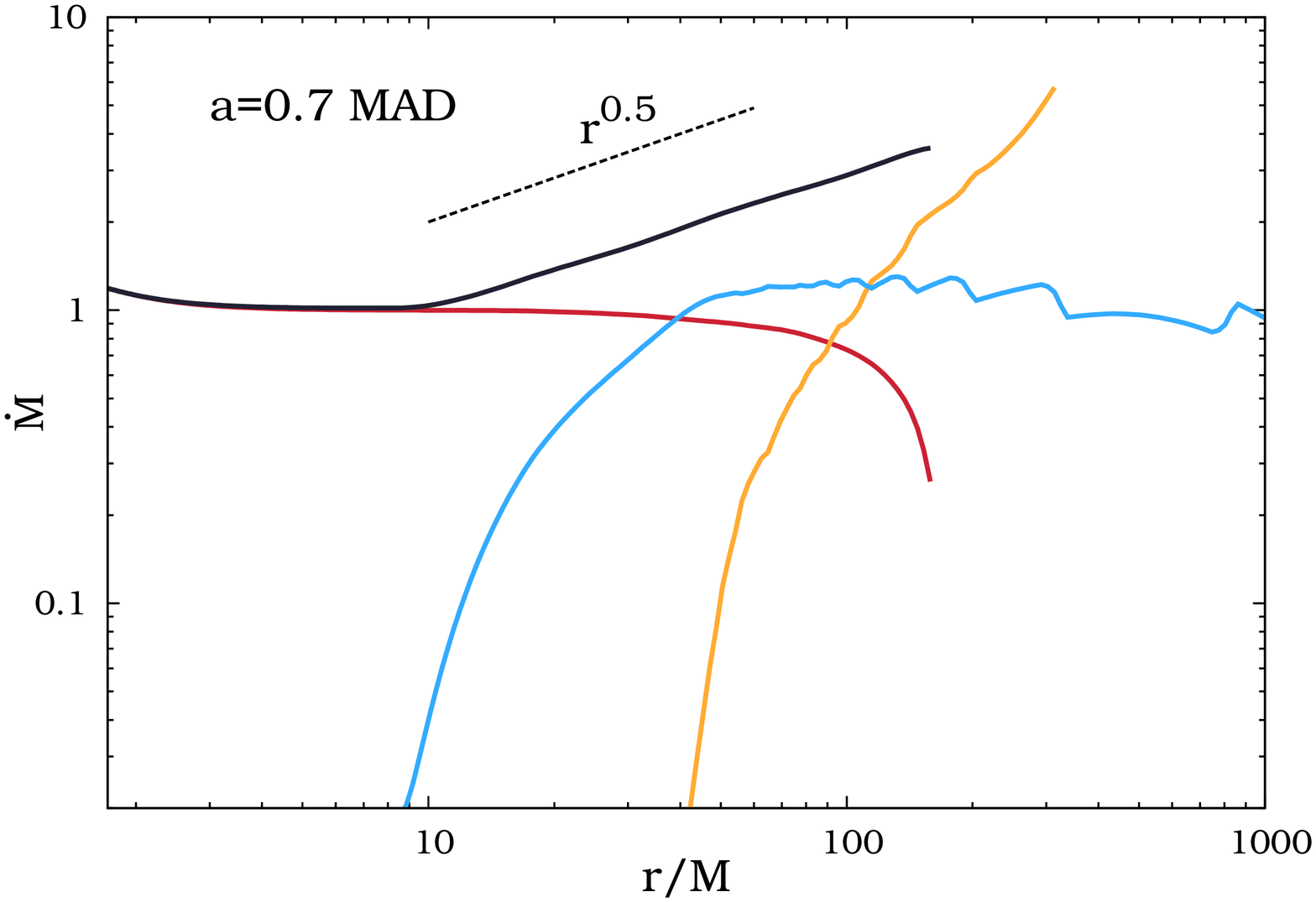}}
\caption{Radial profiles of the mass flux $\dot{M}$ for $a_*=0.7$ SANE
  (top) and MAD (bottom panel) at time chunk T5.  Red, orange and blue
  lines show the net mass accretion rate, mass outflow in the wind,
  and mass outflow in the jet, respectively. All fluxes are normalized
  by $\dot{M}$ at $r=10$, and each line is terminated at the
  corresponding radius of inflow or outflow equlibrium (Table~\ref{t.convradii}).  The
  black lines show the total inflow flux, i.e., the integral over
  sphere of the inflowing rest mass (eq.~\ref{eq:Mdot}).}
  \label{f.mdota7}
\end{figure}

\subsection{Angle-integrated mass and energy fluxes}
\label{s.angleintegrated}

Figure \ref{f.mdota7} shows radial profiles of various mass flow rates
for the $a_*=0.7$ SANE and MAD simulations. The red lines show the
absolute value of the normalized net mass accretion rate
(eq.~\ref{eq:Mdot} with $\theta$ integrated over the full range 0 to
$\pi$) while the black lines show the magnitude of the total inflowing
mass flux (same integral but over inflowing gas only). The curves are
terminated at the limiting inflow equilibrium radius of the disc. As
expected, the net $\dot{M}$ is equal to unity (because of the
normalization) and is independent of radius (indicative of steady
state), though constancy is violated near the outer edge, where
complete equilibrium has not been achieved.

The blue curves show the mass outflow rate in the jet; in this case
the integral in equation (\ref{eq:Mdot}) is limited to those values of
$\theta$ where the Bernoulli parameter $\mu$ exceeds the critical
value $\mu_{\rm crit}$ and $u^r>0$. Since the jet originates close to
the BH, we see that the blue lines asymptote to a constant value at
large radii.  Thus the simulations provide a reliable estimate of the
mass loss rate in the jet. The rate is about 10\% of the accretion
rate for the SANE model and roughly equal to the accretion rate for
the MAD model.

In contrast to the jet, the mass outflow rate in the wind (orange
lines) increases steadily with radius and there is no sign of
convergence at large radii. This confirms that mass loss in the wind
is dominated by large radii. Since the simulations reach inflow equilibrium at
best out to a few hundred $R_G$, whereas real flows in nature extend
to outer radii $R_{\rm out}\sim10^5R_G$ or even larger, this means
that estimates of mass loss rates require considerable
extrapolation. This is discussed below.

\begin{figure}
  \centering
\subfigure{\includegraphics[height=.45\textwidth,angle=270]{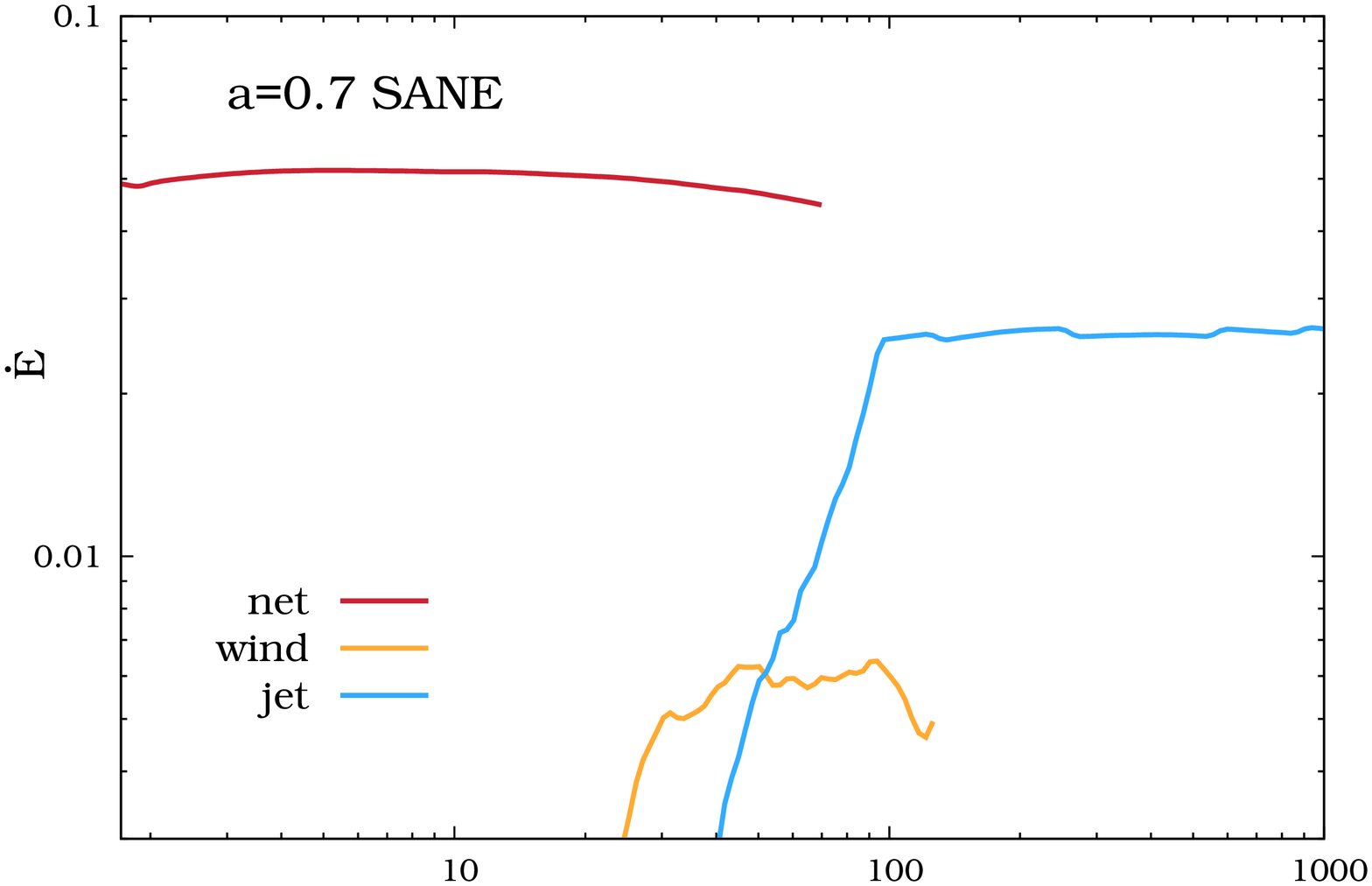}}
\subfigure{\includegraphics[height=.45\textwidth,angle=270]{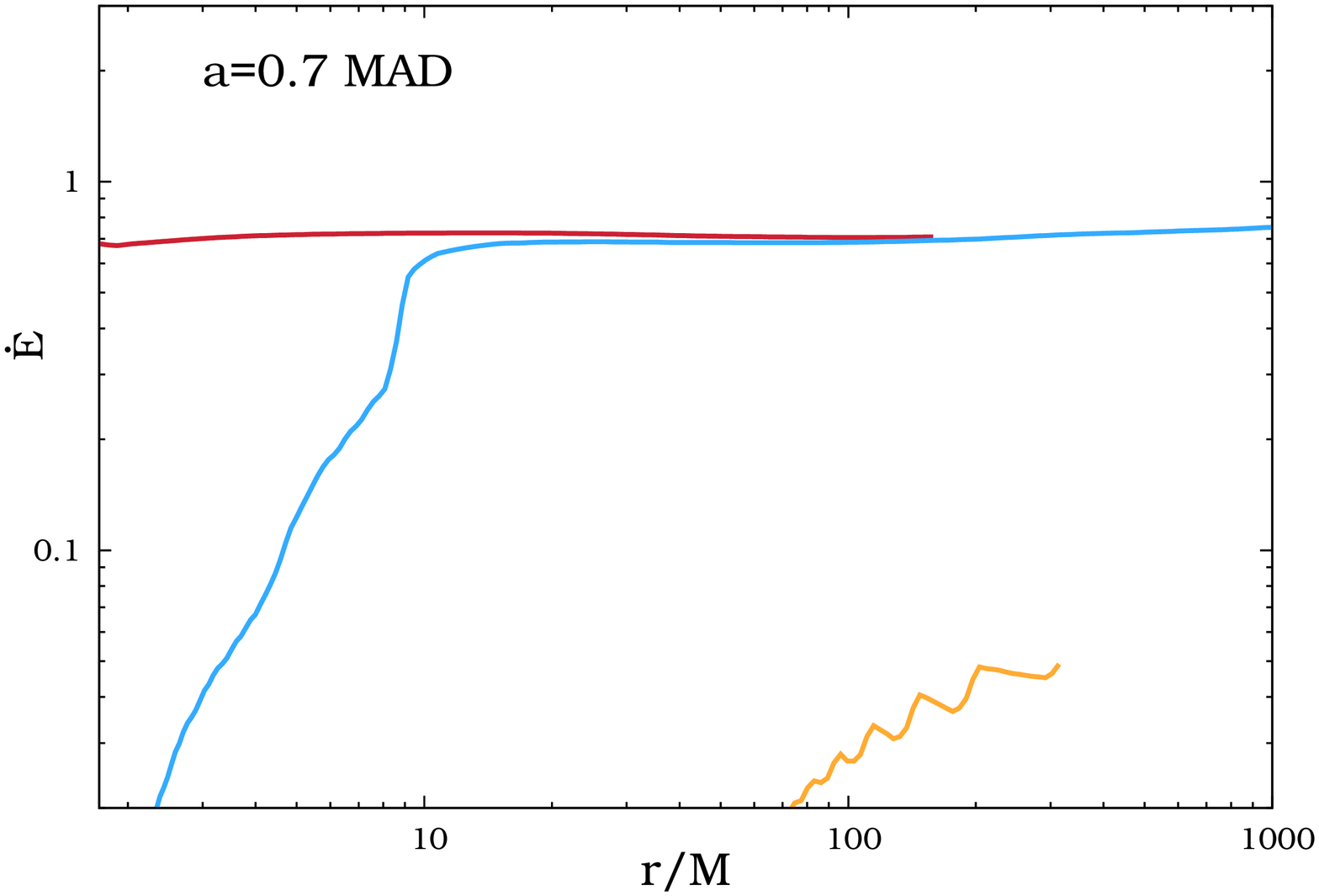}}
\caption{ Similar to Fig.~\ref{f.mdota7} but for the energy flux
  $\dot{E}$.}
  \label{f.edota7}
\end{figure}

Figure \ref{f.edota7} shows corresponding results for the energy flow
rate. Here the red lines show the efficiency of the accretion flow,
i.e., the energy that flows out to infinity normalized by the net mass
accretion rate. We see that the $a_*=0.7$ SANE run has an efficiency
of about 5\% whereas the MAD run has a much larger efficiency of about
70\%. The stronger magnetic flux around the BH in the latter enables
much more efficient tapping of the BH spin energy. The energy outflow
rates in the jet (blue curves) are well converged, just like the mass.
Even the energy outflows rates in the wind (orange curves) seem
reasonably well converged. This is consistent with the simple analysis
we present next.

\subsection{Radial scalings of outflows}
\label{s.scalings}

Let us define the mass inflow rate $\dot M_{\rm in}(r)$ at a given
radius as the sum of the net mass accretion rate, which we call
$\dot{M}_{\rm BH}$, and the mass outflow rate in the jet and wind
$\dot{M}_{\rm wind}$ \citep[e.g.,][]{SPB99, YYOB13}. Thus,
$\dot{M}_{\rm in}$ is the mass accretion rate we would calculate if we
restricted the $\theta$ integral in equation~\ref{eq:Mdot} to the
regions with $u^r<0$. It is reasonable to assume that $\dot{M}_{\rm
  wind}$ scales with radius as a power-law \citep{adios-paper}. Hence
let us assume that
\begin{eqnarray}
\dot M_{\rm wind}(r) &=& \dot M_{\rm BH}\left(\frac {r}{r_{\rm
    in}}\right)^s, \label{e.adios}\\ 
\dot M_{\rm in}(r) &=& \dot M_{\rm BH}
\left[1+\left(\frac {r}{r_{\rm in}}\right)^s\right],
\label{e.adios2}
\end{eqnarray} 
where we expect the index $s$ to lie in the range $0-1$, and $r_{\rm
  in}$ is some characteristic radius, typically of order tens of
$R_G$.  The black curves in Fig.~\ref{f.mdota7} show the variation of
$\dot{M}_{\rm in}$ vs $r$ for the two simulations. Note the
approximate power-law behavior, with $s\gtrsim 0.5$.

From equation (\ref{e.adios}) it is evident that the mass outflow rate
is dominated by large radii. Therefore, unless we have a reliable
estimate of the value of $s$, we cannot hope to obtain an accurate
estimate of the mass loss rate in an ADAF. The situation is better in
the case of energy outflow. The differential mass loss at a given
radius is given by \be d\dot M_{\rm wind}= d\dot M_{\rm
  in}=s\frac{\dot M_{\rm BH}}{r}\left(\frac {r}{r_{\rm in}}\right)^s
dr.  \ee We expect that any mass that flows out at radius $r$ will
carry with it an energy equal to some fraction $\xi$ of the local
potential energy. Thus, we estimate the local energy loss rate in the
wind to be \be d\dot E_{\rm wind}=d\dot M_{\rm wind}\frac{\xi
  c^2}{r}=\xi s \dot M_{\rm BH} c^2\frac{r^{s-2}}{r^s_{\rm in}} dr,
\ee which gives a cumulative energy loss rate of \be \dot E_{\rm
  wind}(r)=\frac{\xi s}{1-s}\frac{\dot M_{\rm BH} c^2}{r_{\rm
    in}}\left[1-\left(\frac{r_{\rm in}}{r}\right)^{1-s} \right]\approx
\frac{\xi s}{1-s}\frac{\dot M_{\rm BH} c^2}{r_{\rm in}}.  \ee We see
that the energy loss in the wind is dominated by the innermost regions
(as confirmed in Fig.~\ref{f.edota7}), hence the simulations ought to
provide reliable estimates of the energy feedback rate into the
surroundings.  The momentum outflow rate has a scaling intermediate
between those of mass and energy outflow.

GRMHD simulations of BH accretion discs are limited by computational
power. Even in the best cases (e.g., the simulations discussed here),
the inflow equilibrium region of a simulation extends only to radii of order a
few hundred $R_G$ (Table~\ref{t.convradii}). Because of this
limitation, we can obtain meaningful estimates of the total mass loss
rate in the wind only if the mass outflow behaves in a self-similar
fashion and the range of inflow equilibrium of the simulation is far enough
from the BH that we can obtain a reliable estimate of the index $s$.
Having this caveat in mind, we discuss first the more reliable energy
outflow.

\subsection{Energy outflow}
\label{s.energy}

\begin{figure*}
\centering
\subfigure{\includegraphics[height=.35\textwidth,angle=270]{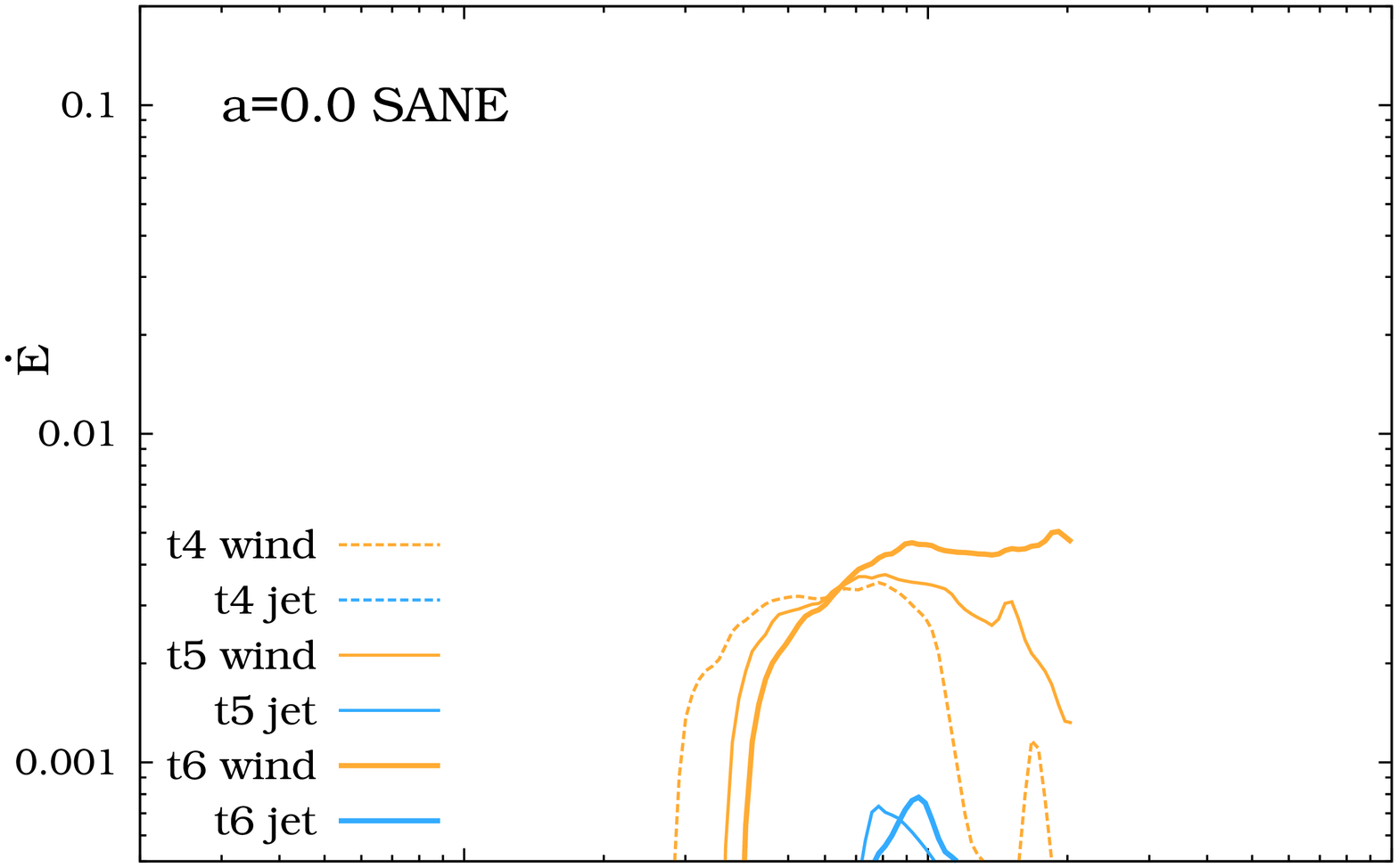}}\hspace{-.8cm}
\subfigure{\includegraphics[height=.35\textwidth,angle=270]{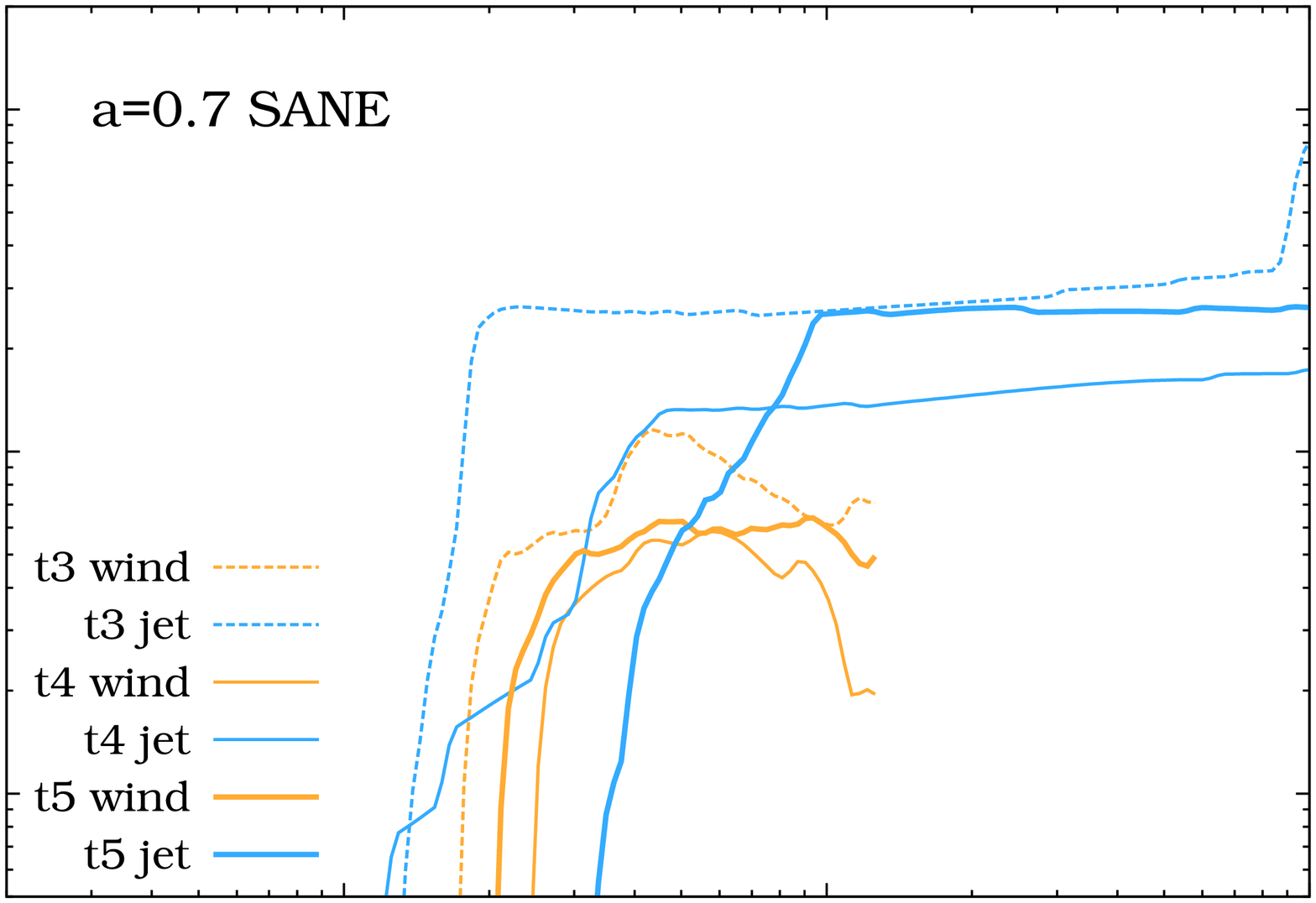}}\hspace{-.8cm}
\subfigure{\includegraphics[height=.35\textwidth,angle=270]{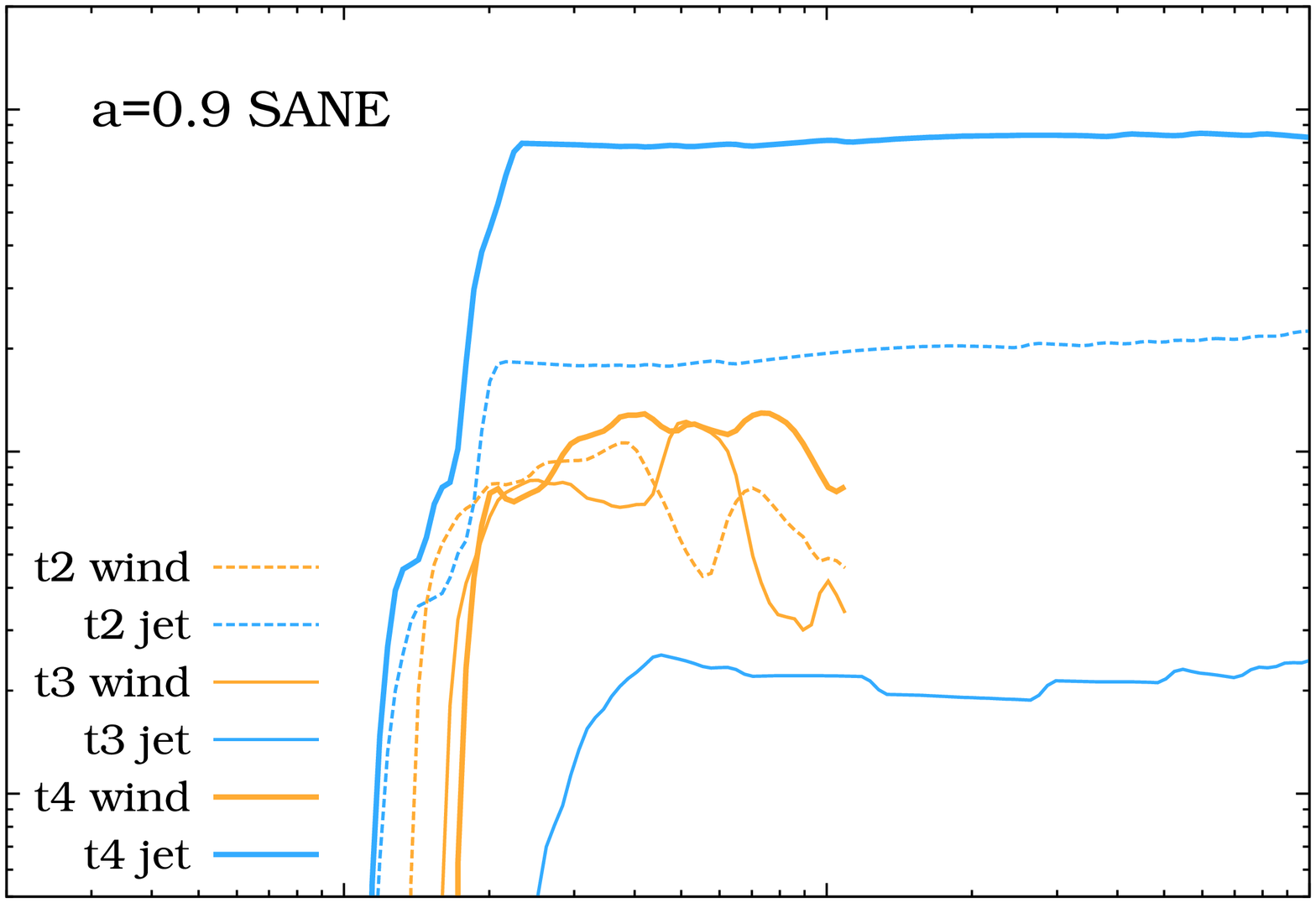}}\vspace{-.8cm}\\
\subfigure{\includegraphics[height=.35\textwidth,angle=270]{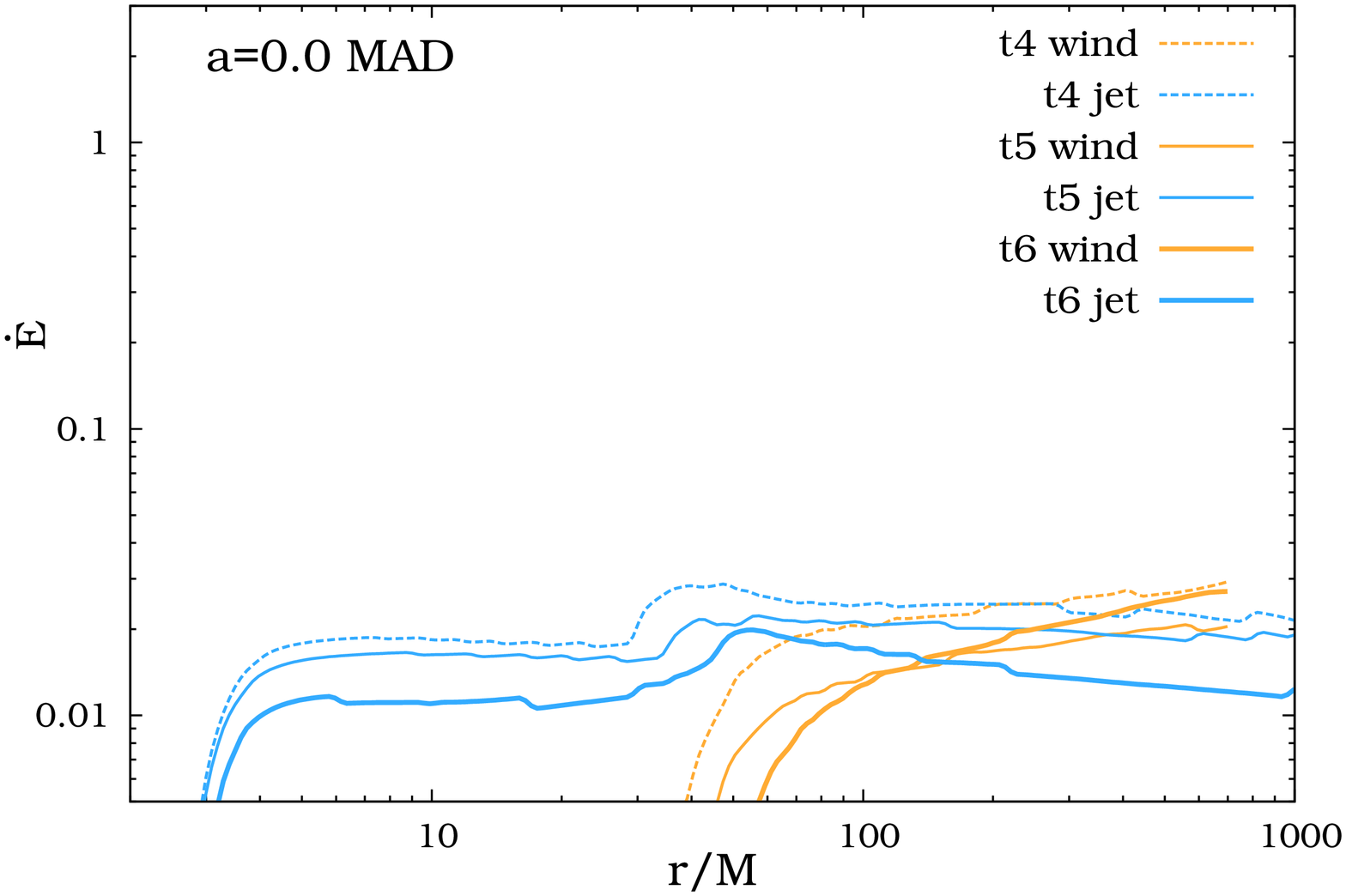}}\hspace{-.8cm}
\subfigure{\includegraphics[height=.35\textwidth,angle=270]{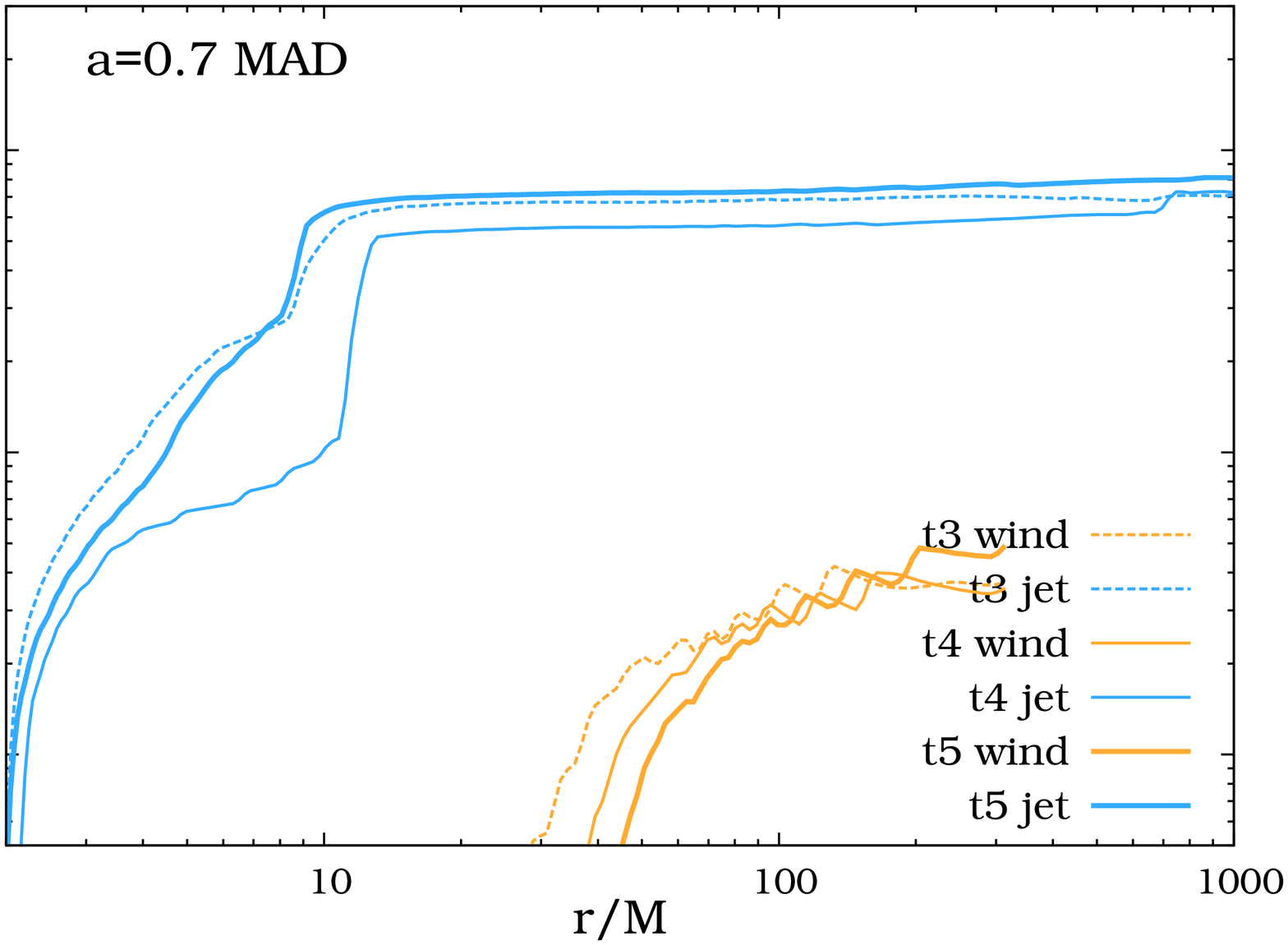}}\hspace{-.8cm}
\subfigure{\includegraphics[height=.35\textwidth,angle=270]{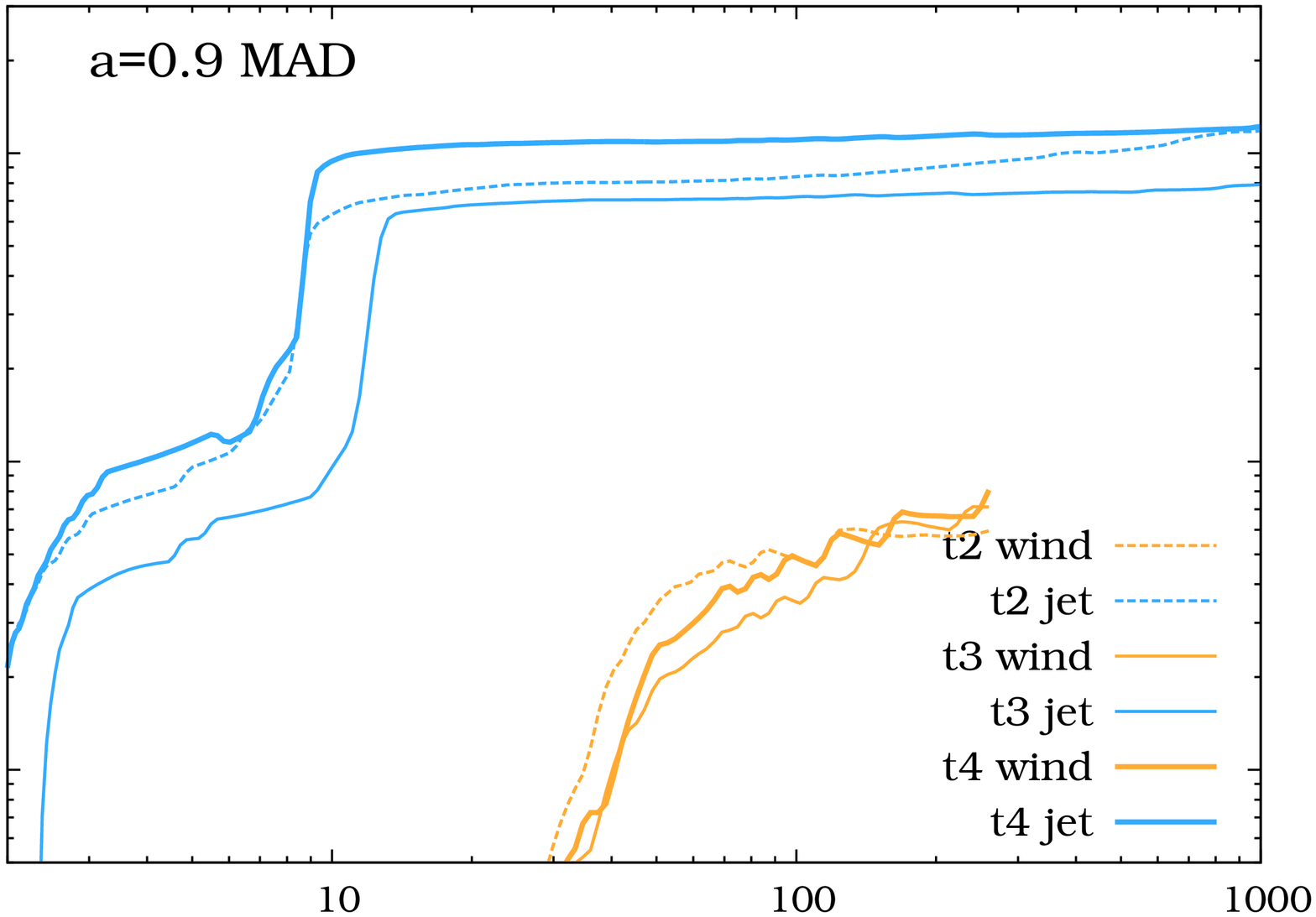}}
\caption{Fluxes of energy ($\dot{E}$) for SANE (top) and MAD
  (bottom) simulations with given value of BH spin. Blue and orange
  lines correspond to the jet and wind regions, respectively. On each
  sub-panel three sets of lines are plotted corresponding to three
  most recent chunks of time for each simulation.}
\label{f.tNedot}
\end{figure*}

Figure~\ref{f.tNedot} shows radial profiles of energy
outflow in the jet and wind regions in the final three time chunks of
our simulations: chunks T4, T5, T6 for the $a_*=0$ SANE and MAD runs,
T3, T4, T5 for the two $a_*=0.7$ runs, and T2, T3, T4 for the two
$a_*=0.9$ runs. We see that the wind energy profiles (orange curves)
are poorly converged, in agreement with \cite{narayan+12a} who found that outflows from discs around non-rotating
BHs showed poor convergence with time. On average, the longer the duration of a given
simulation, the further from BH the winds originate. There is
non-monotonic behavior in the $a_*=0.7$ SANE and $a_*=0.9$ MAD models,
but we believe this is simply because the magnetic flux around the BH
increased in the last time chunk for these two simulations (compare
Fig.~\ref{f.fluxes_aN}).  Note that the energy lost in the wind for
the SANE models, and to a lesser extent for the MAD models, is
independent of radius, i.e., the energy budget is dominated by the
innermost region of the wind, in agreement with the scalings discussed
in Section~\ref{s.scalings}.

The energy profiles $\dot{E}(r)$ of the simulated jets show very good
spatial convergence, asymptoting to a constant value at large
radii. This indicates that the jet criterion we adopted ($\mu>\mu_{\rm
  crit}$) closely follows the energy flux streamlines at large radii.
Variations among different time chunks is due to fluctuations in the
magnetic flux at the horizon, as discussed further below.

\begin{figure}
\centering
\subfigure{\includegraphics[height=.45\textwidth,angle=270]{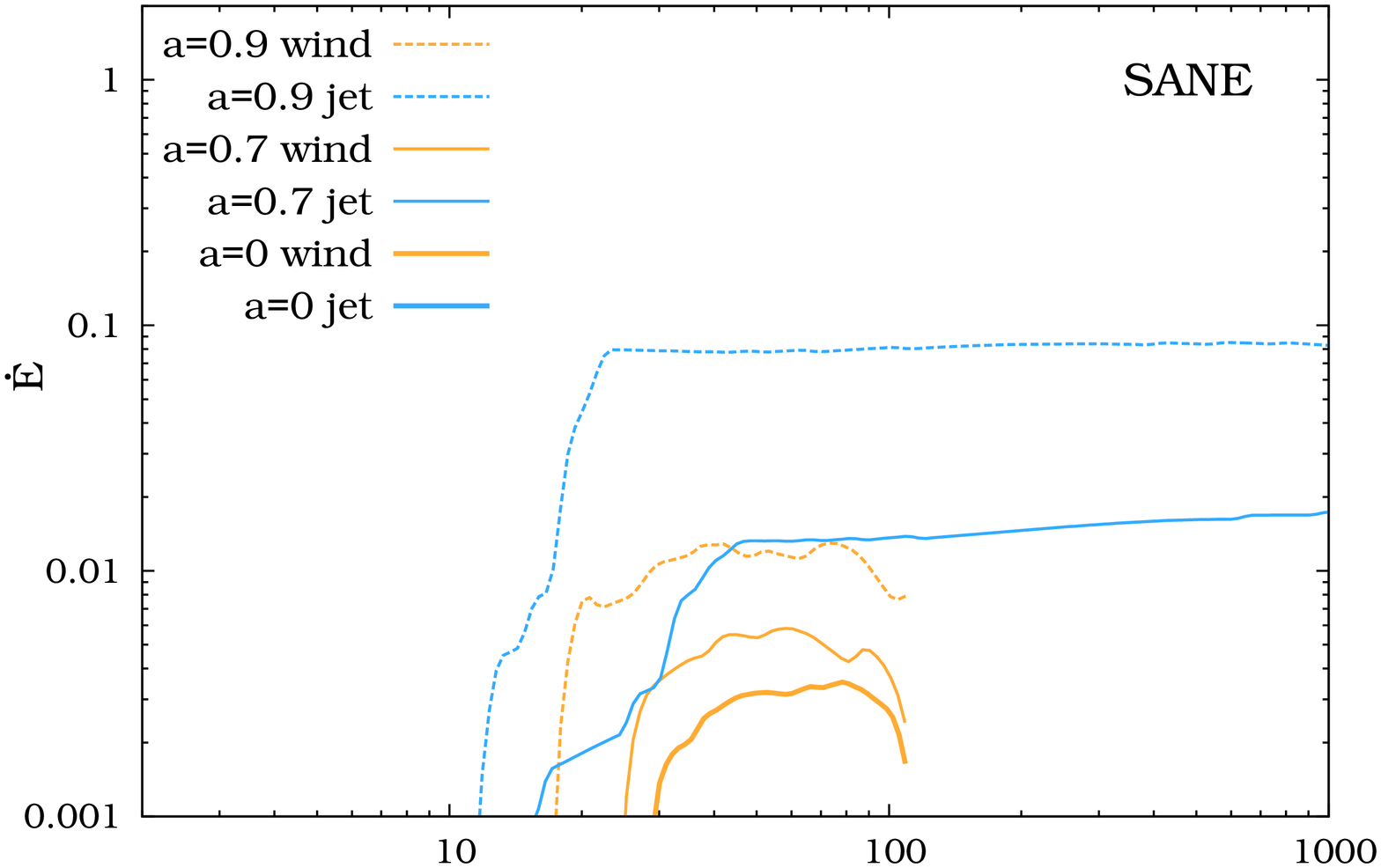}}
\subfigure{\includegraphics[height=.45\textwidth,angle=270]{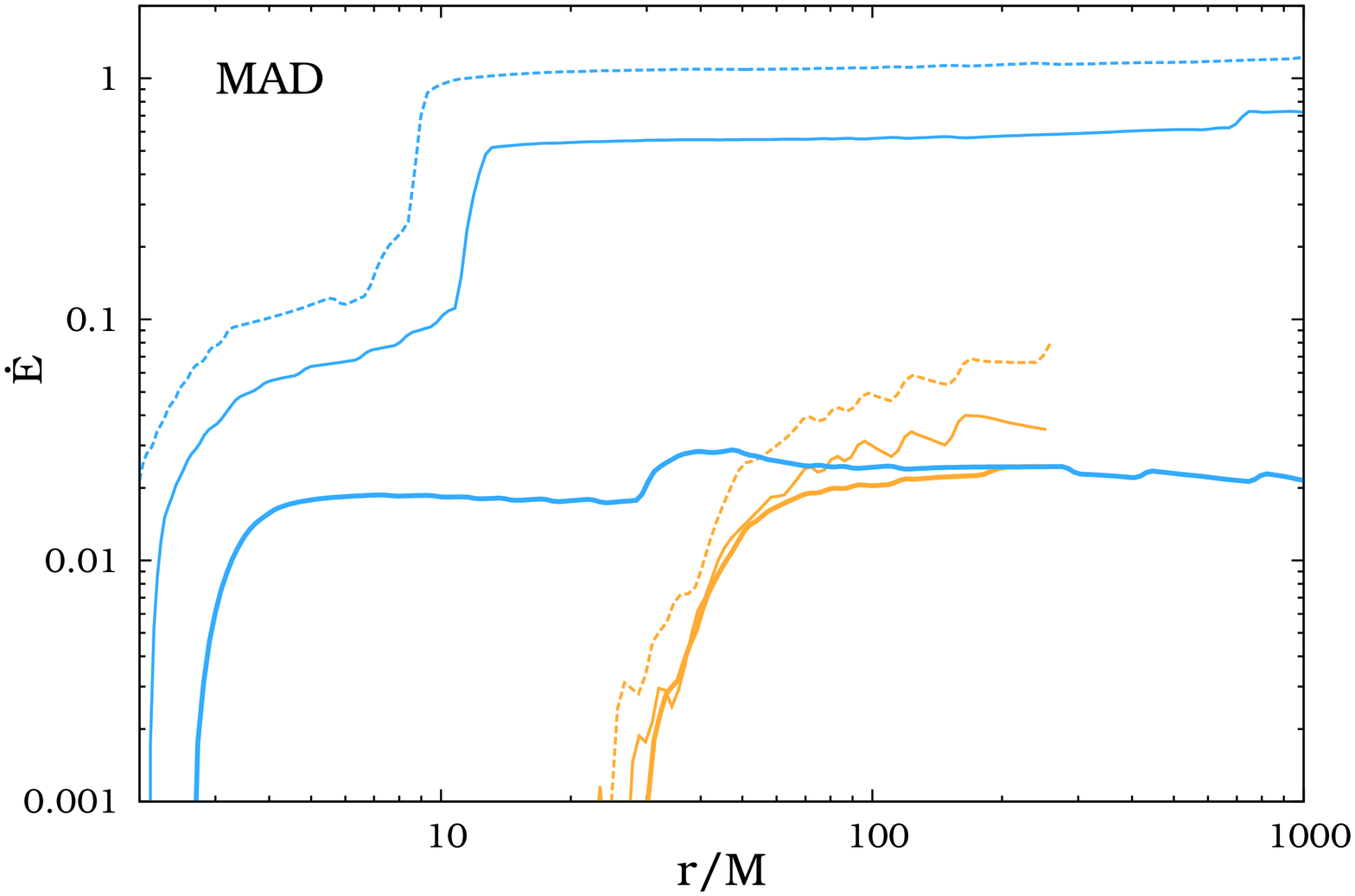}}
\caption{Radial profiles of the energy flux ($\dot{E}$)
  for SANE (top) and MAD (bottom panel) models at time chunk
  T4. Profiles for three values of BH spins are presented. All
  fluxes are normalized to $\dot{M}$ at $r=10$ and the lines are
  terminated at the radius of outflow equilibrium of the corresponding
  $a_*=0.9$ model.}
\label{f.edott4}
\end{figure}

Because of the lack of convergence with time of the wind regions in
the simulations, it is not obvious how one should compare different
simulations to study the effect of BH spin. In the following, we
choose to compare simulations at the same physical time, viz., time chunk T4.
Figure~\ref{f.edott4} shows the results. The power of both the jet and
the wind increases with BH spin for both SANE and MAD simulations. The
variation in the case of the wind is modest, whereas jet power shows a
very strong dependence on BH spin. In fact, the jet power in the
$a_*=0.9$ MAD simulation exceeds $\dot M c^2$, showing that the jet is
powered by more than accretion energy. At least some part of the power
must come directly from the BH \citep{tchekh+11}.

Theoretical jet models indicate that the power extracted from a
spinning accreting BH scales as \citep{blandfordznajek,Tchekhovskoy+12a,penna+13},
\begin{equation}
\label{eq.bz}
\eta_{\rm jet} = \frac{P_{\rm jet}}{\dot{M}c^2} \sim \frac{0.05}{4\pi c}
\Phi_{\rm BH}^2 \Omega_H^2,
\end{equation} 
where $\Phi_{\rm BH}^2$ is the magnetic flux threading the horizon
(eq.~\ref{e.phibh}), and  $\Omega_H$ is the angular velocity of the outer BH horizon (eq.~\ref{eq.omhor}).
 The jet
powers in our simulations are generally consistent with the spin
dependence in this formula; the $a_*=0.9$ has the strongest jet and
$a_*=0$ has essentially no jet.  Note that the non-zero jet power we
find for the $a_*=0.0$ MAD model is an artifact caused by the
activation of numerical floors at the polar axis and the corresponding
injection of mass and energy.

\begin{figure}
  \centering
\subfigure{\includegraphics[height=.45\textwidth,angle=270]{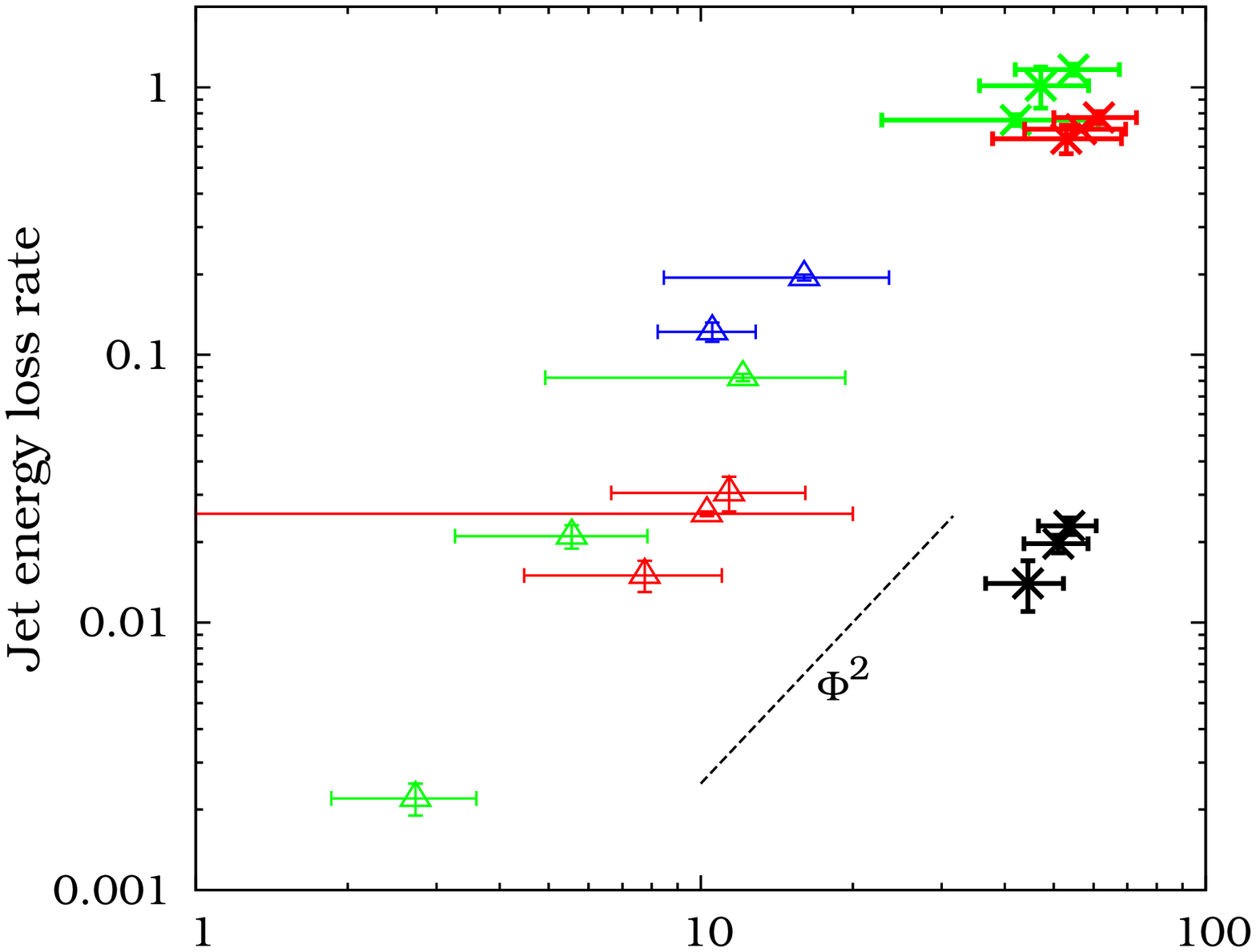}}\vspace{-1.5cm}
\subfigure{\includegraphics[height=.45\textwidth,angle=270]{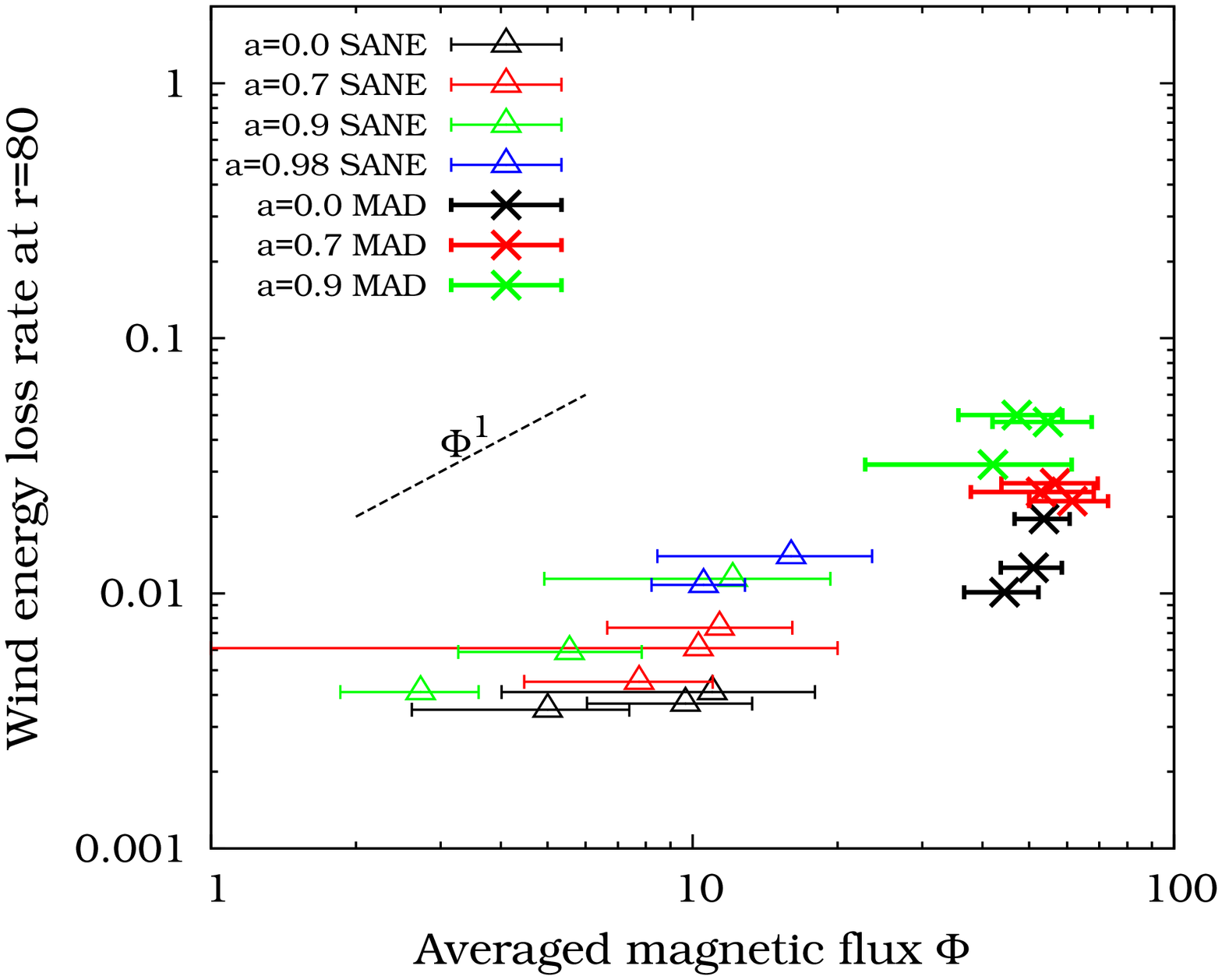}}
\caption{Outflow rates of the energy flux ($\dot{E}$) in the jet (top) and the wind (bottom panel) as
a function of the magnetic flux at BH horizon averaged over the duration of given time chunk.
For each model values corresponding to the three most recent time chunks are presented. Triangles and crosses
are for SANE and MAD models, respectively. Colors denote BH spin. The outflow rate in the wind was measured
at $r=80$ while the jet power is averaged over $r=100\div1000$ and the vertical errorbars show the 
minimum and maximum values in this range.}
  \label{f.edotvsflux}
\end{figure}

In Fig.~\ref{f.edotvsflux} we test the scaling of jet power with
$\Phi_{\rm BH}^2$. The horizontal axes show the magnetic flux
$\langle\Phi_{\rm BH}\rangle$ (see eq.~\ref{e.phibh}) averaged over a
given chunk of time.  For each simulation, three points are shown,
corresponding to the final three time chunks of that simulation (only
two points in the case of the $a_*=0.98$ model because of the very
short duration of this run). The horizontal error bars reflect the
standard deviation of the magnetic flux within the particular time
chunk.  Colors denote BH spin. Triangles and crosses correspond to
SANE and MAD simulations, respectively.

The top panel in Fig.~\ref{f.edotvsflux} shows the energy loss rate in
the jet as a function of magnetic flux at the horizon.  Within each
model and for a given spin, the jet power follows the $\Phi_{\rm
  BH}^2$ scaling quite well, validating this prediction of theory. The
bottom panel of the same figure shows the energy carried away by the
wind measured at a common radius of r=80. This quantity again
increases with the magnetic flux $\Phi_{\rm BH}$, but the dependence
is much weaker than for the jet.  This shows that the wind receives
only some of its energy from the BH, the rest coming directly from the
gravitational energy released by the accreting gas.

\subsection{Mass outflow}
\label{s.massloss}

\begin{figure*}
\centering
\subfigure{\includegraphics[height=.35\textwidth,angle=270]{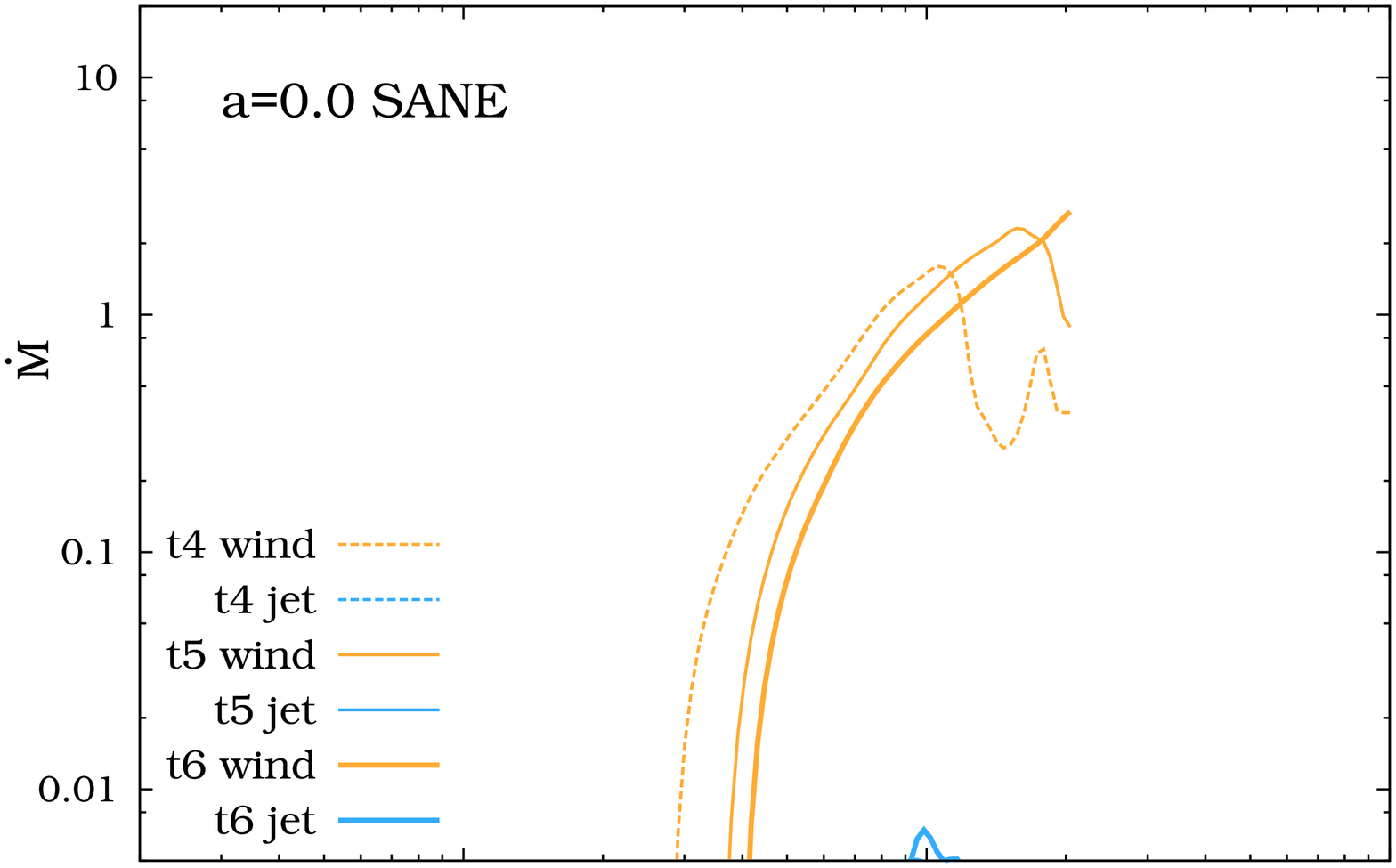}}\hspace{-.8cm}
\subfigure{\includegraphics[height=.35\textwidth,angle=270]{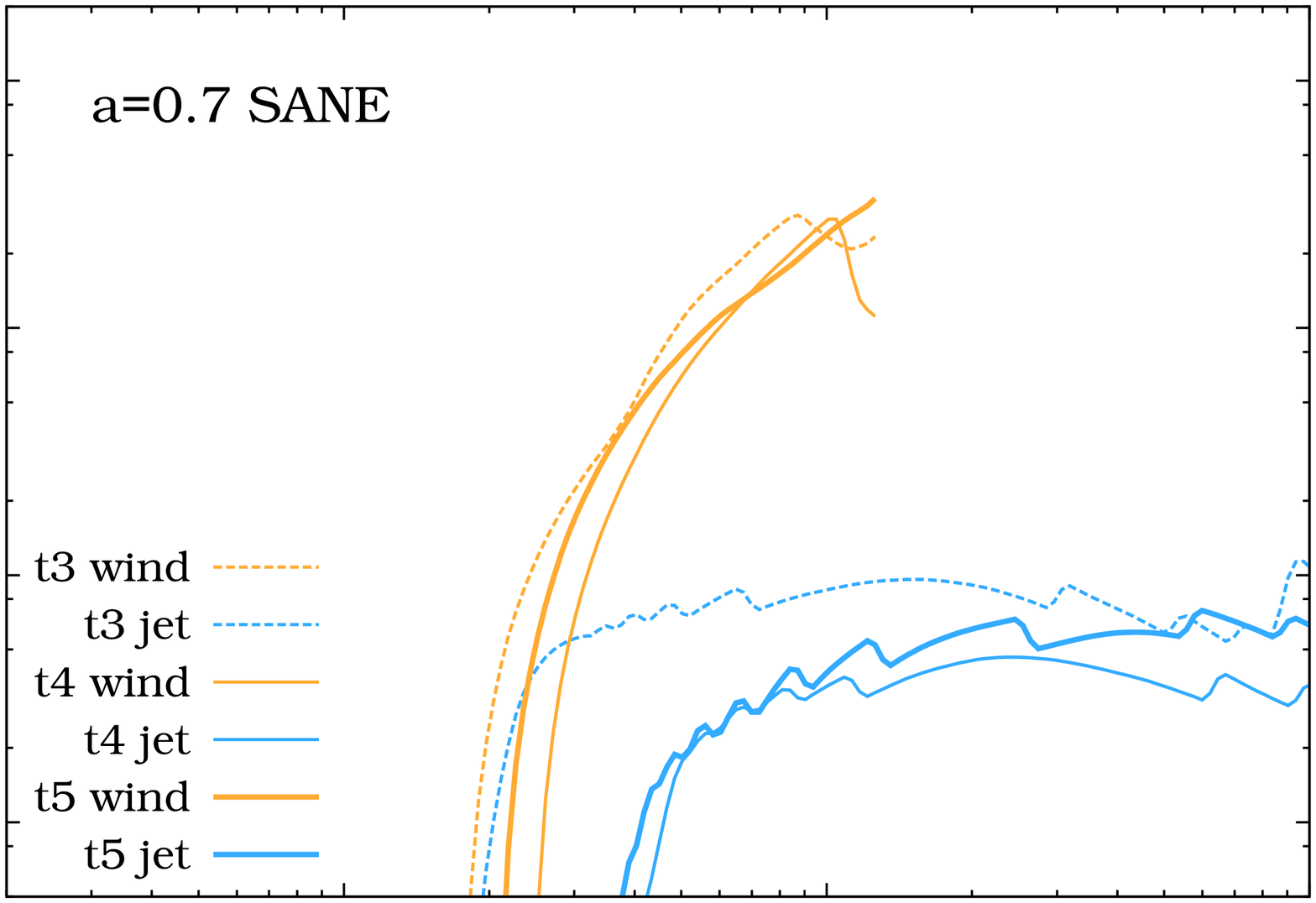}}\hspace{-.8cm}
\subfigure{\includegraphics[height=.35\textwidth,angle=270]{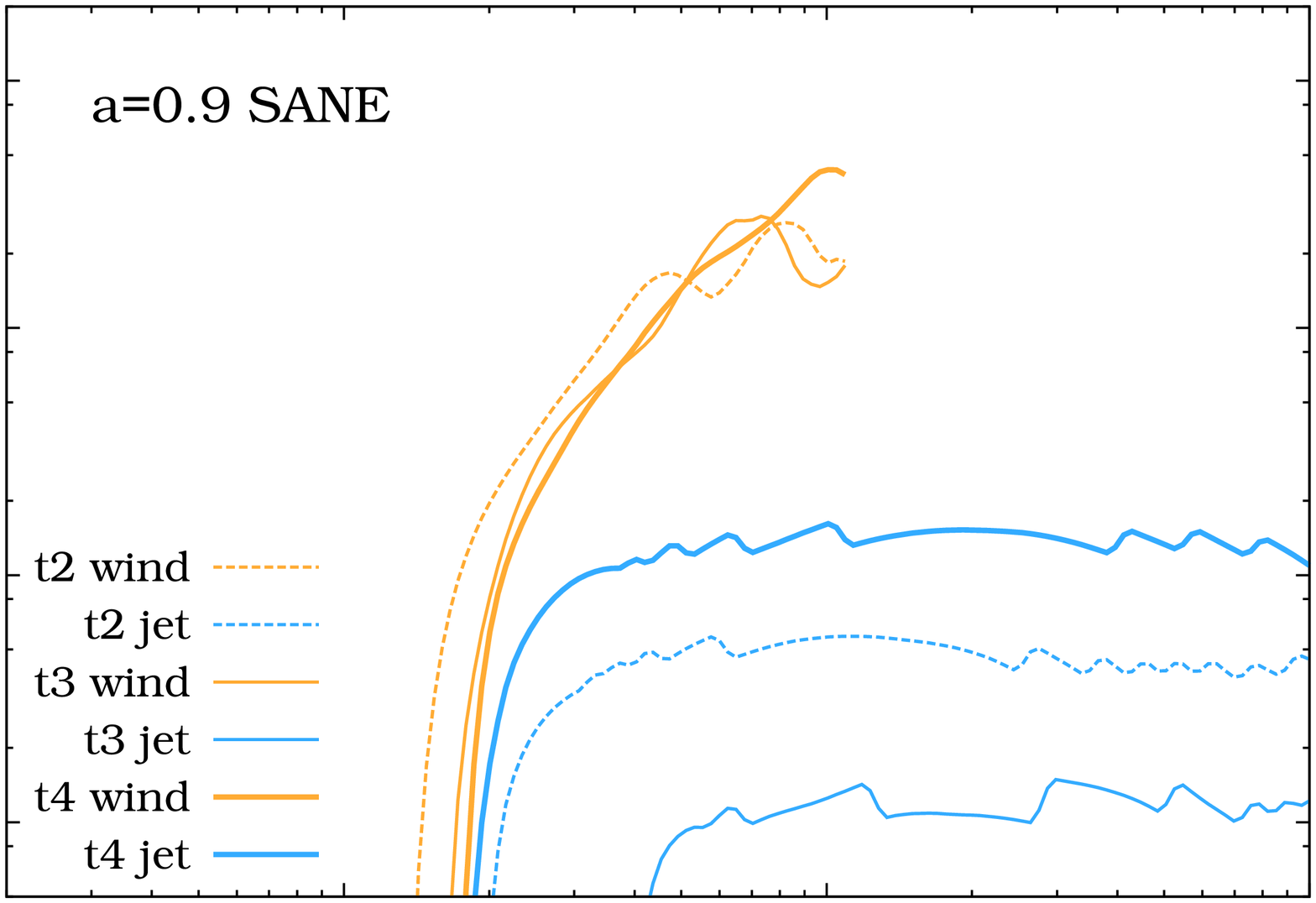}}\vspace{-.8cm}\\
\subfigure{\includegraphics[height=.35\textwidth,angle=270]{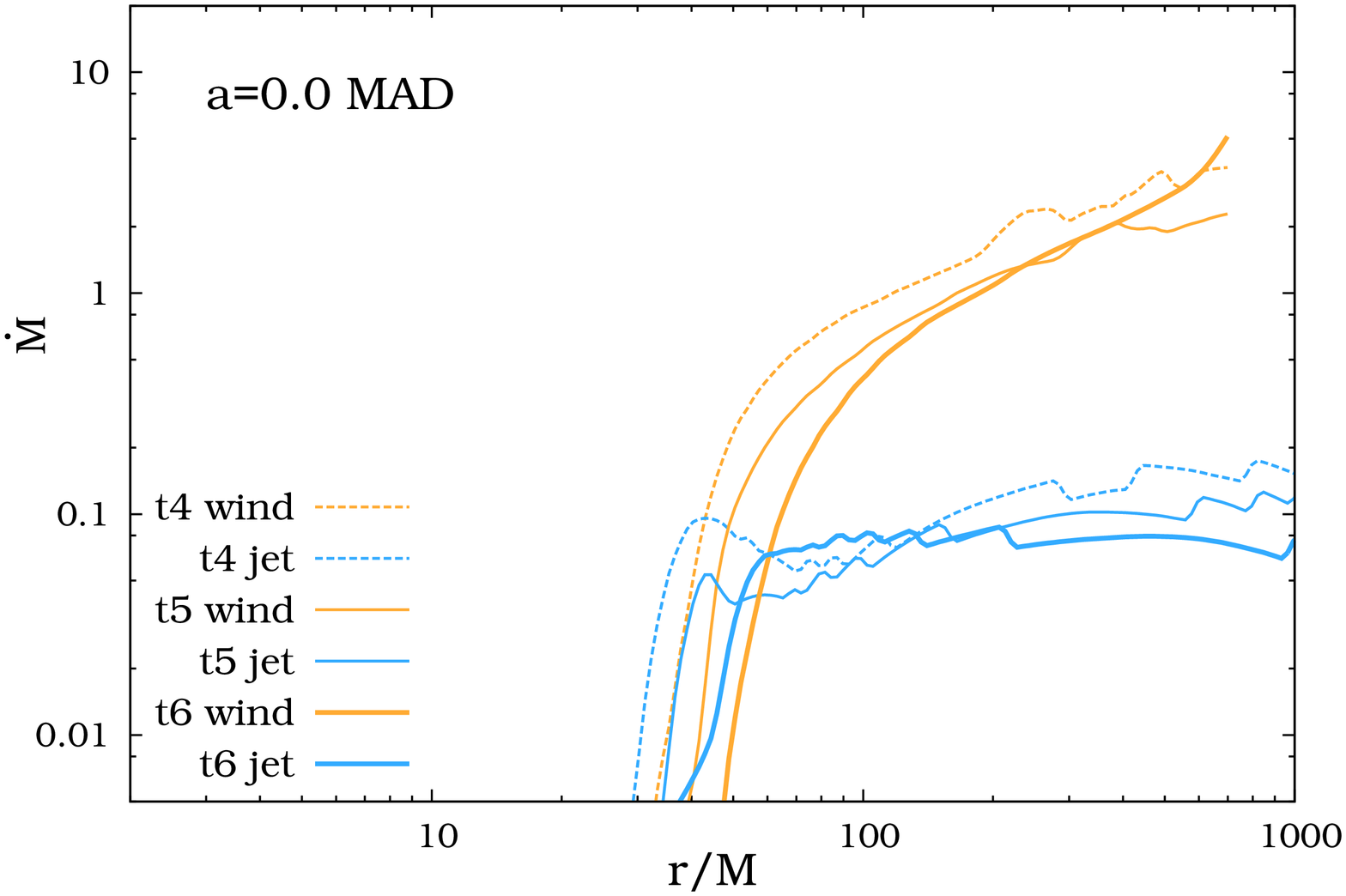}}\hspace{-.8cm}
\subfigure{\includegraphics[height=.35\textwidth,angle=270]{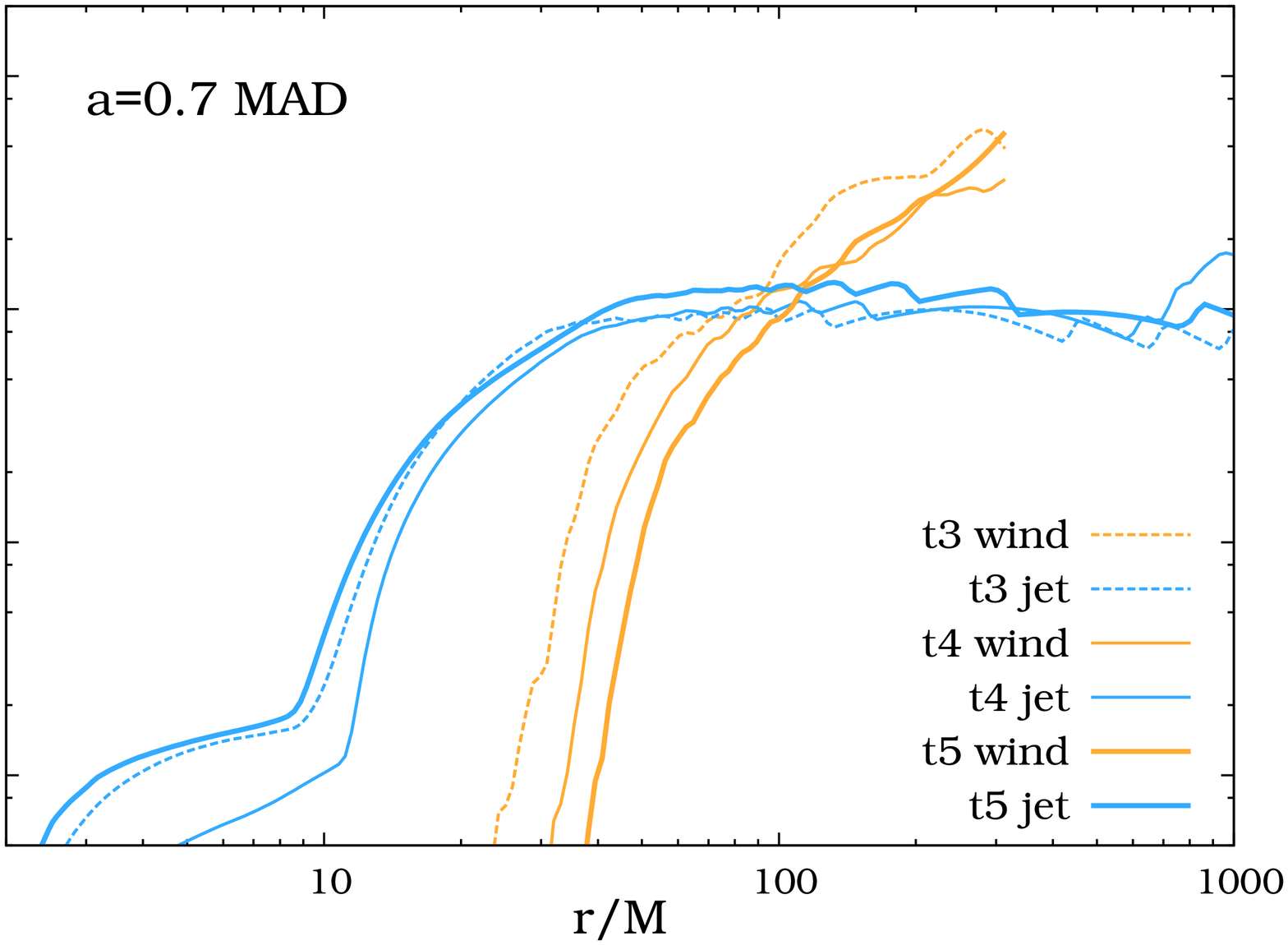}}\hspace{-.8cm}
\subfigure{\includegraphics[height=.35\textwidth,angle=270]{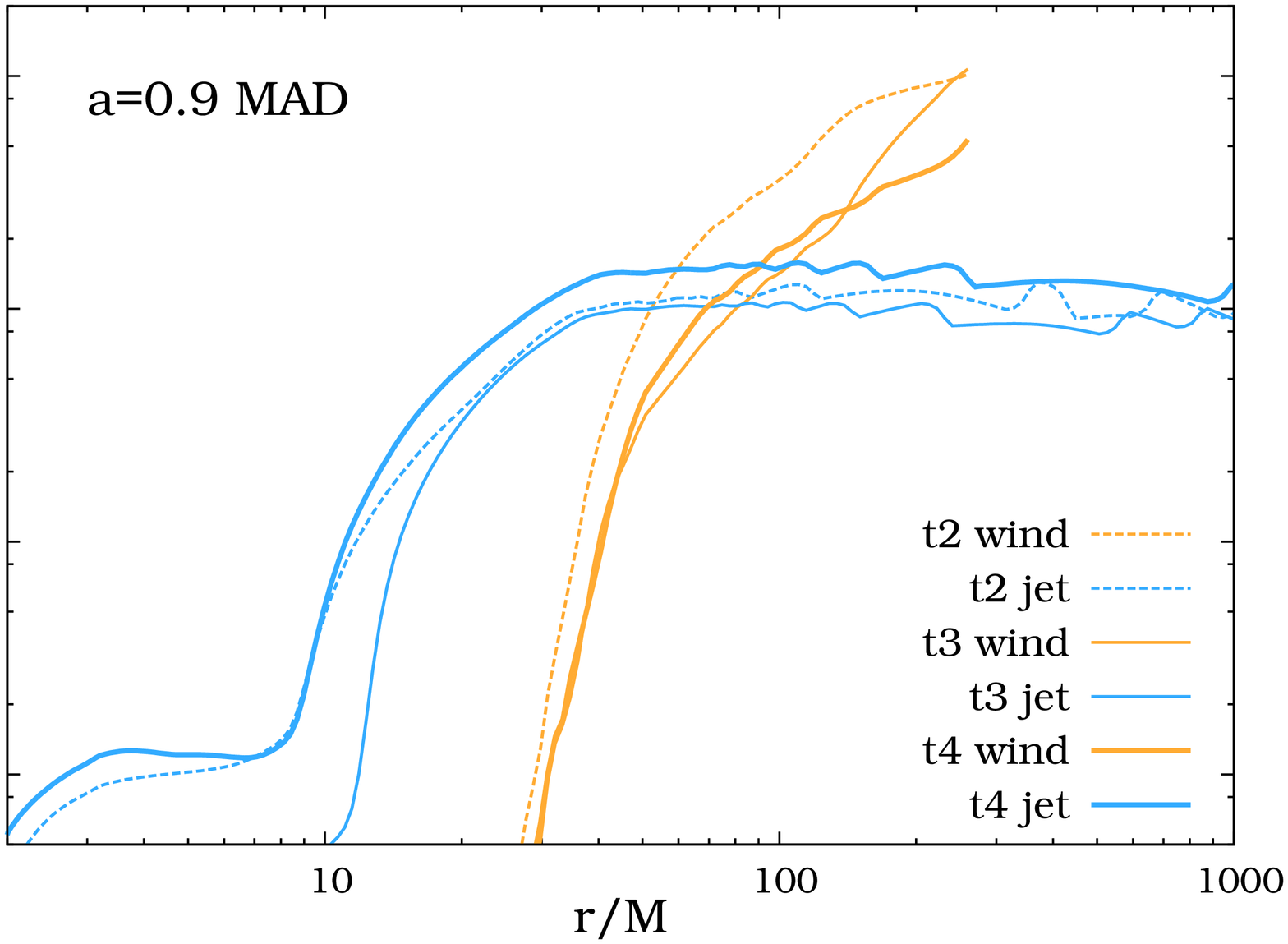}}
\caption{
Similar to Fig.~\ref{f.tNedot} but for the mass flux ($\dot{M}$).
}
\label{f.tNmdot}
\end{figure*}

Figure~\ref{f.tNmdot} shows radial profiles of mass outflow for SANE
and MAD models with BH spin $a_*=0$, $0.7$, and $0.9$. For each model,
the last three time chunks are shown. Lack of convergence of the mass
outflow rate in the wind as a function of time is clearly visible for
most of the models --- the orange lines move steadily outward with
time.  The mass outflow rate in the jet is well converged in space,
i.e., it quickly saturates at a consant value. However, it varies with
time (most profound for $a_*=0.9$ SANE model) as a result of changing
magnetic flux threading the horizon (Fig.~\ref{f.fluxes_aN}).

\begin{figure}
  \centering
\subfigure{\includegraphics[height=.45\textwidth,angle=270]{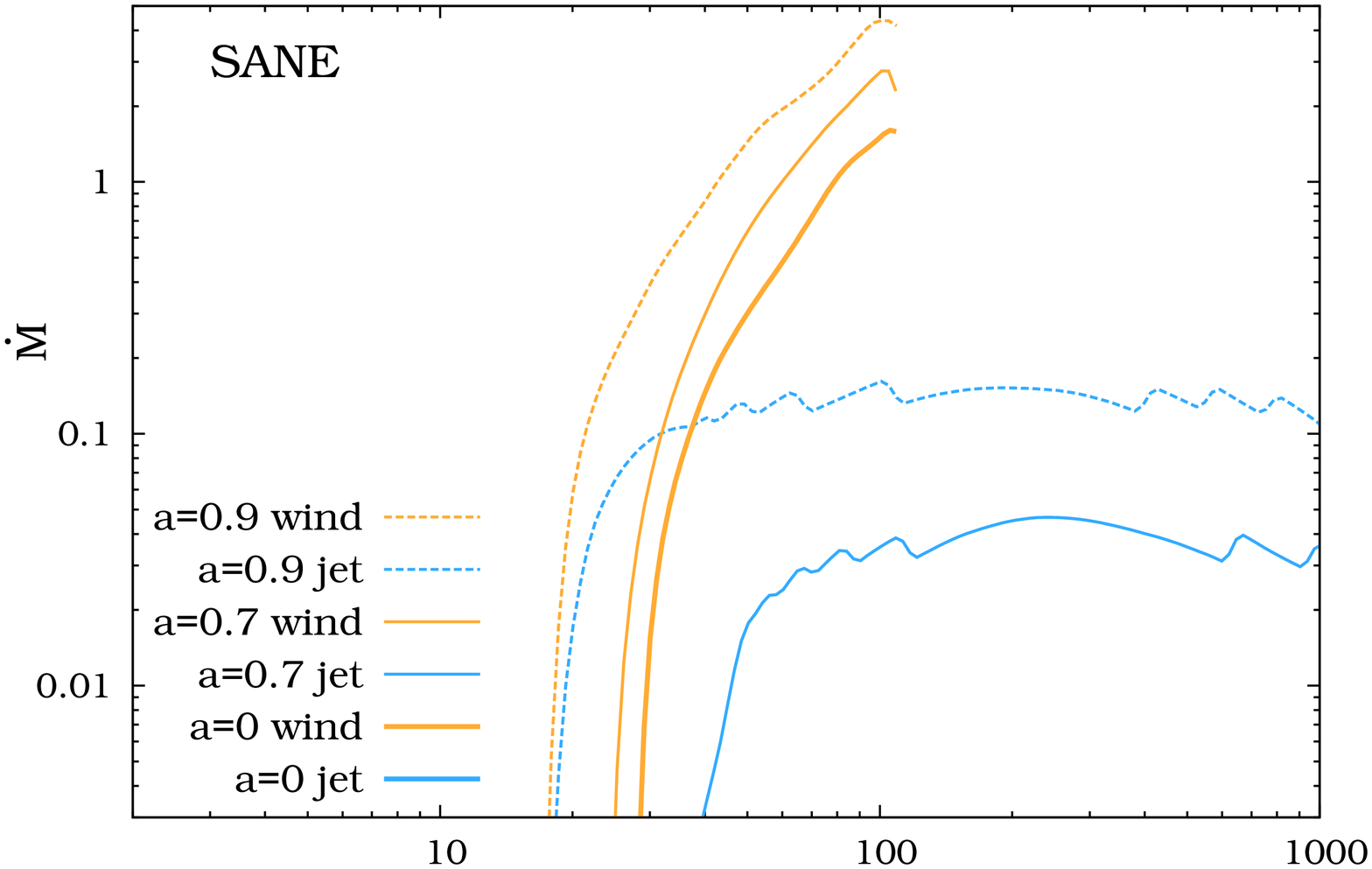}}
\subfigure{\includegraphics[height=.45\textwidth,angle=270]{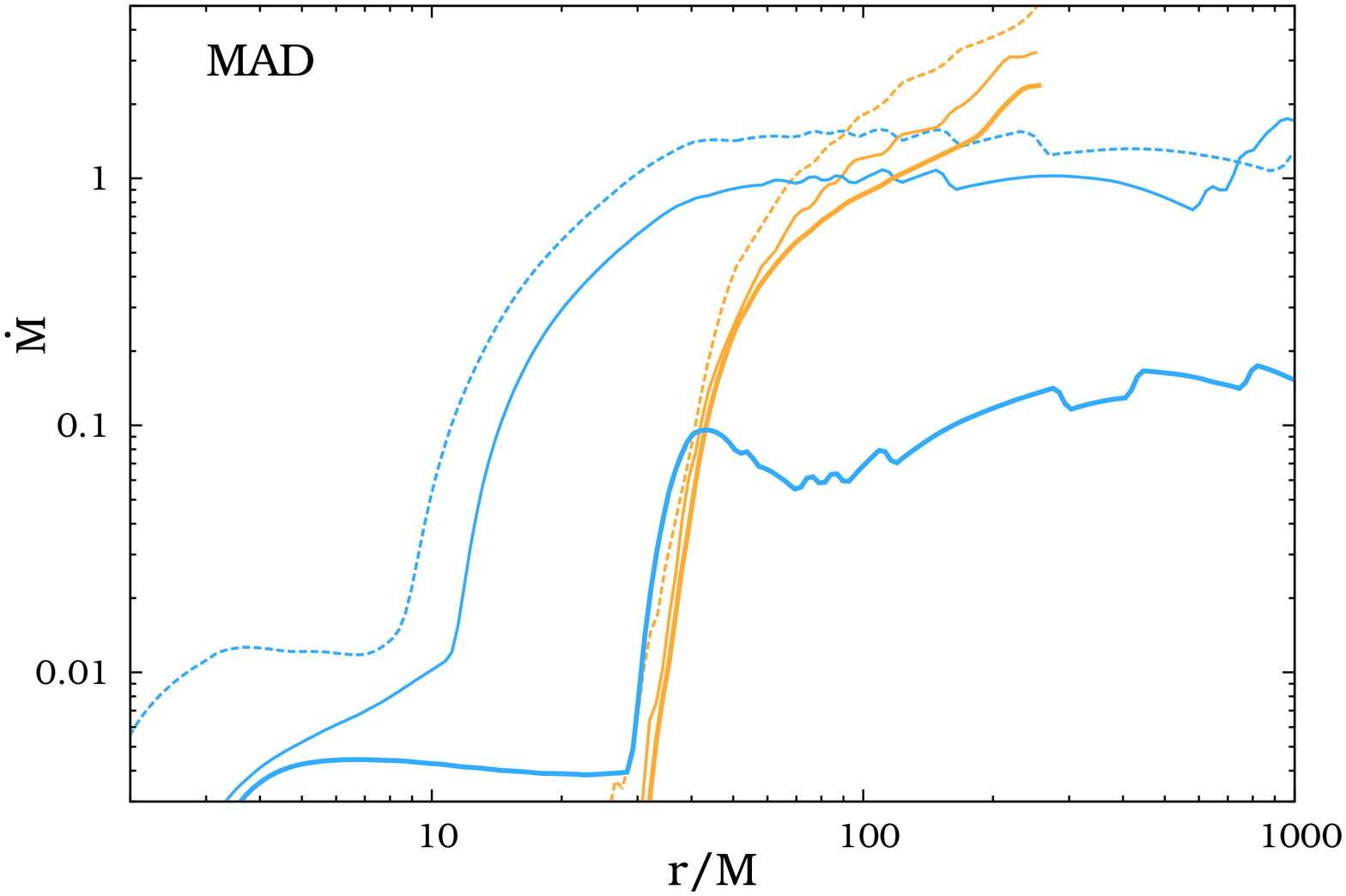}}
\caption{
Similar to Fig.~\ref{f.edott4} but for the mass flux ($\dot{M}$).}
  \label{f.mdott4}
\end{figure}

Fig.~\ref{f.mdott4} shows profiles of mass outflow in the jet and wind
for time chunk T4 and three BH spins, $a_*=0$, $0.7$, and $0.9$. In
the SANE simulations (top panels), where the jet is weak, mass loss is
dominated by the wind at all radii.  Mass loss is strongest and starts
closest to the BH for the highest value of BH spin. In the case of the
MAD solutions (bottom panels), mass outflow in the jet (cocoon)
dominates at smaller radii (except for $a_*=0$), and the wind takes
over only at larger radii. Outflow rates initially overlap each other
and then diverge showing the same dependence on BH spin. The jet mass
loss rate in particular shows noticeable dependence on the BH spin.

\begin{figure}
  \centering
\subfigure{\includegraphics[height=.45\textwidth,angle=270]{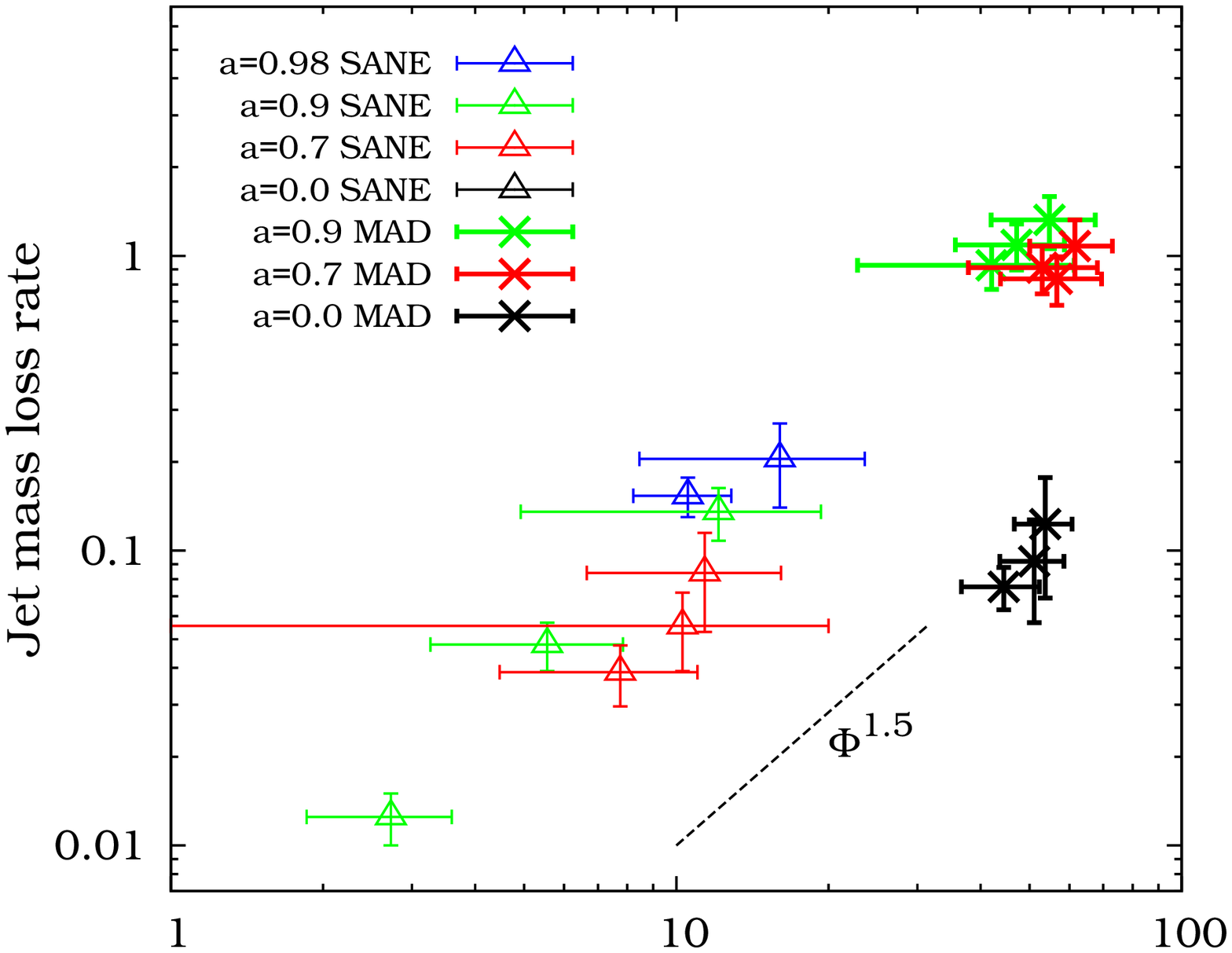}}\vspace{-1.5cm}
\subfigure{\includegraphics[height=.45\textwidth,angle=270]{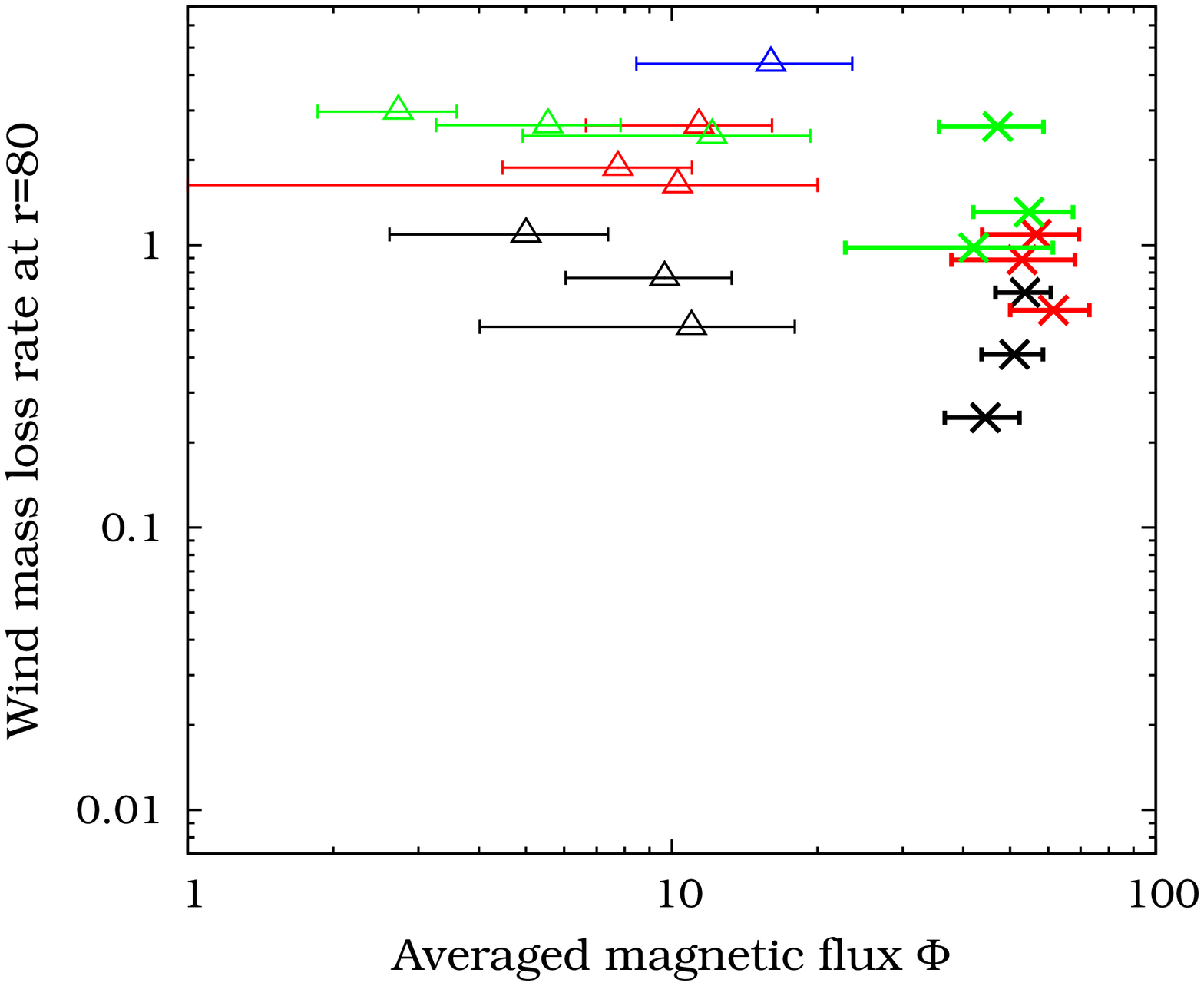}}
\caption{Similar to Fig.~\ref{f.edotvsflux} but for the rest mass flux $\dot{M}$.}
  \label{f.mdotvsflux}
\end{figure}

The top panel of Fig.~\ref{f.mdotvsflux} shows the relation between
the magnetic flux at the horizon $\Phi_{\rm BH}$ and the mass loss
rate in the jet. There is a strong correlation, with a dependence
approximately $\propto \Phi_{\rm BH}^{1.5}$, which is similar to the
scaling of the jet energy loss rate ($\Phi_{\rm BH}^2$) but a little
shallower.  The slightly different slope is because the terminal
Lorentz factor of the jet ($u^t$) scales with the magnetic flux
roughly as $\Phi_{\rm BH}^{0.5}$, i.e., jets in MAD solutions are
typically more relativistic than those in SANE solutions.

Thus the mass in the jet
cocoon is driven essentially by energy outflow from the BH
horizon. There is also a clear dependence on BH spin, with larger
spins giving stronger mass loss in the jet.  The mass loss in the jet
for the $a_*=0$ MAD model is an artifact, and represents leakage of
mass from the the disc wind into the jet region.

The bottom panel shows the mass loss rate in the wind as measured at
$r=80$ (which lies inside the wind outflow equilibrium region for all the
simulations) as a function of $\Phi_{\rm BH}$.  It is clear that, for
a given BH spin, there is essentially no correlation between the mass
loss in the wind and the magnetic flux. This is reasonable. Winds flow
out from relatively large radii in the disc, where the gas is not
sensitive to the magnetic flux near the BH.  Note, however, that there
is a correlation between the wind mass loss rate and BH spin --- on
average higher spin leads to stronger mass outflow.

\subsection{Outflows of momentum}
\label{s.momentum}

The relativistic momentum flux given by eq.~(\ref{eq:Pdot}) reduces for particles to,
\be \dot P=\frac{\dot Mv}{\sqrt{1-v^2}}, \ee
 and in the non-relativistic limit gives $\dot P=\dot Mv$. Therefore, one could expect
that the profiles of outflowing momentum are similar to the profiles of lost rest-mass but
normalized by a coefficient related to the total gas velocity $v$.

\begin{figure*}
\centering
\subfigure{\includegraphics[height=.35\textwidth,angle=270]{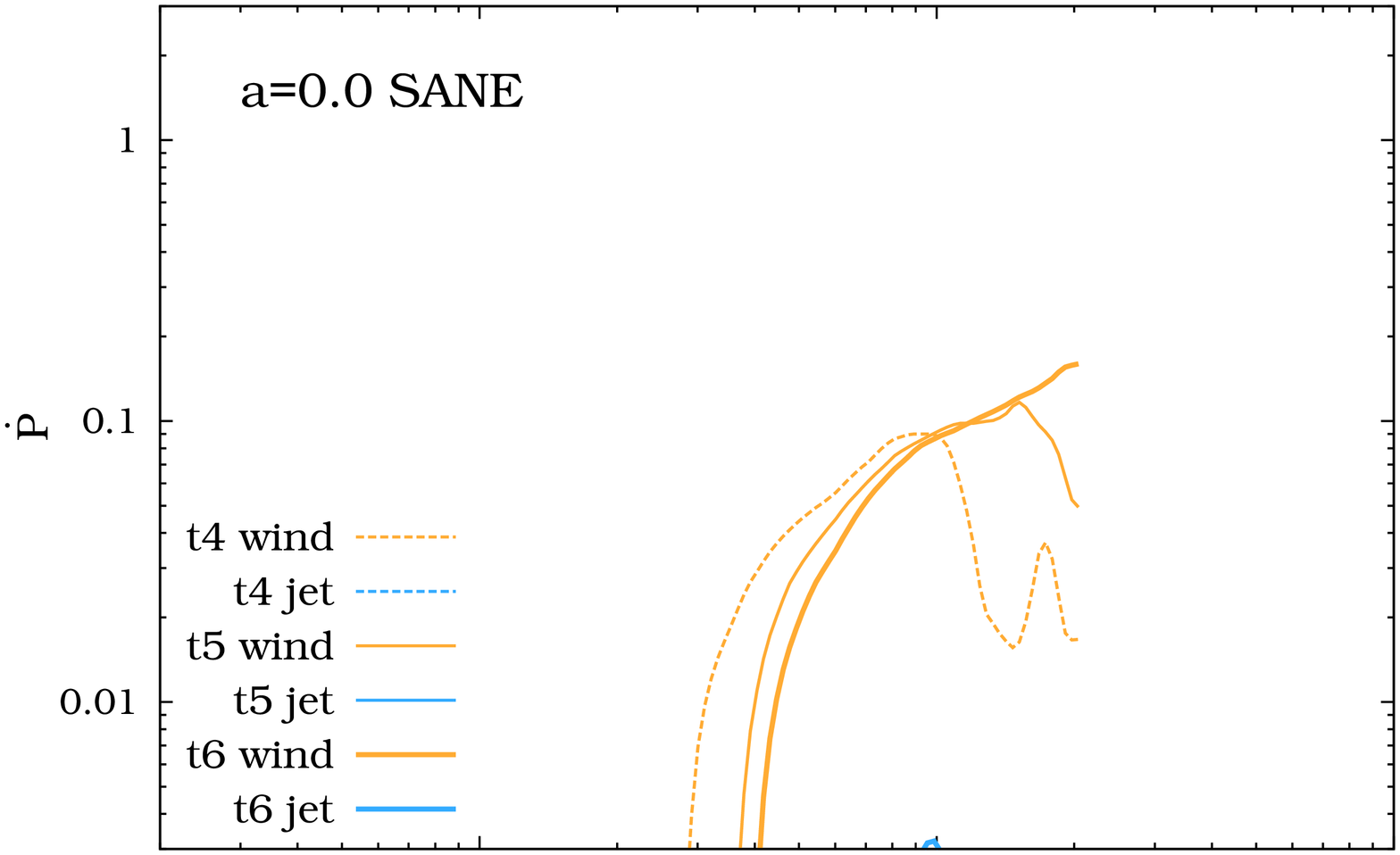}}\hspace{-.8cm}
\subfigure{\includegraphics[height=.35\textwidth,angle=270]{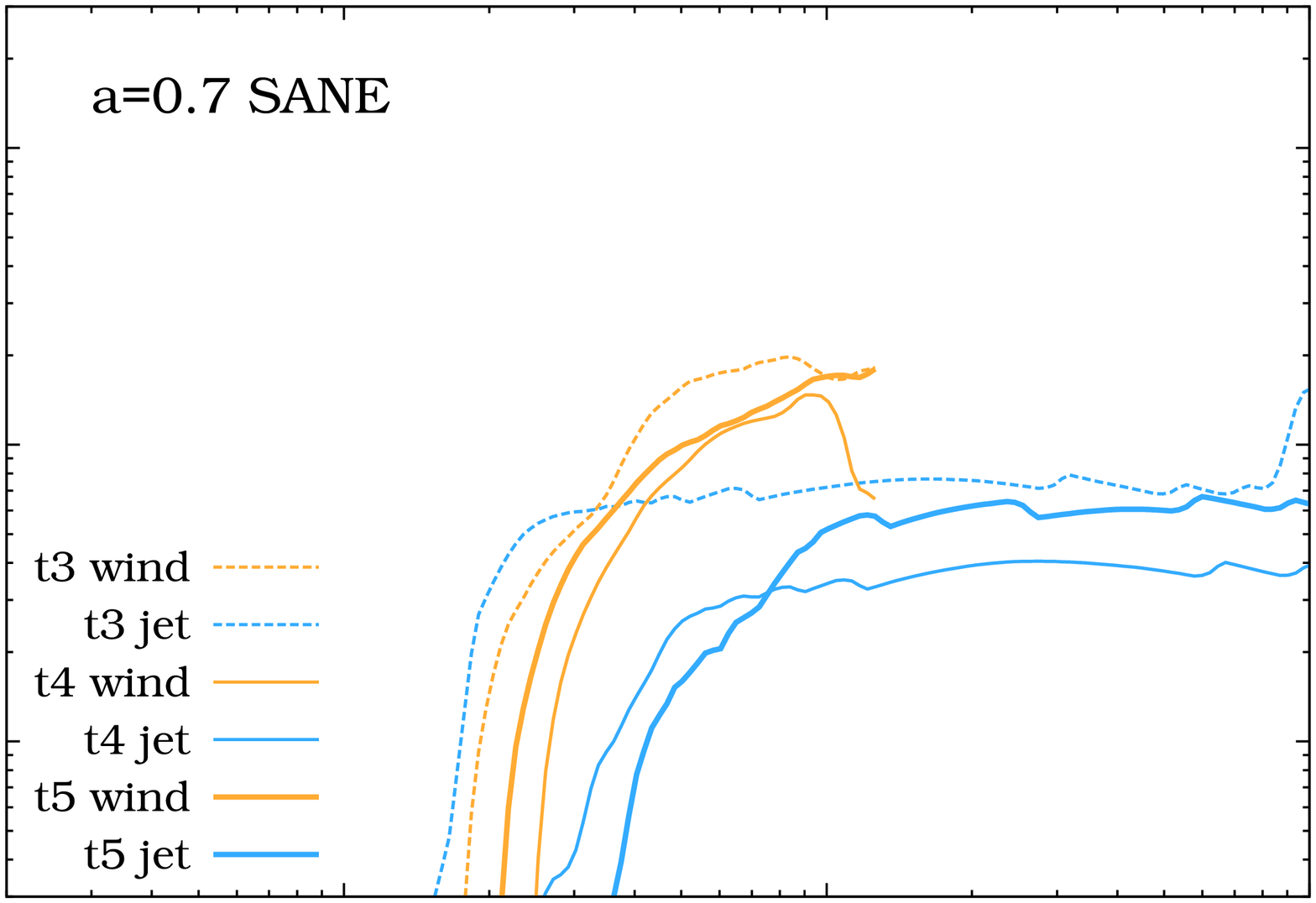}}\hspace{-.8cm}
\subfigure{\includegraphics[height=.35\textwidth,angle=270]{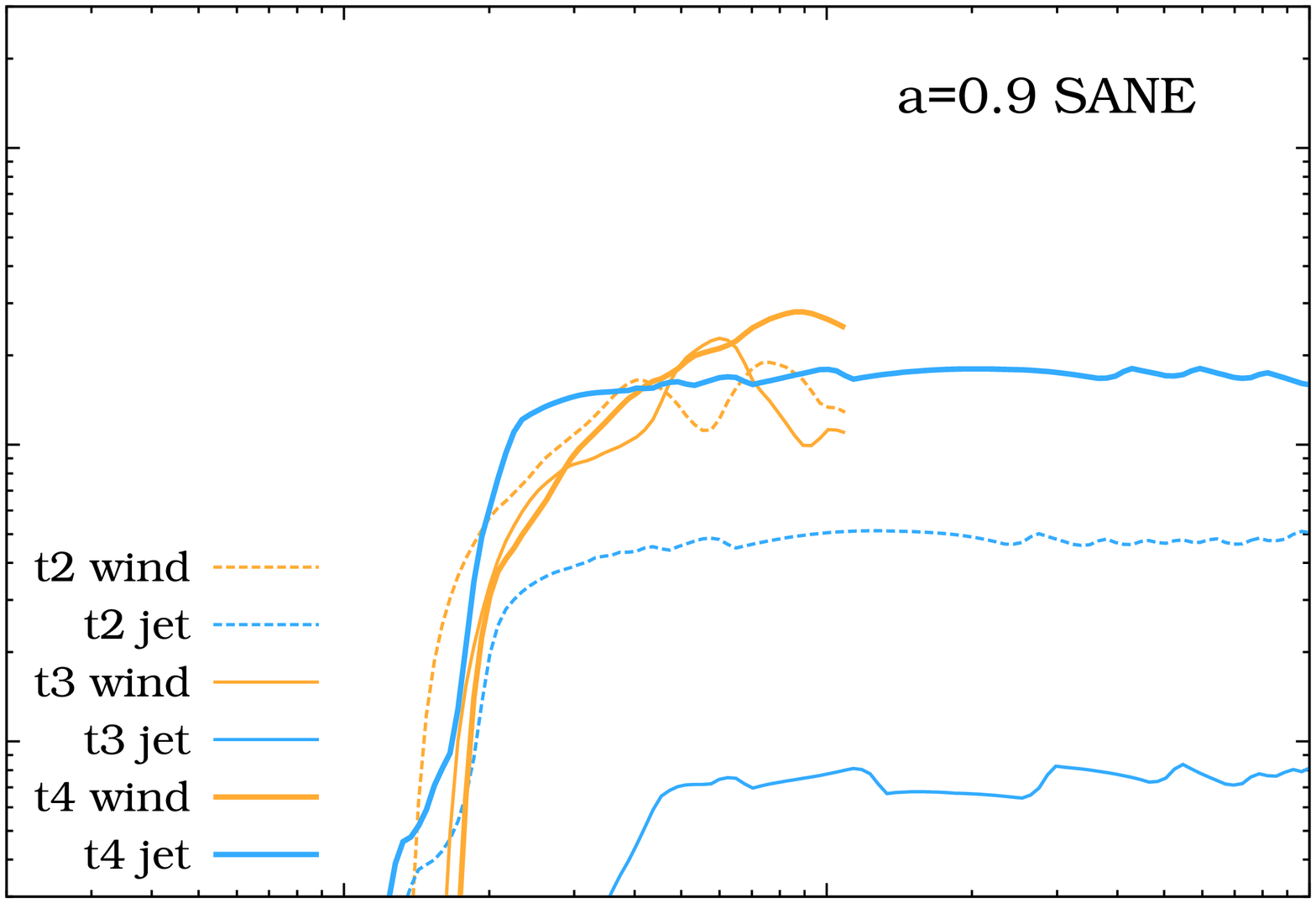}}\vspace{-.8cm}\\
\subfigure{\includegraphics[height=.35\textwidth,angle=270]{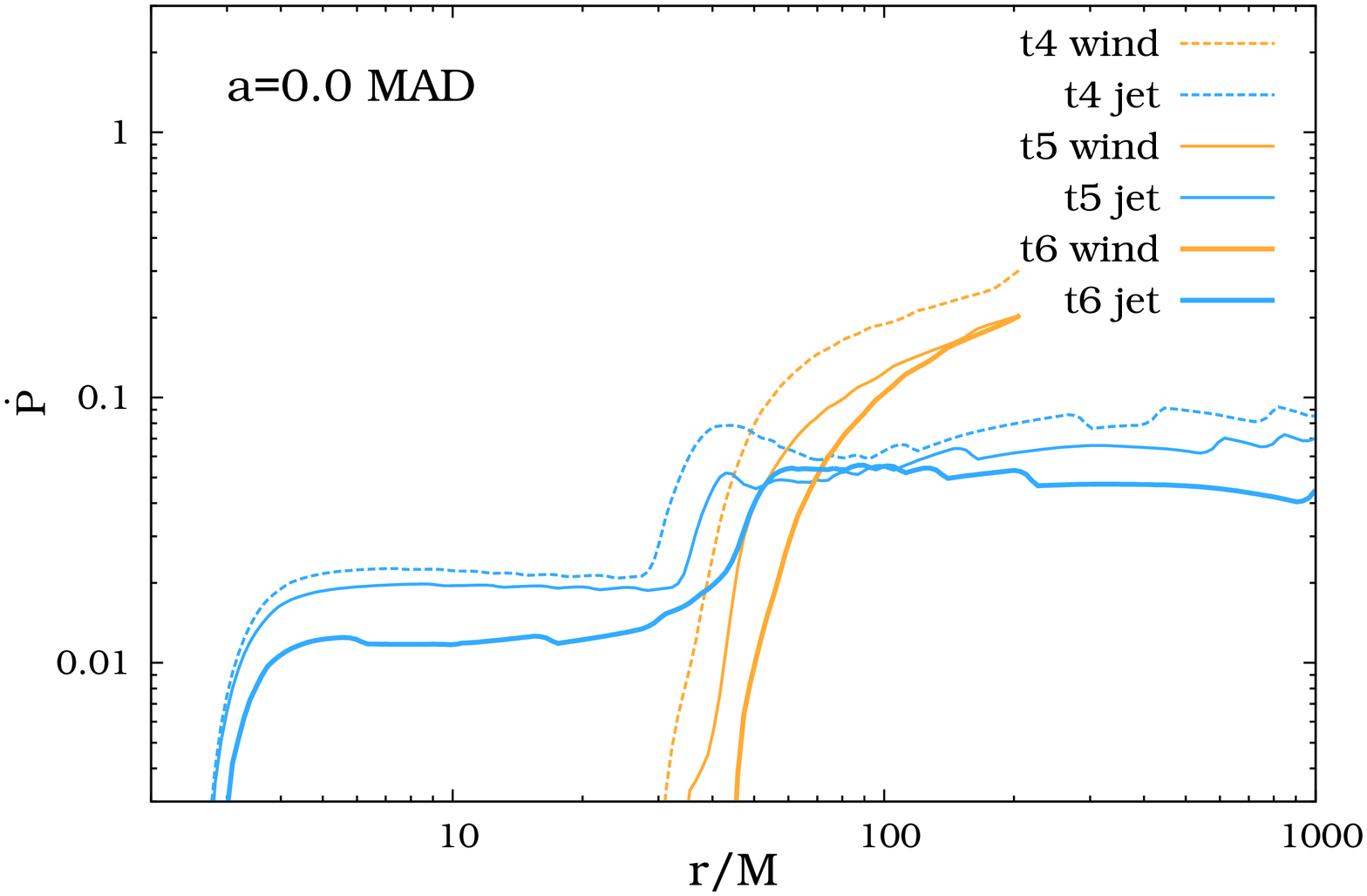}}\hspace{-.8cm}
\subfigure{\includegraphics[height=.35\textwidth,angle=270]{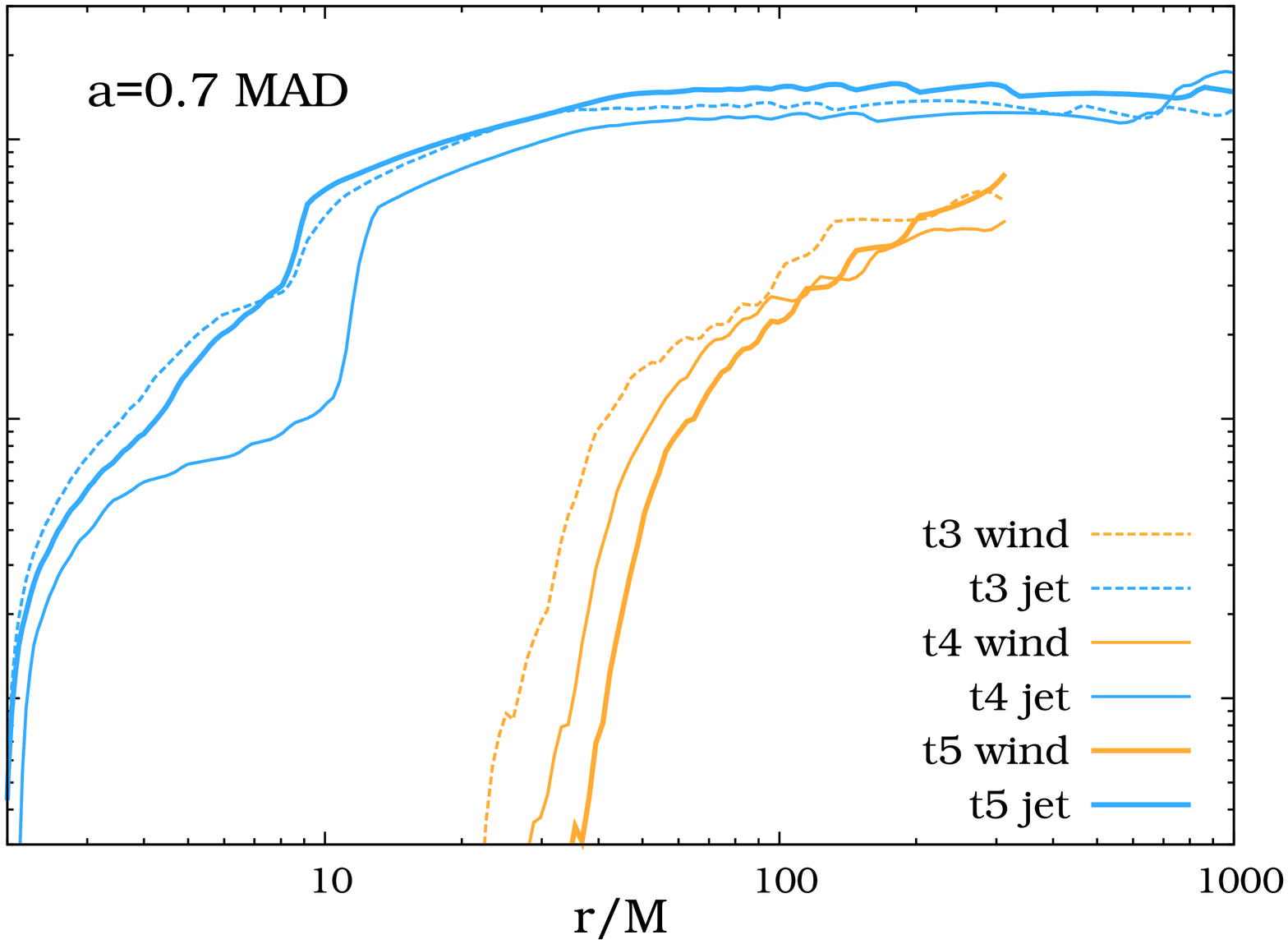}}\hspace{-.8cm}
\subfigure{\includegraphics[height=.35\textwidth,angle=270]{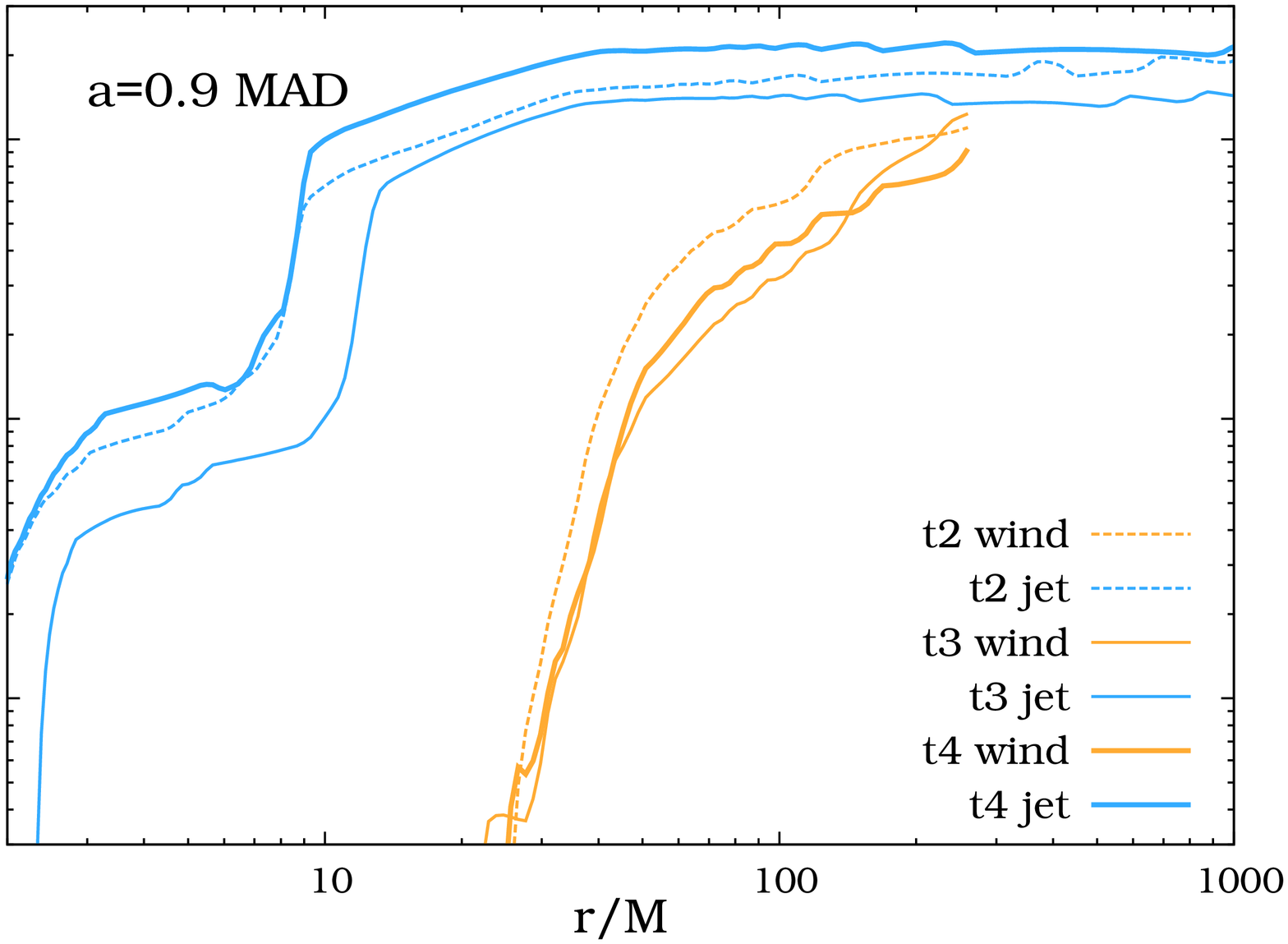}}
\caption{Similar to Figs.~\ref{f.tNedot} and \ref{f.tNmdot} but for fluxes of momentum ($\dot P$) for SANE (top) and MAD
  (bottom) simulations.}
\label{f.tNpdot}
\end{figure*}

For the mildly-relativisitic wind region the dominant component of gas velocity is
the azimuthal component ($v\approx r^{-1/2}$). 
Fig.~\ref{f.tNpdot} shows the radial profiles of the integrated
momentum flux $\dot P$. Indeed, the momentum lost
in the wind (orange lines) resembles profiles of the rest-mass lost in that region
(Fig.~\ref{f.tNmdot}). Its magnitude is noticeably smaller
and follows the rescaling with characteristic gas velocity, e.g., at $r=100$ is a factor of $v=0.1$ lower. 
As a result of the radial dependence of the azimuthal velocity the profiles are less steep than for the rest-mass flux.

The gas in the jet region is highly relativistic (Fig.~\ref{f.dengam}) 
and its velocity is dominated by the radial component. In a similar way, the radial 
profiles of momentum lost in the jet region should resemble the rest-mass loss rates
with the scaling factor depending on the gas velocity. For the jet region with the characteristic Lorentz factor
$u^t\approx 2$ this factor is close to unity. Fig.~\ref{f.tNpdot} shows that indeed the momentum
lost in the jet region (blue lines) is quantitatively similar to the profiles of rest-mass lost in this region.
The fact that both are constant in radius for $r\gtrsim 50$ proves that the characteristic velocity
in the jet region does not change.

\subsection{Approximate model of outflows}
\label{s.fits}

\begin{figure*}
  \centering
\includegraphics[width=.75\textwidth]{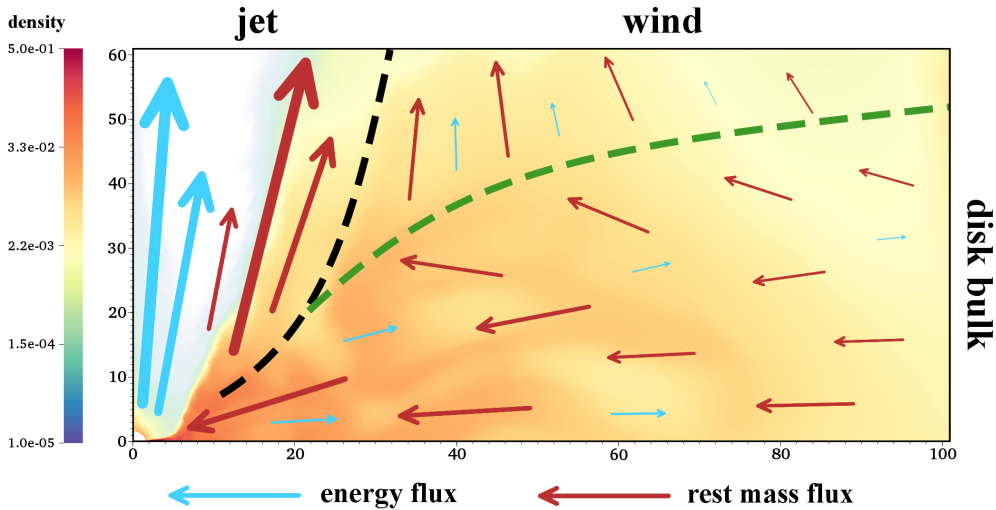}
\caption{Schematical picture of the disc structure near a rotating BH.}
  \label{f.scheme}
\end{figure*}

Outflows in a typical strongly magnetized ADAF around a rotating BH
can be divided into a slow wind at large radii and a relativistic jet
along the poles.  Such a structure is schematically shown in
Fig.~\ref{f.scheme}. Most of the energy is lost via the jet; for large
enough BH spins, the power of the jet may even exceed the total energy
budget $\dot{M}c^2$ of the accretion flow. Mass is lost through both
the wind and the ``cocoon'' surrounding the jet.  At large radii, mass
loss is dominated by the wind.

In the case of the jet, as discussed in previous sections, the energy
and mass outflow are sensitive both to the BH spin as well as the
magnetic flux at the horizon.  Outflows in the wind also depend on
these parameters, but less less sensitively.

\begin{figure}
\centering
\subfigure{\includegraphics[height=.45\textwidth,angle=270]{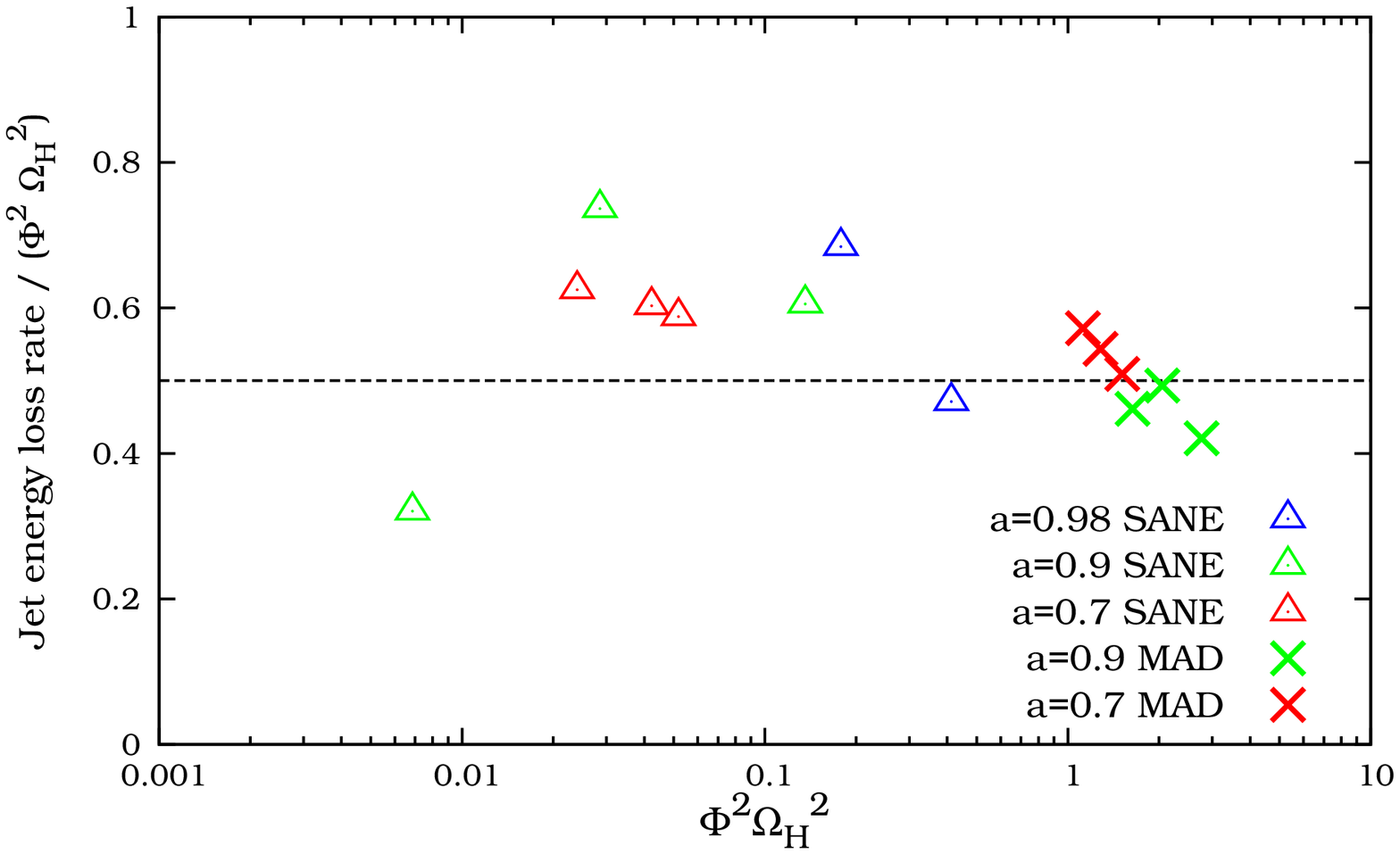}}
\subfigure{\includegraphics[height=.45\textwidth,angle=270]{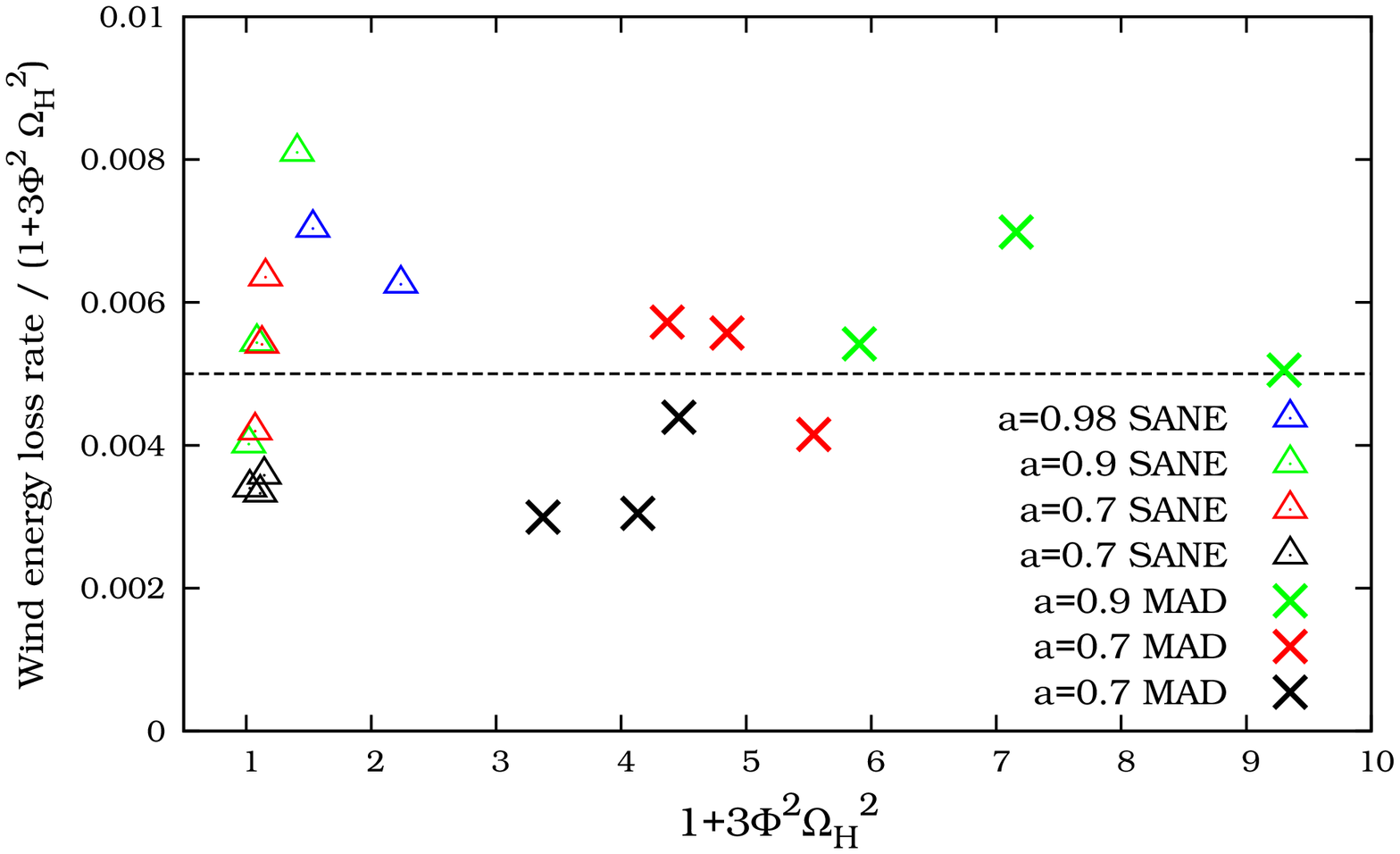}}
\subfigure{\includegraphics[height=.45\textwidth,angle=270]{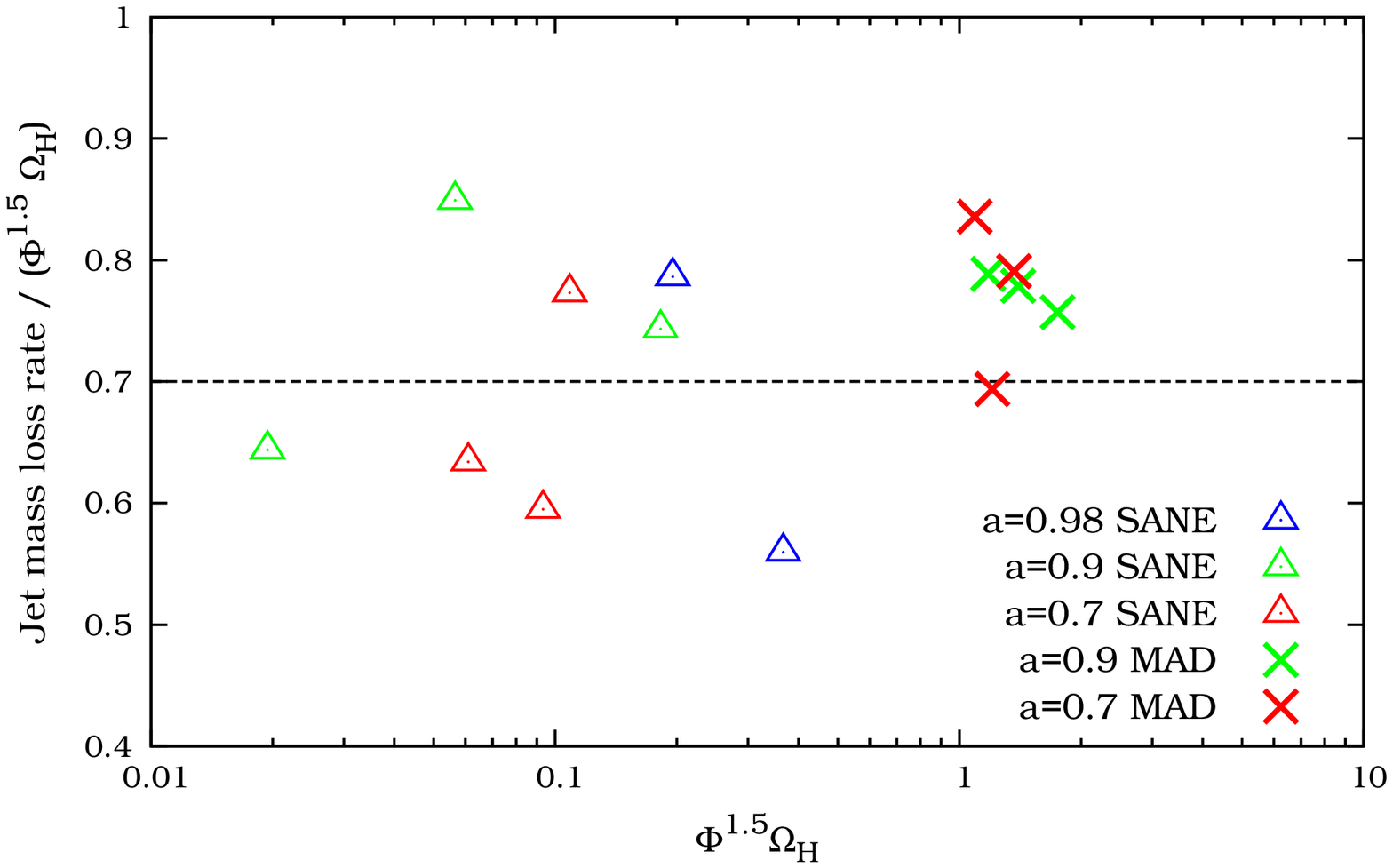}}
\caption{Coefficients in fitting formulae (top to
  bottom: eqs.~(\ref{e.ejet}), (\ref{e.ewind}) and (\ref{e.mjet}) for various models. The horizontal dashed lines
show the values chosen for the fits.}
\label{f.fitcoeffs}
\end{figure}

According to the BZ model (Section \ref{s.bz}) the power extracted
from the BH (eq. \ref{eq.bz}), $P_{\rm BH}$, is proportional to the
square of the magnetic flux threading the horizon $\Phi_{\rm BH}$ and
the square of the angular velocity of the BH horizon $\Omega_{\rm
  H}$. Using this scaling as a guide and keeping in mind the
discussion in Sections \ref{s.energy}--\ref{s.momentum}, we construct
here simple empirical fits which reasonably reproduce the simulation
results. Except for $\dot{M}_{\rm wind}$, all
the other outflow rates are good to within a factor of order unity
(Fig.~\ref{f.fitcoeffs}). This is true even for $\dot{P}_{\rm wind}$
so long as $s$ is not very different from 1/2.

All the estimates given in this subsection are normalized by the mass
accretion rate on the BH, $\dot{M}_{\rm BH}$. Thus, energy outflow
rates are given in units of $\dot{M}_{\rm BH}c^2$, momentum outflow
rates in units of $\dot{M}_{\rm BH}c$, and mass outflow rates
in units of $\dot{M}_{\rm BH}$. 

The cumulative energy, mass and momentum outflow at radius $r$ may be
written as a sum of contributions from the jet and the wind, 
\bea
\hspace{2.5cm} \dot E&=&\dot E_{\rm jet}+\dot E_{\rm wind},\\ \dot
M&=&\dot M_{\rm jet}+\dot M_{\rm wind},\\ \dot P&=&\dot P_{\rm
  jet}+\dot P_{\rm wind}.  \eea As discussed in
Section~\ref{s.scalings}, our simulations provide reliable estimates
of the two energy fluxes. The energy loss from the system is dominated
by the outflow in the jet (Fig.~\ref{f.tNedot}) which is driven by
energy extracted from the BH through the BZ process. We approximate
the jet energy as \be
\label{e.ejet}
\dot E_{\rm jet} \approx 0.5\, \tilde \Phi_{\rm BH}^2\, \tilde
\Omega_{\rm H}^2,\\ \ee where \be \tilde\Phi_{\rm BH}=\Phi_{\rm BH}/50
\ee is the magnetic flux threading the horizon, normalized by the
characteristic value for a MAD disc, and \be \tilde\Omega_{\rm
  H}=\Omega_{\rm H}/0.2 \ee is the horizon angular velocity
(eq.~\ref{eq.omhor}), normalized by the angular velocity for a spin
$a_*=0.7$ BH $\Omega_{\rm H}(a_*=0.7)\approx0.2$. As the top panel of
Fig.~\ref{f.fitcoeffs} shows, the coefficient 0.5 in
eq.~(\ref{e.ejet}) provides a good fit to the results presented in
Section~\ref{s.energy}.\footnote{In Section~\ref{s.angleintegrated} we
  discussed energy flux in jet for $a_*=0.7$ MAD disc having
  efficiency $\eta=70\%$. This is consistent with eq.~(\ref{e.ejet})
  because the magnetic flux for this simulation $\Phi_{\rm BH}\approx
  60$ and the $\tilde \Phi_{\rm BH}^2$ factor is not unity.}
 
The energy loss in the wind is much weaker (Fig.~\ref{f.tNedot}) and
we estimate it as, \be \dot E_{\rm wind} \approx 0.005(1+ 3 \tilde
\Phi_{\rm BH}^2\tilde \Omega_{\rm H}^2).
\label{e.ewind}
\ee where the empirical factor $(1+3 \tilde \Phi_{\rm BH}^2\tilde
\Omega_{\rm H}^2)$ accounts for the increase of the wind power with BH
spin and magnetic flux (see the bottom panel of
Fig.~\ref{f.edotvsflux}). The middle panel of Fig.~\ref{f.fitcoeffs}
validates the choice of the coefficient.

The mass loss rates we measure are robust only in the jet
region. Following the discussion in Section~\ref{s.energy}, we
approximate it as, \be
\label{e.mjet}
\dot M_{\rm jet} \approx 0.7\, \tilde \Phi_{\rm BH}^{1.5}\, \tilde
\Omega_{\rm H}, \ee where the coefficient $0.7$ is our best guess from
the simulation results (see bottom panel of Fig.~\ref{f.fitcoeffs}).

The radial profiles of the mass loss rate in the wind in the various
simulations are not converged in time and are therefore less
reliable. Even estimating the power-law index $s$ and the
characteristic radius $r_{\rm in}$ (eq.~\ref{e.adios}) is difficult
because the region where the profiles show power-law behavior is close
to the radius $r_{\rm conv}$ defined in eq.~(\ref{eq.rconv}). Roughly,
it appears that $0.5<s<0.7$ \citep[in agreement with][]{YWB12b}.
Let us write, \be
\label{e.mwind}
\dot M_{\rm wind} \approx \left(\frac{r}{r_{\rm in}}\right)^s, \ee
where $r_{\rm in}$ is the radius where the mass flux in the wind
equals the net accretion rate on the BH: $\dot M_{\rm wind}(r_{\rm
  in})=1$.  From Fig.~\ref{f.tNmdot} it appears that $r_{\rm
  in}\gtrsim 100$ and that it tends to be smaller (outflows stronger)
for rotating BHs. Better estimation of $r_{\rm in}$ is not possible
due to the limitations of our simulations.

Finally, the fluxes of momentum in the jet and wind are approximately
related to the corresponding mass loss rates by, \bea
\hspace{2.5cm}\dot P_{\rm jet}&\approx&\dot E_{jet},\\ \dot P_{\rm
  wind}&\approx&\dot M_{\rm wind}r^{-0.5}, \eea where we have assumed
$u^t\approx 2$ in the jet, that the characteristic velocity of the wind
originating at radius $r$ is
$v\approx v_\phi = r^{-0.5}$, and that $s>0.5$.

\section{Comparison with thin discs}
\label{s.thin}

So far we have shown that outflows of mass and energy are common in
thick accretion discs. This class of discs corresponds to radiatively
inefficient flows forming at very low \citep[$L\lesssim 0.01L_{\rm
    Edd}$,][]{nm08} or very high ($L\gtrsim L_{\rm Edd}$) accretion
rates; the latter regime is known as the slim disc
\citep{abr88}. However, many BHs in microquasars and galactic nuclei
are known to accrete with moderate accretion rates which correspond to
a radiatively efficient, geometrically thin disc. Do such discs
produce outflows similar to those found in the case of thick discs?

Jets in Galactic BHs are quenched around the time they change state
from the low-hard, presumably geometrically thick, to the high-soft
state corresponding to a geometrically thin disc
\citep[e.g.,][]{remillard+mcclintock06}. This fact may suggest that
relativistic outflows are not characteristic for such class of discs.

Numerical studies of geometrically thin discs have been limited so far
because of the requirement of including radiative transfer. All GRMHD
simulations of thin discs performed so far
\citep[e.g.,][]{2008ApJ...687L..25S,2010MNRAS.408..752P,2011ApJ...743..115N,2012MNRAS.424.2504Z}
are based on an artificial cooling function which drives discs to an
arbitrarily chosen entropy corresponding to a required given disc
thickness. More sophisticated treatment of radiation in the context of
thin accretion flows is being developed
\citep[e.g.,][]{2012ApJS..201....9F,sadowskietal12}.

To compare the power of outflowing mass and energy between thick and
thin discs we have performed two additional simulations which use the
cooling function as decribed in \cite{2012MNRAS.424.2504Z} to drive
the disc towards $h/r\approx 0.1$ which corresponds to $L\approx0.3
L_{\rm Edd}$. We tested two values of BH spin $a_*=0.0$ and
$a_*=0.9$. The initial magnetic field was set in a similar way to the
SANE simulations descibed earlier, i.e., its poloidal component formed
a set of counter-orientated loops. Details of the simulations are
given in Table~\ref{t.thinmodels}.

\begin{table}
\caption{Thin disc models}
\label{t.thinmodels}
\centering\begin{tabular}{@{}ccccc}
\hline
 Model & BH spin & Initial & Resolution & $t_{\rm final}$ \\
& ($a_*$) & magnetic field & &\\
\hline
 $a_*=0.0$ thin   & $0.0$    & multi-loop     & 264x64x64 & 50000\\
 $a_*=0.9$ thin   & $0.9$    & multi-loop     & 264x64x64 & 50000\\ 
\hline
\end{tabular}
 \end{table}

\begin{figure}
  \centering
\includegraphics[height=.95\columnwidth,angle=270]{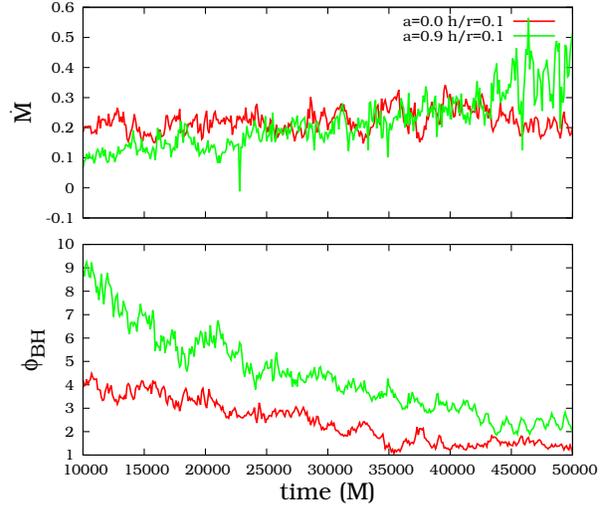}
\caption{Accretion rate (top) and vertical magnetic flux threading the
  horizon (bottom) history for thin disc simulations with $a_*=0.0$
  and $a_*=0.9$.}
  \label{f.fluxes_aN_thin}
\end{figure}

Fig.~\ref{f.fluxes_aN_thin} shows the history of the accretion rate
and the magnetic flux through the BH horizon for the thin disc
simulations. For both values of BH spin the accretion rate is roughly
contant. The magnetic flux, however, decreases with time. For both
simulations the non-dimensional flux parameter $\Phi_{\rm BH}$
(eq.~\ref{e.phibh}) is well below the MAD-characteristic
value\footnote{$\Phi_{\rm BH,MAD} \propto (h/r)^{0.5}$ \citep{sasha_swiftmad}.}  $\Phi_{\rm BH,MAD}=25$ reflecting
the fact that the magnetic pressure does not saturate and that the
discs are in the SANE state.

\begin{figure*}
  \centering
\renewcommand{\arraystretch}{0.5}
\begin{tabular}{MMMM}
& $|\dot m|$ &$|\dot e|$ & $|\dot e_{\rm tot}|$  \\
\begin{sideways}$a_*=0.0$ thin\end{sideways}&
\subfigure{\includegraphics[width=.3\textwidth]{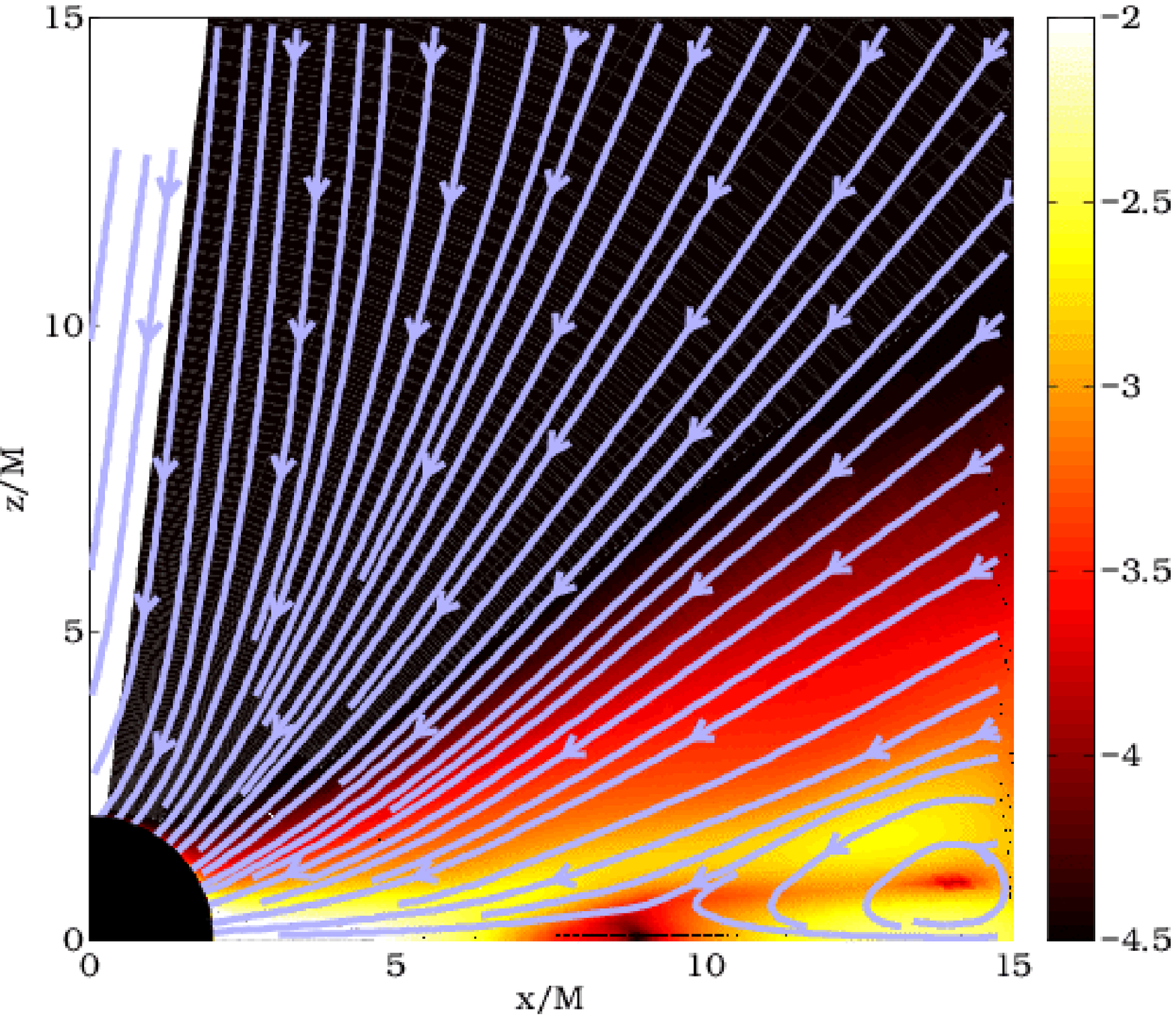}}&
\subfigure{\includegraphics[width=.3\textwidth]{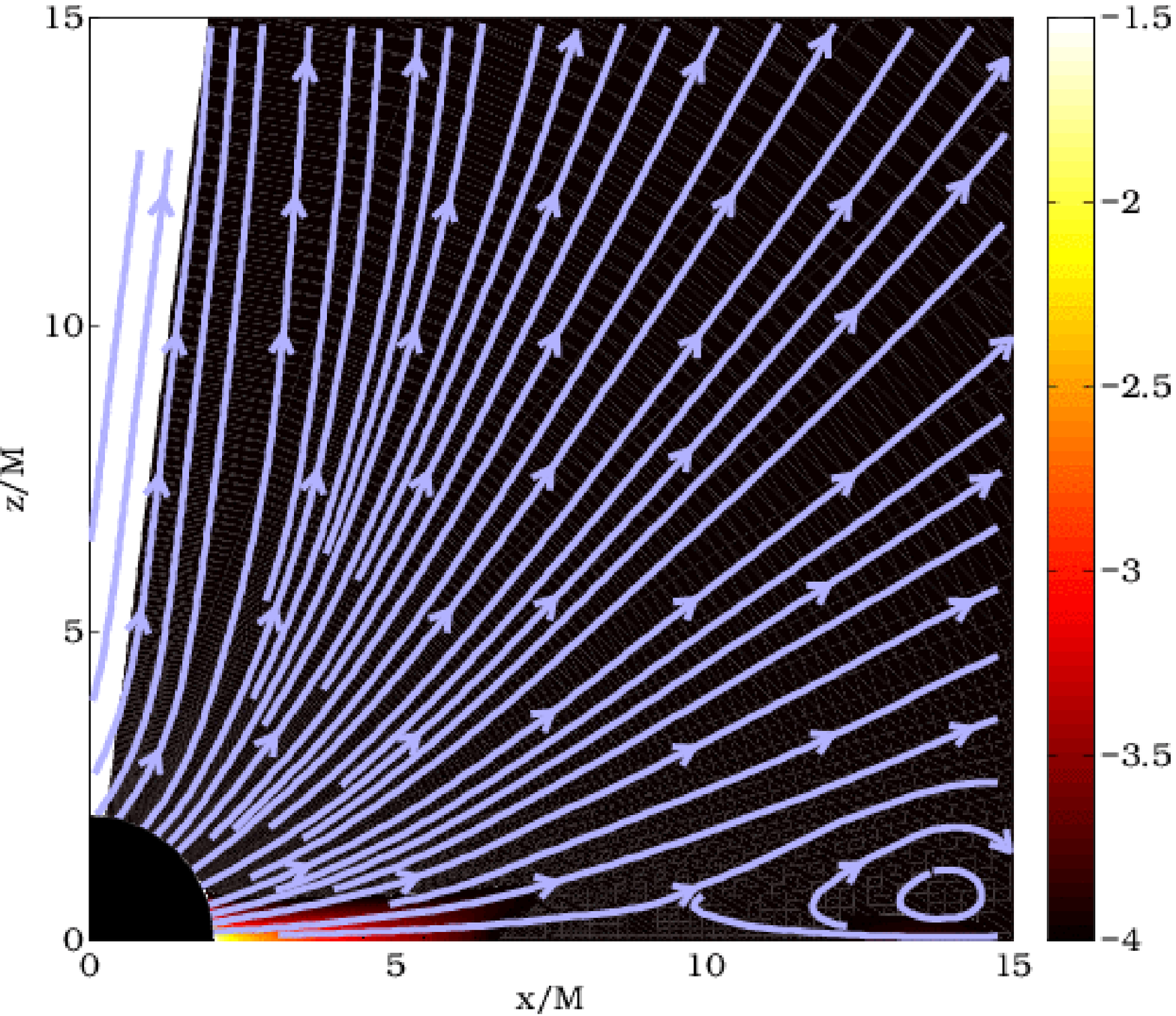}}&
\subfigure{\includegraphics[width=.3\textwidth]{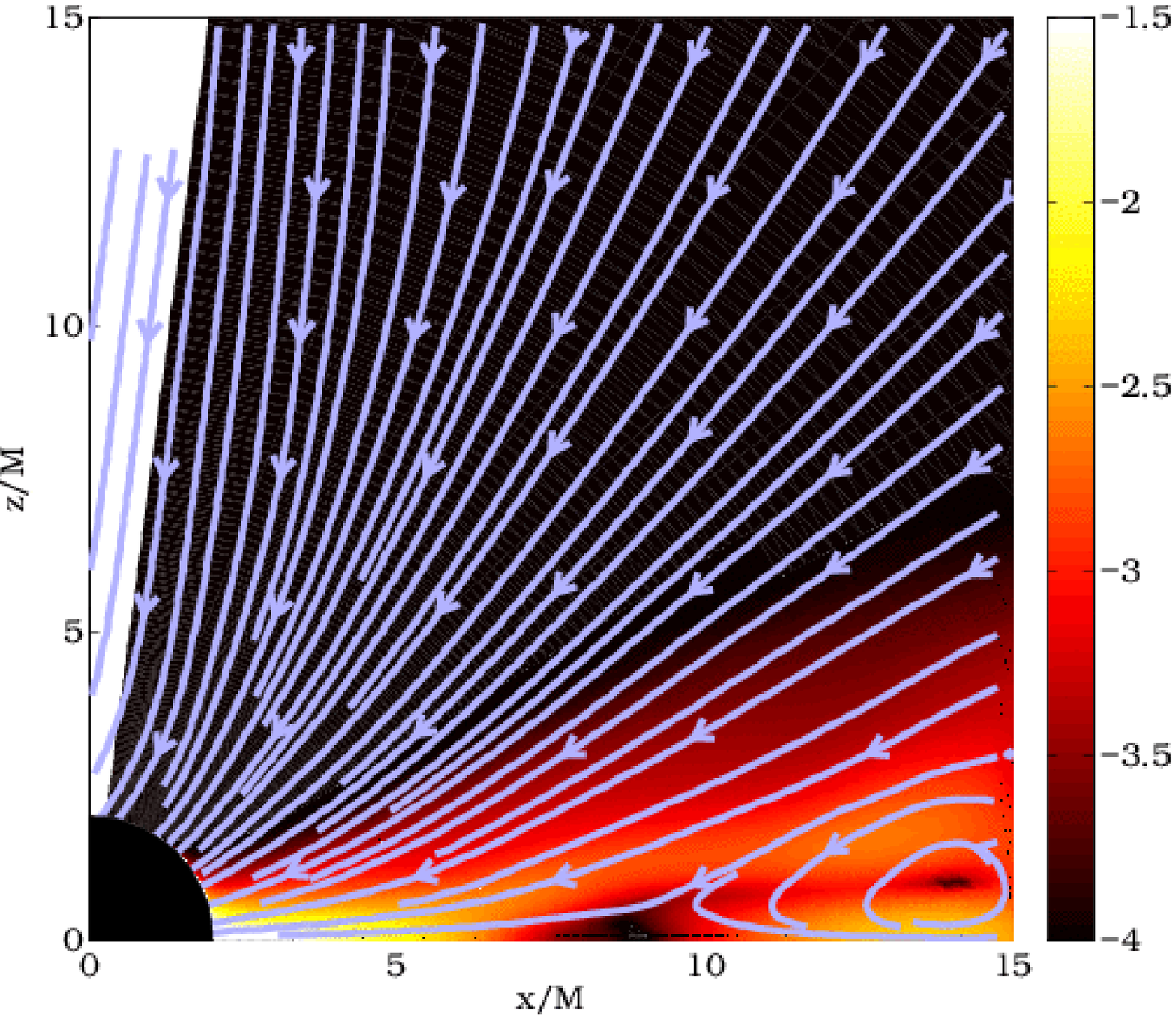}}\\
\begin{sideways}$a_*=0.9$ thin\end{sideways}&
\subfigure{\includegraphics[width=.3\textwidth]{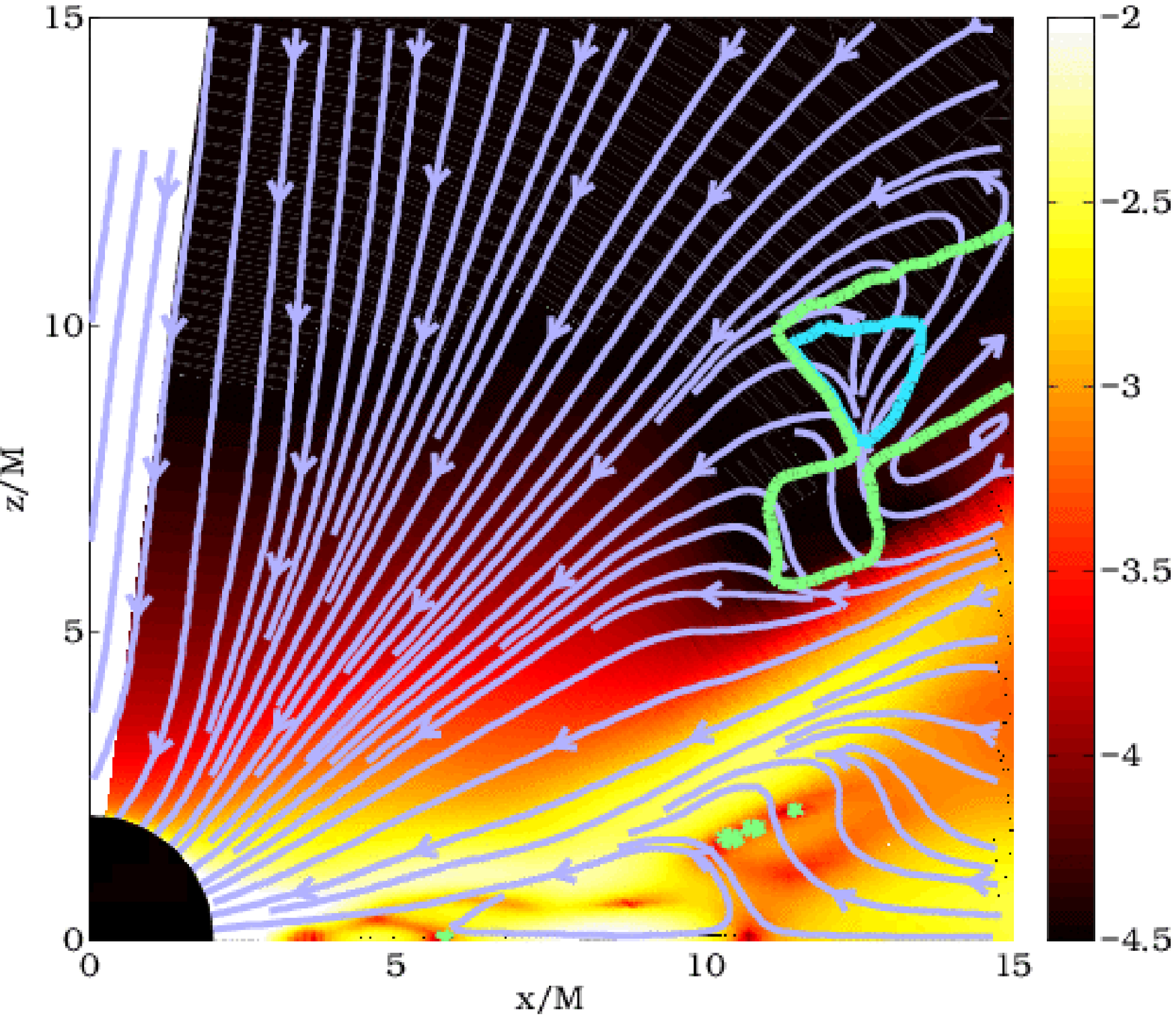}}&
\subfigure{\includegraphics[width=.3\textwidth]{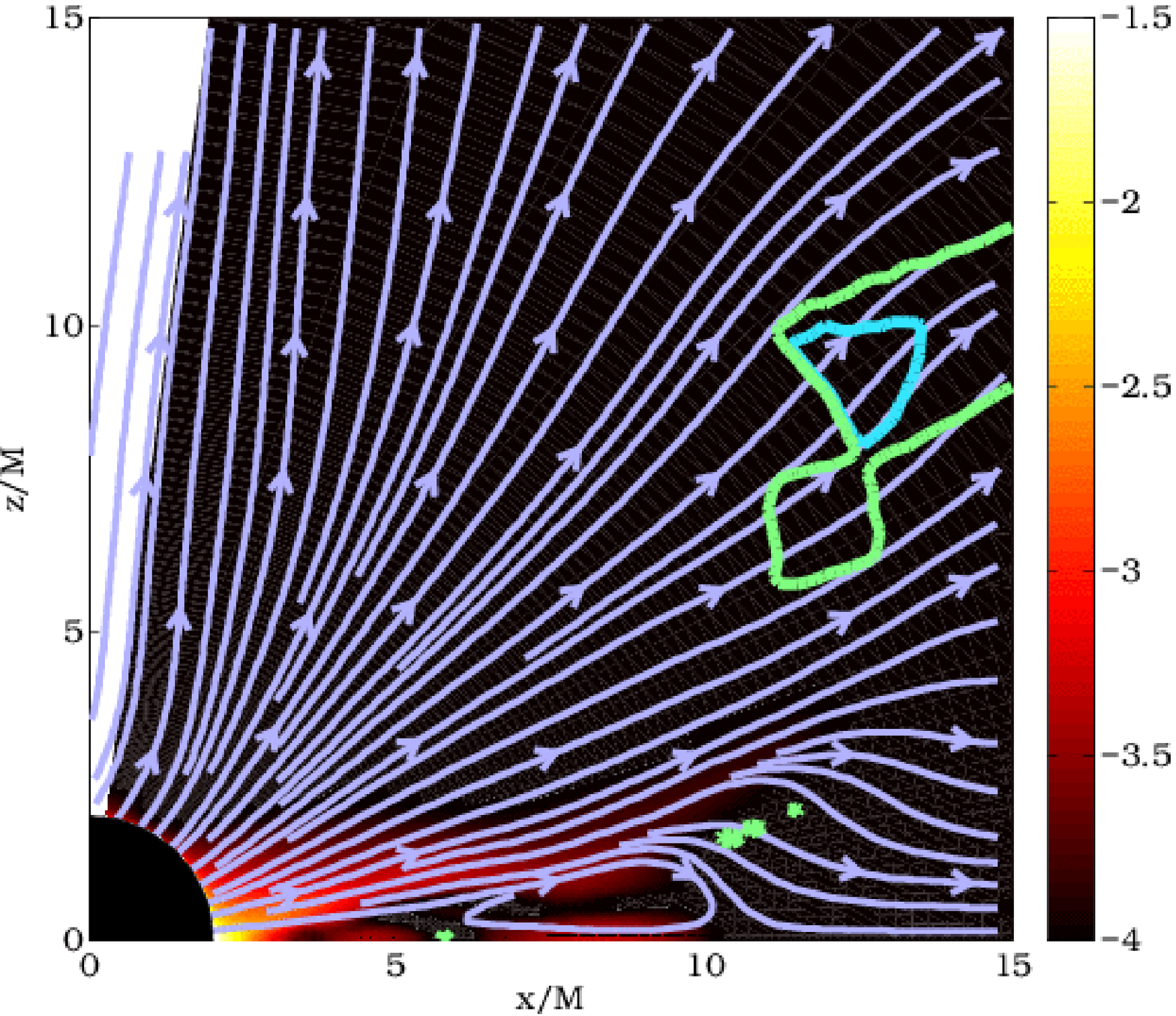}}&
\subfigure{\includegraphics[width=.3\textwidth]{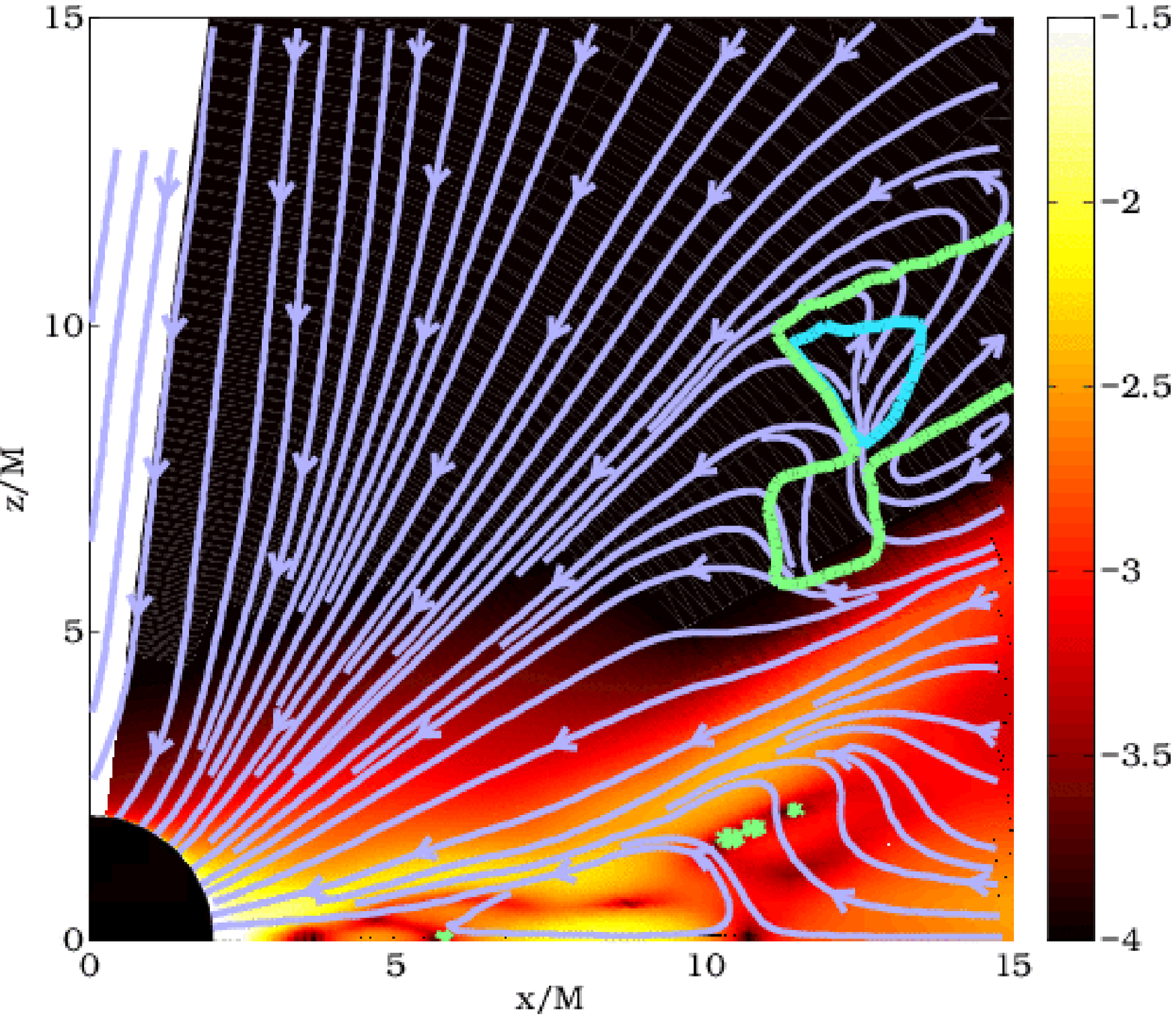}}\\
\end{tabular}
\caption{Similar to Fig.~\ref{f.streams} but for thin disc models with
  $a_*=0.0$ (top) and $a_*=0.9$ (bottom panels). }
  \label{f.streams.thin}
\end{figure*}

Fig.~\ref{f.streams.thin} shows the magnitude of the 
local rest mass flux
($\dot{m}$, Eq.~\ref{e.mdot}), the energy flux ($\dot{e}$, Eq.~\ref{e.edot}), and the total energy flux
($\dot{e}_{\rm tot}$, Eq.~\ref{e.Trt}) in the poloidal plane, averaged over $t=25000
\div 50000$, for thin disc simulations with $a_*=0.0$ and $0.9$,
respectively. Because of the much lower radial velocity in thin discs,
the range of inflow equilibrium in the disc region is limited to $r\lesssim
15$. Even with this limitation, we believe the simulations should be
sufficient to provide useful estimates of the outflow in a jet.  It is
thus significant that we see no sign of a jet in either of the two
simulations, suggesting that thin discs are not conducive to producing
relativistic jets. We also do not see any outflowing mass or energy in
a wind. However, this result is less significant since, even in the
geometrically thick ADAF runs described earlier, the wind usually
begins only at relatively large radii. Our thin disc simulations have not reached inflow equilibrium at such radii.

The thin disc simulations described here were set up with multiple
poloidal magnetic field loops which, as explained earlier, results in
unsaturated magnetic field around the BH and thus are destined to
produce weaker jets. Initializing the magnetic field with a single
loop, as in the MAD ADAF runs, should result in stronger magnetic flux
at the horizon. This would be the thin disc equivalent of the MAD
state. (Note that, historically, the original conceptual paper on the
MAD solution by \cite{narayan+mad} considered thin discs, whereas the
numerical simulations by \cite{igu03} which motivated this
paper dealt with ADAFs.) Such an experiment is worthwhile and is left
for future work. However, comparing the flow structure of the two thin
disc simulations discussed here (Fig.~\ref{f.streams.thin}) with the
SANE ADAF simulations discussed earlier (e.g., third row of
Fig.~\ref{f.streams}) already suggests that geometrically thin discs
are less efficient in generating relativistic outflows of rest mass
and energy than equivalent thick discs. Because of the very limited
region of inflow equilibrium in Fig.~\ref{f.streams.thin}, we cannot address
the efficiency of generating wind-like outflows except in the
innermost regions (where there is no wind).

\section{Discussion and Summary}
\label{s.discussion}

\begin{figure*}
  \centering
\includegraphics[height=.75\textwidth,angle=0]{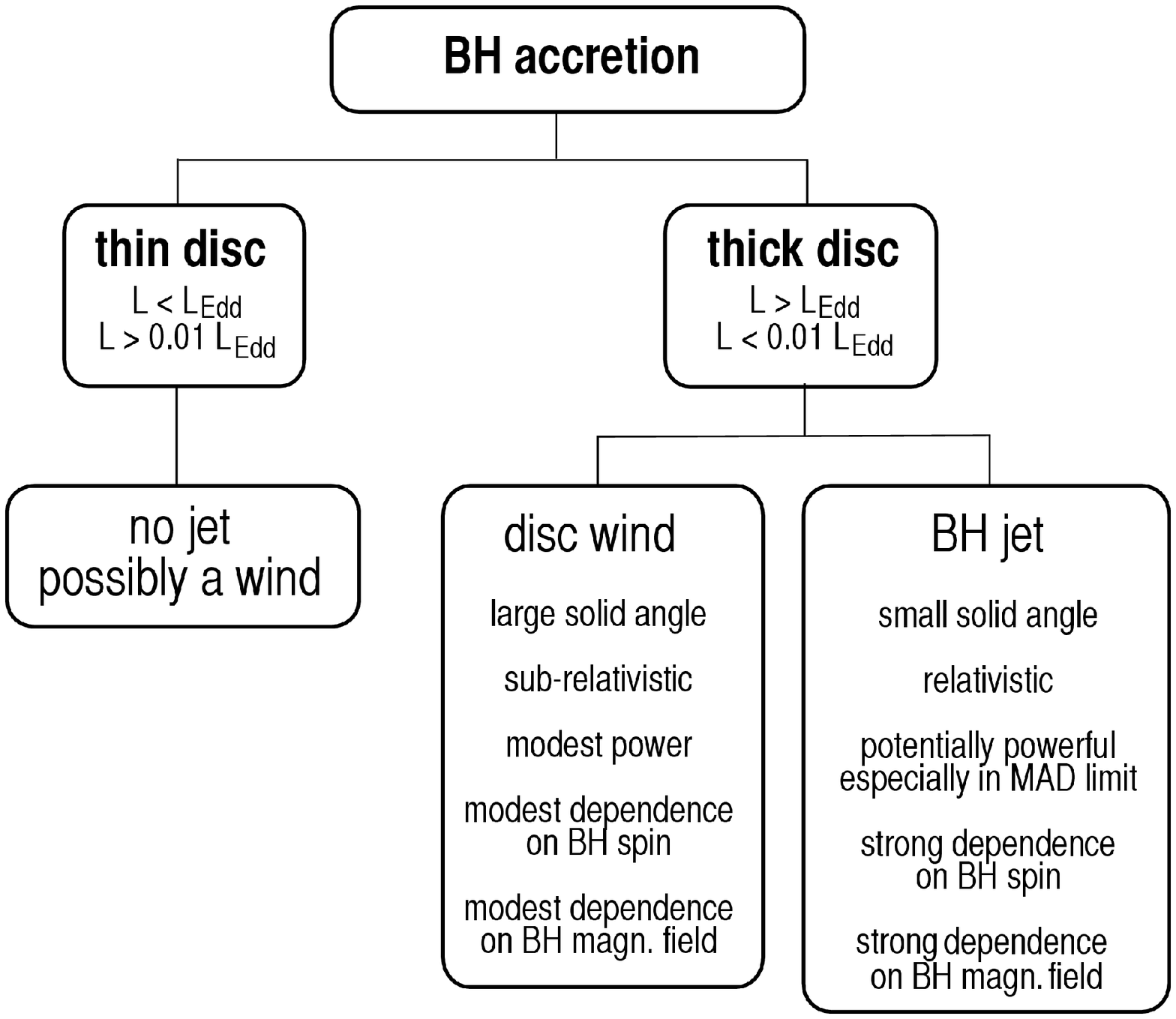}
\caption{Qualitative description of outflows in BH accretion discs.}
  \label{f.scheme.bhflows}
\end{figure*}

In this paper we presented a number of GRMHD simulations of magnetized
non-radiative geometrically-thick accretion discs around BHs, and used
them to investigate the outflow of energy, momentum and mass from
these systems.  The simulations covered a range of BH spin parameters
$a_*=a/M$ or, equivalently, BH horizon angular frequencies
$\Omega_{\rm H}$, and thus probed the effect of this important
parameter on outflows.  We also tested two initial configurations of
the magnetic field, which led respectively to configurations with a
weak magnetic field around the BH (SANE configuration) and a saturated
magnetic field (MAD configuration). This enabled us to study the
dependence of outflow properties on the magnetic flux $\Phi_{\rm BH}$
around the BH.

Each simulation was run for a long time in order to reach steady state
over as wide a range of radius as possible (see
Table~\ref{t.convradii} for estimates of the size of the inflow and outflow equilibrium region). In each
simulation, we divided the flow into an inflowing ``disc'', and an
outflowing ``wind'' and ``jet''. The latter two zones were
distinguished on the basis of their velocity at infinity. The jet
consists of all matter that flows out with $v\,\gtrsim\,0.3$ at
infinity and the wind is the rest of the outflowing matter with
asymptotic $v\,\lesssim\,0.3$. More precisely, the boundary is defined
by a critical Bernoulli parameter, $\mu_{\rm crit}=0.05$
(eq.~\ref{eq:mu}). Figure \ref{f.scheme} shows the three zones
schematically.

Our results are summarized in Section \ref{s.fits} in the form of
approximate fitting functions. In the case of the jet, we are able to
provide fairly reliable estimates of its outflow properties.
Specifically, we give fitting functions for the jet energy outflow
$\dot{E}_{\rm jet}$, momentum outflow $\dot{P}_{\rm jet}$ and mass
outflow $\dot{M}_{\rm jet}$. The energy and momentum outflow rates
vary proportional to $\Phi_{\rm BH}^2 \Omega_{\rm H}^2$, as predicted
by the BZ model.

The situation is less clear in the case of the wind. We believe our
estimates of the wind energy outflow $\dot{E}_{\rm wind}$ and, to a
lesser extent, the wind momentum outflow, $\dot{P}_{\rm wind}$, are
fairly reliable. However, the mass outflow in the wind is highly
uncertain. This is because mass outflow is dominated by large radii
and our simulations do not go out to a large enough radius to enable
us to extrapolate reliably to the putative outer edge of the accretion
flow at (say) the Bondi radius.  Specifically, in equation
(\ref{e.mwind}), we cannot determine the value of $s$ with any
certainty. The situation is exacerbated by the fact that, as the
simulation progresses in time, the wind launch radius appears to track
the limiting radius of inflow equilibrium in the disc. That is, the wind seems always to begin
within a factor of about two of that radius, but this is
where the results are least reliable. Modulo this serious uncertainty,
it appears that the energy, momentum and mass outflow rates in the
wind have some dependence on the BH angular velocity $\Omega _{\rm H}$
and the BH magnetic flux $\Phi _{\rm BH}$.

SMBH ``feedback'' in galaxy clusters is believed to play an important
role in keeping the cluster gas hot and preventing catastrophic star
formation \citep{CiottiOstriker01,
  BruggenKaiser02, RuszkowskiBegelman02, ChurazovEtAl04}. Observational evidence suggests that the feedback
occurs via relativistic jets. The scaling relations given in this
paper for $\dot{E}_{\rm jet}$ and $\dot{P}_{\rm jet}$ are likely to be
useful in this context.

SMBH feedback also appears to occur inside AGN host galaxies
\citep{SilkRees98, King03, HopkinsEtAl09}, but in this case it is less
obvious that the jet is important.  Jets are much too collimated to
have much of an effect, e.g., our jets subtend less than 10\% of the
solid angle around the BH at a radius $r\sim10^2$, and the solid angle
at larger radii is much less because of continued collimation.  On the
other hand, the wind in our simulations covers nearly 50\% of the
solid angle around the BH (the other 50\% being covered by the thick
disc). Thus, the wind is likely to have a strong effect on the host
galaxy, especially since, unlike radiation which can escape through
optically thin regions of the galaxy, a wind is certain to deposit all
its energy and momentum in the interstellar medium of the galaxy.  The
scaling relations given in this paper for $\dot{E}_{\rm wind}$ and
$\dot{P}_{\rm wind}$ (of which the energy scaling is more reliable)
are thus relevant.

We also presented simulations of artifically cooled thin accretion
discs ($h/r\approx 0.1$) for two BH spins: $a_*=0$, 0.9.  Neither
simulation showed any evidence for either a jet or a wind. We note
that both simulations were in the SANE regime and the radius of
inflow equilibrium was only $r\approx 15$. Nevertheless, our preliminary
conclusion from these simulations is that thin discs are significantly
less efficient in producing relativisic jets. We cannot say anything
about winds from thin discs.

The flowchart in Fig.~\ref{f.scheme.bhflows} summarizes our
qualitative conclusions from this study of outflows from accreting
BHs. Geometrically thin discs appear not to produce relativisitic
outflows, nor do they have magnetically-driven winds from small
radii. Mass loss through magnetically- or line-driven winds at larger
radii is certainly possible, but is beyond the scope of this
work. Geometrically thick ADAFs exhibit two kinds of outflows: wind
and jet. The wind originates from relatively large radii in the flow.
It is responsible for most of the mass outflow, and carries a modest
amount of energy and momentum. The properties of the wind depend
relatively weakly on the BH spin and the magnetic flux at the
horizon. The wind covers a large solid angle and is an excellent
agency for feedback.  The jet is a relativistic outflow which emerges
from radii close to the BH horizon and is highly collimated along the
poles. Its properties depend strongly on both the BH spin and the
magnetic flux at the horizon. In favorable cases --- rapid BH spin and
large magnetic flux (MAD limit) -- the flux of energy and momentum in
the jet are very much larger than the corresponding fluxes in the
wind. However, because of the strong collimation, jet feedback is
probably important only on the largest scales, e.g., galaxy clusters.

Apart from the mass accretion rate $\dot{M}_{\rm BH}$ and the
BH angular velocity $\Omega_{\rm H}$, the present study confirms the
importance of a third parameter, the dimensionless magnetic flux
$\Phi_{\rm BH}$ at the BH horizon. In principle, different systems
might have different values of $\Phi_{\rm BH}$, making it much more
difficult to come up with useful predictions for individual systems.
One mitigating factor is that the numerical simulations described here
as well as other recent work
\citep{tchekh+11,Tchekhovskoy+12b,narayan+12a,MTB+13} suggest that
ADAFs easily reach the MAD state in which the magnetic flux saturates
at its limiting value $\Phi_{\rm BH}\sim50-60$. Thus, it is
conceivable that the majority of observed systems would be in the MAD
limit and have a similar value of $\Phi_{\rm BH}$, thus eliminating
this uncertainty. Note that plenty of magnetic flux is available in
the external medium, more than enough to saturate the field at the
horizon \citep{narayan+mad}, and theoretical arguments suggest that a
geometrically thick ADAF is likely to transport the magnetic field
inward efficiently \citep{guilet12, guilet13}.

We conclude with two caveats.  First, our formulae for the energy and
momentum outflow in the jet and wind are expressed in terms of the net
mass accretion rate on the BH, $\dot{M}_{\rm BH}$. In practice, to use
these relations, one needs to be able to estimate $\dot{M}_{\rm BH}$
given the mass supply rate at the feeding zone of the BH, say the
Bondi radius $R_{\rm B}\,\gtrsim\,10^5R_G$.  Unfortunately, there is
considerable uncertainty on this issue.  In the language of
equation~\ref{e.adios}, if the Bondi accretion rate at $R=R_{\rm B}$
is $\dot{M}_{\rm B}$, then very roughly we expect
\begin{equation}
\dot{M}_{\rm B} = \dot{M}_{\rm in}(r_{\rm B})
=\dot{M}_{\rm BH} \left[1+\left(\frac{r_{\rm B}}{r_{\rm in}}
\right)^s\right].
\end{equation}
Depending on the values of $s$ and $r_{\rm in}$, the ratio
$\dot{M}_{\rm BH}/\dot{M}_{\rm B}$ could vary over a wide range.  What
can be done about this uncertainty? There is hope that $r_{\rm in}$
could be determined via simulations. However, it might be much harder
to obtain a reliable estimate of $s$, at least through GRMHD simulations. 
Some authors have obtained fairly good estimates of $s$ via large dynamic range hydro
simulations \citep[e.g.,][obtain $s\sim 0.5$]{YWB12b}, but it is uncertain if their
results will carry over to MHD. This problem needs to be resolved before
accretion disc simulation results can be used for galaxy feedback
studies.

Second, the simulations presented here correspond to non-radiative
flows, whereas ADAFs in nature do produce at least some
radiation. \cite{dibi+12} included radiative cooling in their simulations and found that
it introduces significant differences in the structure of the flow
near the black hole for accretion rates above $10^{-7}$ Eddington.
Their result, however, depends on assumptions regarding the electron
temperature in the two-temperature plasma. More conservatively, one
might expect noticeable effects from radiation for accretion rates
$\gtrsim 10^{-4}$ Eddington. Radiation can also have a strong effect at and beyond the Bondi radius. Specifically, X-rays
produced by inverse Compton scattering near the black hole can heat
gas near the Bondi radius and significantly modify its thermal
properties \citep{sazonov+04,yuan+09}. In turn, this can
significantly modify the mass supply at the Bondi radius, thereby
forming a feedback loop.

\section{Acknowledgements}
A.S. and R.N. were supported in part by NASA grant NNX11AE16G. 
We acknowledge NSF support via XSEDE
resources, and NASA support via High-End
Computing (HEC) Programme through the NASA Advanced Supercomputing (NAS)
Division.

\bibliographystyle{mn2e}
\bibliography{mybib}

\end{document}